\documentclass[%
reprint,
nofootinbib,
nobibnotes,
amsmath,amssymb,
aps,
]{revtex4-2}
\tighten

\def\gsim{\mathrel{\rlap{\raise 0.511ex \hbox{$>$}}{\lower 0.511ex\hbox{$\sim$}}}}
\def\lsim{\mathrel{\rlap{\raise 0.511ex \hbox{$<$}}{\lower 0.511ex\hbox{$\sim$}}}}
 
\usepackage{color}
\usepackage{graphicx}% Include figure files
\usepackage{dcolumn}% Align table columns on decimal point
\usepackage{bm}% bold math
\usepackage{amsmath, amsthm, amssymb, amsfonts}
\usepackage[normalem]{ulem} % Provide \sout
\usepackage{hyperref}% add hypertext capabilities
\usepackage{diagbox}
\usepackage{multirow}
\usepackage{array, booktabs}
\usepackage{float}
\usepackage{soul}
\usepackage[normalem]{ulem}

\definecolor{violet}{rgb}{0.62, 0.0, 1.0}
\definecolor{teal}{rgb}{0., 0.5, 0.5}

\newcommand{\STAB}[1]{\begin{tabular}{@{}c@{}}#1\end{tabular}}
\begin{document}

\preprint{APS/123-QED}

\title{Systematic errors due to \emph{quasi}-universal relations in binary neutron stars and their correction for unbiased model selection}

\author{Rahul Kashyap$^1$}
\email{rkk5314@psu.edu}
\author{Arnab Dhani$^1$}
\email{aud371@psu.edu }
\author{Bangalore Sathyaprakash$^{1,2,3}$}%
\email{bss25@psu.edu}
 \affiliation{$^1$Institute for Gravitation and the Cosmos, Department of Physics, The Pennsylvania State University, University Park PA 16802 USA}
\affiliation{$^2$Department of Astronomy and Astrophysics, The Pennsylvania State University, University Park PA 16802 USA}
\affiliation{$^3$School of Physics and Astronomy, Cardiff University, Cardiff, CF24 3AA, United Kingdom}

\date{\today}% It is always \today, today,

% Abstracts should be easy to read and comprehend. Keep them free of acronyms or complicated jargon.
\begin{abstract}
Inference of the equation-of-state (EOS) of dense nuclear matter in neutron-star cores is a principal science goal of X-ray and gravitational-wave observations of neutron stars. In particular, gravitational-wave observations provide an independent probe of the properties of bulk matter in neutron star cores that can then be used to compare with theoretically derived equations of state. In this paper, we quantify the systematic errors arising from the application of EOS-independent \emph{quasi-universal relations} in the estimation of neutron star tidal deformabilities and radii from gravitational-wave measurements and introduce a strategy to correct for the systematic biases in the inferred radii. We apply this method to a simulated population of events expected to be observed by future upgrades of current detectors and the next-generation of ground-based observatories. We show that our approach can accurately correct for the systematic biases arising from approximate universal relations in the mass-radius curves of neutron stars. Using the posterior distributions of the mass and radius for the simulated population we infer the underlying EOS with a good degree of precision.  Our method revives the possibility of using the universal relations for rapid Bayesian model selection of the dense matter EOS in gravitational-wave observations.
\end{abstract}

\maketitle

\section{\label{sec:intro}Introduction}

Gravitational wave observations of compact binary coalescences over three observing runs  \cite{LIGOScientific:2017vwq, LIGOScientific:2018hze, LIGOScientific:2018mvr, LIGOScientific:2020ibl, LIGOScientific:2021djp} of the Advanced Laser Interferometer Gravitational-wave Observatory (LIGO) \cite{TheLIGOScientific:2014jea} and Advanced Virgo \cite{TheVirgo:2014hva} have determined that binary neutron star mergers are among the most abundant sources of transient gravitational waves in the Universe \cite{LIGOScientific:2021psn}. Imprinted in the gravitational waves from binary neutron stars (BNS) is the bulk deformation of the stars due to the tidal field of their companions, quantified in terms of their tidal deformability parameter $\Lambda.$  The measurement of $\Lambda$ from gravitational-wave observations can provide insight into the thermodynamic properties of the high-density nuclear matter in their cores \cite{Flanagan:2007ix, Hinderer:2007mb, Vines:2011ud} as described by their equation of state (EOS) \cite{Baym:1971pw}.  

A zero temperature EOS is a curve with a functional relationship between the pressure $p$ and energy density $\epsilon$, i.e. $p =  p(\epsilon).$ Several formulations of the EOS have been used in the literature such as parameterizations of $p$ as a function of $\epsilon$ in the form of piecewise polytropes (see, e.g., \cite{Read:2008iy, Lattimer:2015nhk}), spectral representations \cite{Lindblom:2010bb, Lindblom:2022mkr}, or the speed of sound \cite{OBoyle:2020qvf}. In this work, we use functional relationships among the mass, radius, and tidal deformability of a non-spinning neutron star (NS), obtained by solving the Tolman-Oppenheimer-Volkoff (TOV) equations \cite{Oppenheimer:1939ne, Tolman:1939jz}, for a given pressure-density curve $p(\epsilon)$, as a mathematical model for an EOS. The magnitude of the tidal deformability is determined by the neutron star mass and their EOS. 

Tidal effects alter the orbital dynamics of a binary neutron star and hence the emitted gravitational waves. The phase evolution of gravitational waves depends predominantly on a certain linear combination of the individual tidal deformabilities $\Lambda_i$ of the companion stars called the \emph{reduced tidal deformability} $\tilde\Lambda$ (see Eq.\,(\ref{eq:def_lams}) for a definition of $\tilde\Lambda$ in terms of individual tidal deformabilities).  The reduced tidal deformability is a measure of the sum of the individual tidal deformabilities and enters the phase evolution of gravitational waves at the fifth post-Newtonian (PN) order \cite{Flanagan:2007ix}. The dual parameter $\delta\tilde{\Lambda},$ which is a measure of the difference in individual tidal deformabilities, enters the phase evolution at a higher, sixth PN order \cite{Flanagan:2007ix,Favata:2013rwa,Wade:2014vqa,Vines:2011ud}. The high PN orders imply that the tidal effects contribute significantly to the phase evolution at gravitational-wave frequencies greater than about $100$ Hz \cite{Harry:2021hls} and most significantly just before the two neutron stars merge at frequencies of 1 kHz--1.5 kHz, depending on the EOS \cite{Dietrich:2020eud}. The noise spectral density of the current detector network is at its lowest around 200 Hz and raises quadratically at larger frequencies \cite{aLIGO:2020wna}. Consequently, there is currently no hope of accurately measuring both the parameters but only the reduced tidal deformability, that too with significant errors. While future observatories might measure symmetric mass ratio ($\eta=m_1m_2/(m_1+m_2)^2$), chirp mass ($\mathcal{M}=\eta^{3/5}(m_1+m_2)$) and $\delta\tilde\Lambda$ better than now, the sensitivity will not be good enough to accurately measure both of the tidal parameters (see, e.g., \textcite{Smith:2021bqc}). The individual tidal deformabilities of the component NSs can be determined only if both $\tilde{\Lambda}$ and $\delta\tilde{\Lambda}$ are measured. In the absence of such measurements, \textcite{Yagi:2015pkc} found approximate correlations between combinations of the individual tidal deformabilities that depend only on the mass ratio of the companion stars and are largely independent of the equations of state (EOSs). These quasi-universal relations can be used to infer the individual tidal deformabilities $\Lambda_k,$ ($k=1,2$) from a measurement of $\tilde{\Lambda}$ alone. 

In addition to the above universal relations, \textcite{Damour:2009vw} discovered a strong sensitivity of the quadrupolar tidal deformability on NS compactness ${\cal C}_k = GM_k/c^2,$ irrespective of the properties of the nuclear matter making up the cores. New universal relations have been proposed recently by \textcite{Saes:2021fzr}, where the ratio of central pressure and densities are found to be correlated with the compactness. It should be pointed out that these universal relations are also `quasi' in nature because the correlations among the various quantities are only approximately true across the different EOS models. 

The universal relations are currently one way to deduce the radii of component NSs without making any assumption about the underlying EOS. This is because the measurement of the radius requires knowledge of the EOS which is currently unknown. 
Universal relations were used to infer the radii of companion neutron stars in GW170817 \cite{LIGOScientific:2017vwq, LIGOScientific:2018hze, LIGOScientific:2018cki}---the first BNS merger ever observed. The residuals on the systematic errors due to the approximate nature of the universal relations were marginalised, assuming a Gaussian distribution of the residuals \cite{Chatziioannou:2018vzf, Kumar:2019xgp, Carson:2019rjx, Biswas:2021pvm}. \citet{Kumar:2019xgp} construct their own universal relation for combining constraints obtained from multiple events. We note an important limitation of model selection based on universal relations. A set of TOV sequences for individual EOSs were used to find quasi-universal features. It, then, seems contradictory to calculate the evidence for the same set of models using the thence inferred quantities. The resolution of such a fallacy is that the use of universal relations is only as good as the fits for them which are 20\% and 2\% accurate for the universal relations given in equations~\ref{eq:unirel1} and~\ref{eq:unirel2}, respectively~\cite{Yagi:2016bkt}. The procedure of marginalising over the residuals was meant to alleviate some of the systematic errors present in the universal relations. However, this process is not infallible since marginalization can bring its own set of systematic errors in the presence of any non-Gaussian feature in the residuals. Moreover, the residuals are not random numbers, they are known quite precisely. Hence, assuming that residuals are described by a normal distribution and marginalizing over it can potentially introduce systematic errors. \citet{LIGOScientific:2019eut} avoid these systematic biases sampling tidal deformability of both components independently without using universal relations. The inferred posteriors of the mass-tidal deformability of both neutron stars were then compared with a wide collection of EOSs. 

A more direct approach to model selection is to parameterize $p =  p(\epsilon)$ curve to obtain the posterior in the space of an assumed set of (e.g., piecewise polytropic, spectral or nuclear) parameters, capable of describing a wide variety of \textit{ab-initio} zero temperature EOS models. The posterior probability of these parameters is inferred for a set of events, which are then combined to develop an effective model selection method. \citet{Lackey:2014fwa} have used a piecewise polytropic method while \citet{Wade:2014vqa} argue that spectral parameterization provides a better constraint when stacking multiple events in model selection. \citet{Biswas:2021pvm}, on the other hand, use a set of nuclear parameters to constrain and combine events for the same purpose.

While we break the degeneracy contained in the posterior PDF, $p(\mathcal{M},\eta,\tilde{\Lambda})$ before model selection, one can also obtain the evidence of an EOS model directly from such a PDF. Among such approaches, \citet{Pacilio:2021jmq} uses the joint posterior PDF to calculate the Bayes factor for each EOS model against a particular EOS by integrating the TOV equations for each value of central densities in a prior range. \citet{Ghosh:2021eqv} follows the method similar to \cite{LIGOScientific:2019eut} but, in the space of $(\mathcal{M},\eta,\tilde{\Lambda})$ rather than $(m_1,m_2,\Lambda_1,\Lambda_2)$. However, we believe that there are larger degeneracies among model EOSs that might be affecting the approach to model selection using the effective tidal deformability compared to breaking the degeneracy to individual tidal deformabilities.

In this study, we will describe a different and improved method of taking the systematic errors of universal relations into account for model selection which works equally well for all binary configurations. Our proposal is as follows. We recognised that at the time of model selection, we have an estimate for the component masses and the assumption of a fiducial model. This gives us the expected distribution of the tidal/radial parameters. Together, given a measurement of the tidal/radial terms, we do a piece-wise shift of the samples using the precisely known residuals with respect to the fiducial model to construct an unbiased inference. The evidence in favor of a model is calculated from this distribution. We find that if systematic errors are unaccounted for, it can lead to an incorrect model being preferred. Predictably, correcting for them results in an unbiased model selection.

The rest of this paper is organised as follows. In Sec.~\ref{sec:net_pop}, we describe the BNS population assumed in this study and present the distributions of errors in key inferred parameters for one of the samples from our simulation. In Sec.~\ref{sec:method}, we describe the universal relations and the EOS-agnostic analysis pipeline that is currently used in the study of dense matter EOS. We also show that the use of universal relations in inferring neutron star tidal deformabilities and radii lead to systematic biases and propose a simple algorithm to correct for the bias during model selection. Sec.~\ref{sec:results} summarizes the main results of this study concluding that the next-generation gravitational-wave observatories will have the ability to precisely measure the dense matter EOS. Sec. \ref{sec:conclusion_discussion} concludes with a brief summary of the paper and future directions.

\section{Networks and population}
\label{sec:net_pop}
In this section, we will outline the two gravitational-wave detector networks considered in this study and describe the population of BNS mergers accessible to them. We will then discuss the parameter estimation capabilities of these networks for gravitational waves from binary neutron star mergers, and in particular, how well they can measure the chirp mass, mass ratio, and tidal deformability that are relevant to measuring the dense matter EOS. 

\subsection{Networks}
\label{subsec:networks}
We consider two different detector networks in this study and they are acronymed and elucidated as follows:
\begin{itemize}
    \item \textit{HLVKI+}: This is a planned global network of five gravitational-wave detectors operating at A+ sensitivity. It consists of three LIGO detectors~\cite{LIGOScientific:2014pky} at Hanford (H), Livingston (L), and India (I) operating at A+ sensitivity, the Virgo (V) detector at AdVirgo+~\cite{VIRGO:2014yos} sensitivity, and the KAGRA detector at KAGRA+~\cite{KAGRA:2018plz} sensitivity. 
    \item \textit{ECS}: This is a proposed next-generation gravitational-wave detector network consisting of two Cosmic Explorer detectors~\cite{Reitze:2019iox}, one in the US and the other in Australia, each with an arm length of 40km, optimized for low-frequency, and an Einstein Telescope~\cite{Punturo:2010zza} in Europe. 
\end{itemize}
Additional details about the networks such as the technologies to be used in them and their sensitivities to different populations of compact binaries can be found in~\textcite{Borhanian:2022czq} and the references therein. We do not consider the detector noise realizations which could further impact the recovery of correct parameters.

\subsection{Population characteristics}
\label{subsec:population}
\subsubsection{Redshift distribution}
\label{subsubsec:redshift_distribution}
We simulate a population of BNS mergers up to a redshift of $z=1$. The redshift distribution of the population is given by
\begin{equation}
    p(z) = \frac{R_z(z)}{\int_0^{10} R_z(z) dz}
\end{equation}
where $R_z(z)$ is the merger rate density in the observer frame and can be expressed as
\begin{equation}
    R_z(z) = \frac{R_m(z)}{1+z}\frac{dV(z)}{dz}.
\end{equation}
Here $dV(z)/dz$ is the comoving volume element and $R_m(z)$ is the merger rate per comoving volume in the source frame which, in turn, is assumed to be proportional to the star formation rate (SFR), $R_f(t)$ and takes the form, \cite{deFreitasPacheco:1997fr}
\begin{equation}
    R_m(z) = \int_{t_{\rm min}}^{t_{\rm max}} R_f(t(z) - t_d) P(t_d) dt_d
\end{equation}
where the $t_z$ is given by 
\begin{equation}
    t(z) = \frac{1+z}{H_o}\int_{z}^{z_f} \frac{dz'}{E(z')}.
\end{equation}
This equation signifies that the binaries that form at time $t(z)-t_d$ merge at time $t(z)$ (i.e. redshift $z$) after a delay time $t_d$. Here, we choose the cosmic SFR to follow~\textcite{Vangioni:2014axa}. The probability distribution for the time to coalesce for a binary after formation is taken to be $P(t_d) \propto 1/t_d$ \citep{Beniamini:2019iop} with a minimum merger time of $t_{\rm min} = 20 \;\rm Myr$ and a maximum of $t_{\rm max} = 1/H_0 = 14.4$ Gyr using $H_0=67.7~km~s^{-1}~Mpc^{-1}$ \citep{Aghanim2020-tl} ($\approx z=30$). Using the local merger rate $R_m(z=0)$ from the second LIGO-Virgo Gravitational-Wave Transient Catalog, \textit{GWTC-2}~\cite{LIGOScientific:2020kqk} to be
\begin{equation}
    R_m(z=0) = 320 \; \rm Gpc^{-3} yr^{-1},
\end{equation}
we find the total number of BNS mergers up to $z=1$ comes out to be approximately 80,000 per year and we take this to be the annual rate. 

\subsubsection{Equations of state simulated}
\label{subsubsec:eos}
We simulate the gravitational-wave signal for our population of BNS for 3 different EOS: ALF2 \cite{Alford:2004pf}, APR3, APR4 \cite{Akmal:1998cf} -- with sufficiently different mass-radius curves. These three EOSs are representative of different regions of the mass-radius space as well as a range of maximum masses for NSs as shown in Fig. \ref{fig:L2eos}. In addition to these we use several other EOSs from literature (BHB \cite{Banik:2014qja}, DD2 \cite{Typel:2009sy}, H3 \cite{Lackey:2005tk}, H4 \cite{Lackey:2005tk}, LS220 \cite{Lattimer:1991nc}, SFHo \cite{Steiner:2012rk}, SLy \cite{Douchin:2001sv}) for the model selection process to be discussed in sec \ref{sec:method}. Two EOSs have been constructed by randomly selecting piecewise polytropic indices from \citep{Godzieba:2020bbz} marked here as PP2 and PP5.

\subsubsection{Distributions of intrinsic and extrinsic parameters}
\label{subsubsec:parameter_distr}
We consider the individual neutron stars to be non-spinning and distributed uniformly in masses between $1 \,\rm M_{\odot}$ and the maximum mass allowed by the corresponding EOS. The extrinsic parameters -- cosine of the inclination angle $\cos\iota$, location of the source in the sky (cosine of the declination angle $\cos\delta$ and right ascension $\alpha$), polarization angle $\psi$, and the phase of coalescence $\phi_0$, of the fiducial BNS population are drawn from a uniform distribution across their domains. The luminosity distances are calculated from the redshifts of the binaries using \textit{Planck18}~\cite{Planck:2018vyg} cosmology.

\begin{figure*}[ht]
    \begin{center}
        \includegraphics[width=0.45\textwidth]{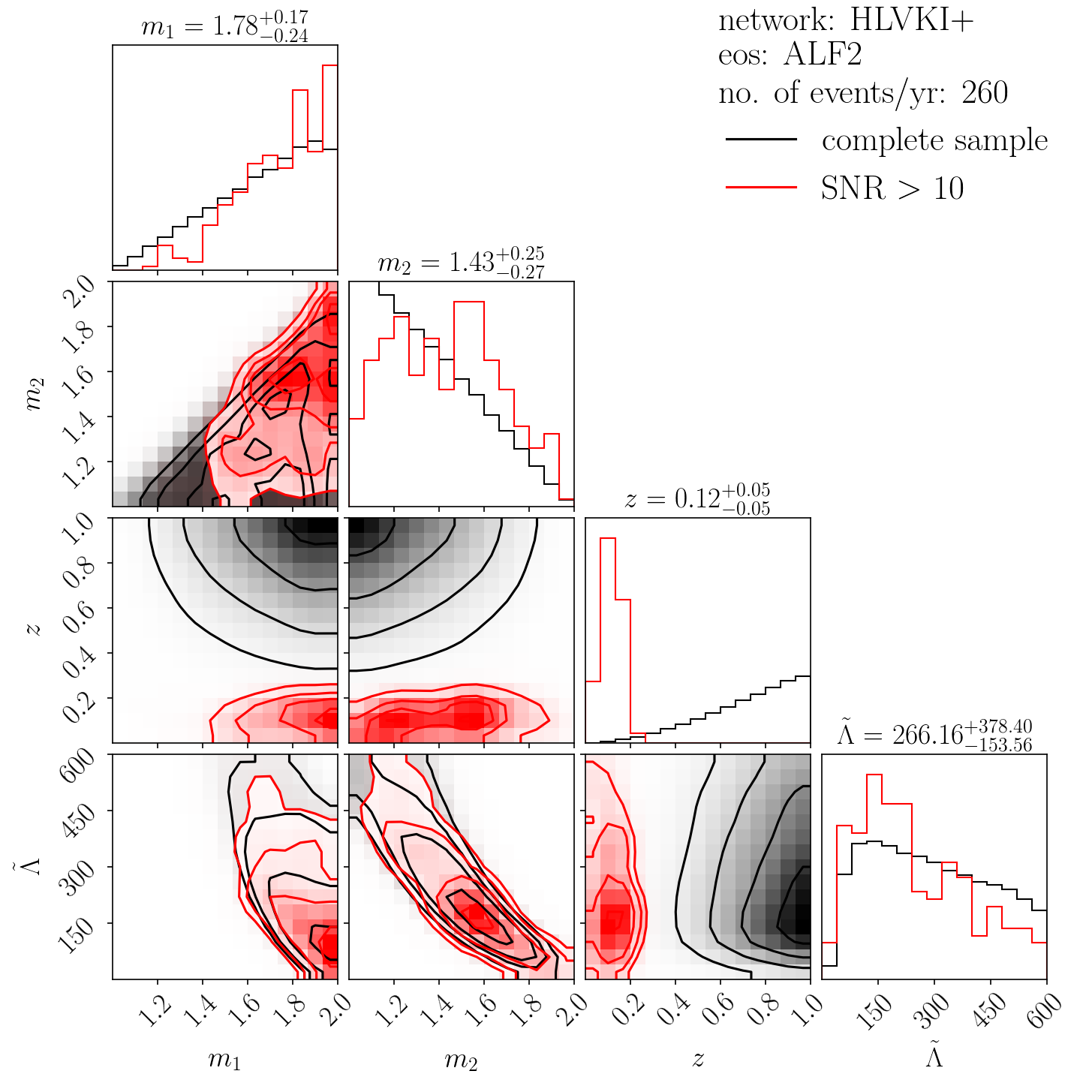}
        \includegraphics[width=0.45\textwidth]{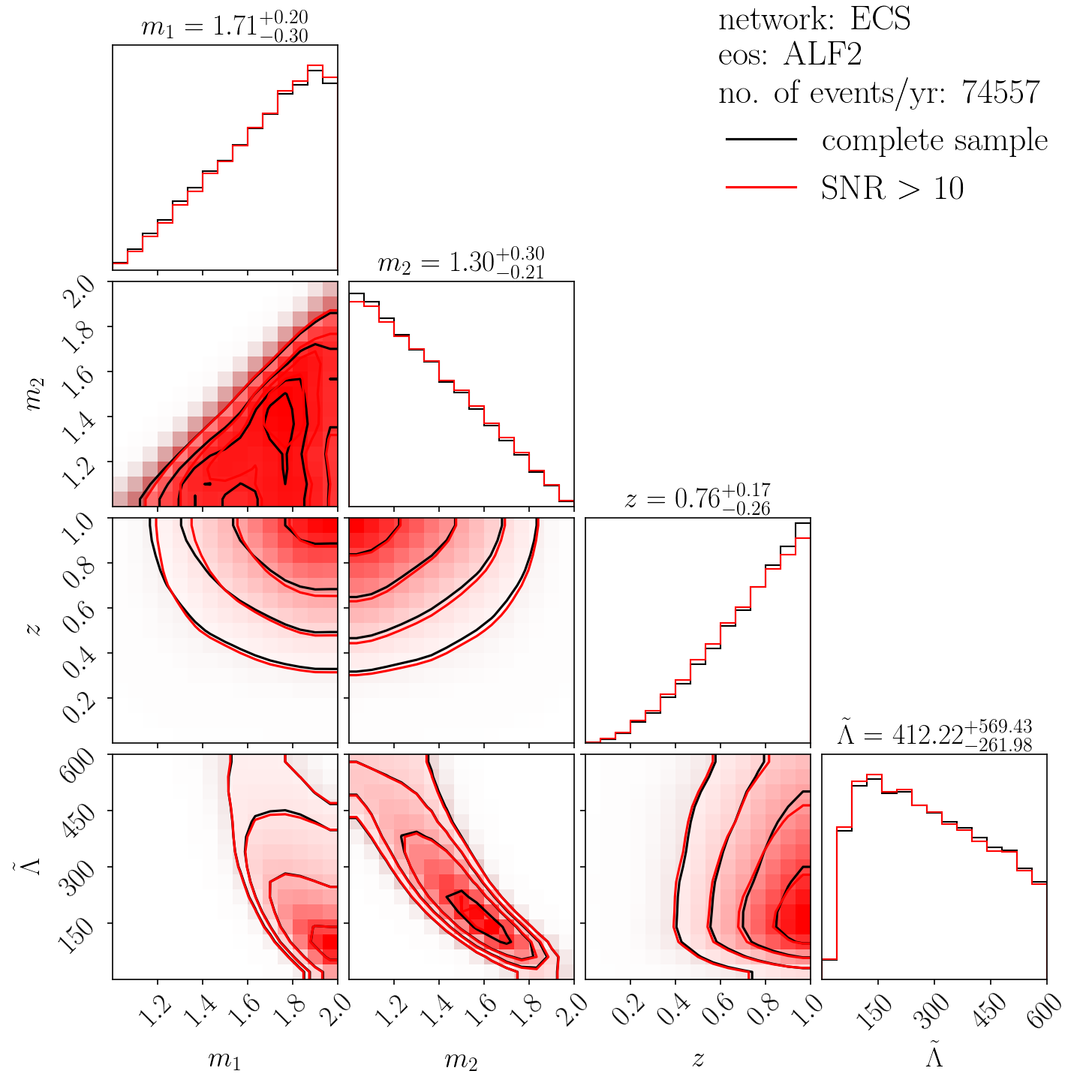}
        \includegraphics[width=0.45\textwidth]{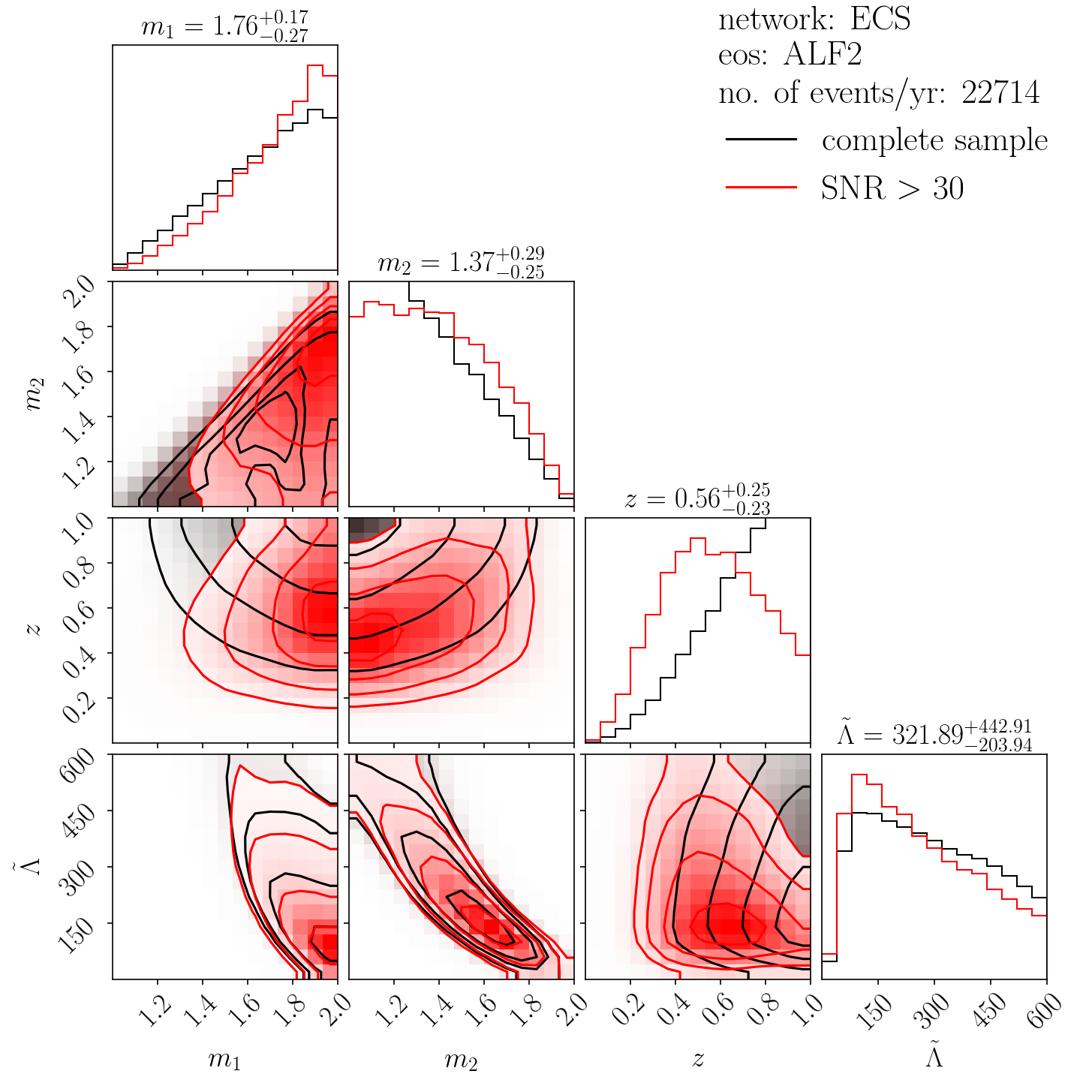}
        \includegraphics[width=0.45\textwidth]{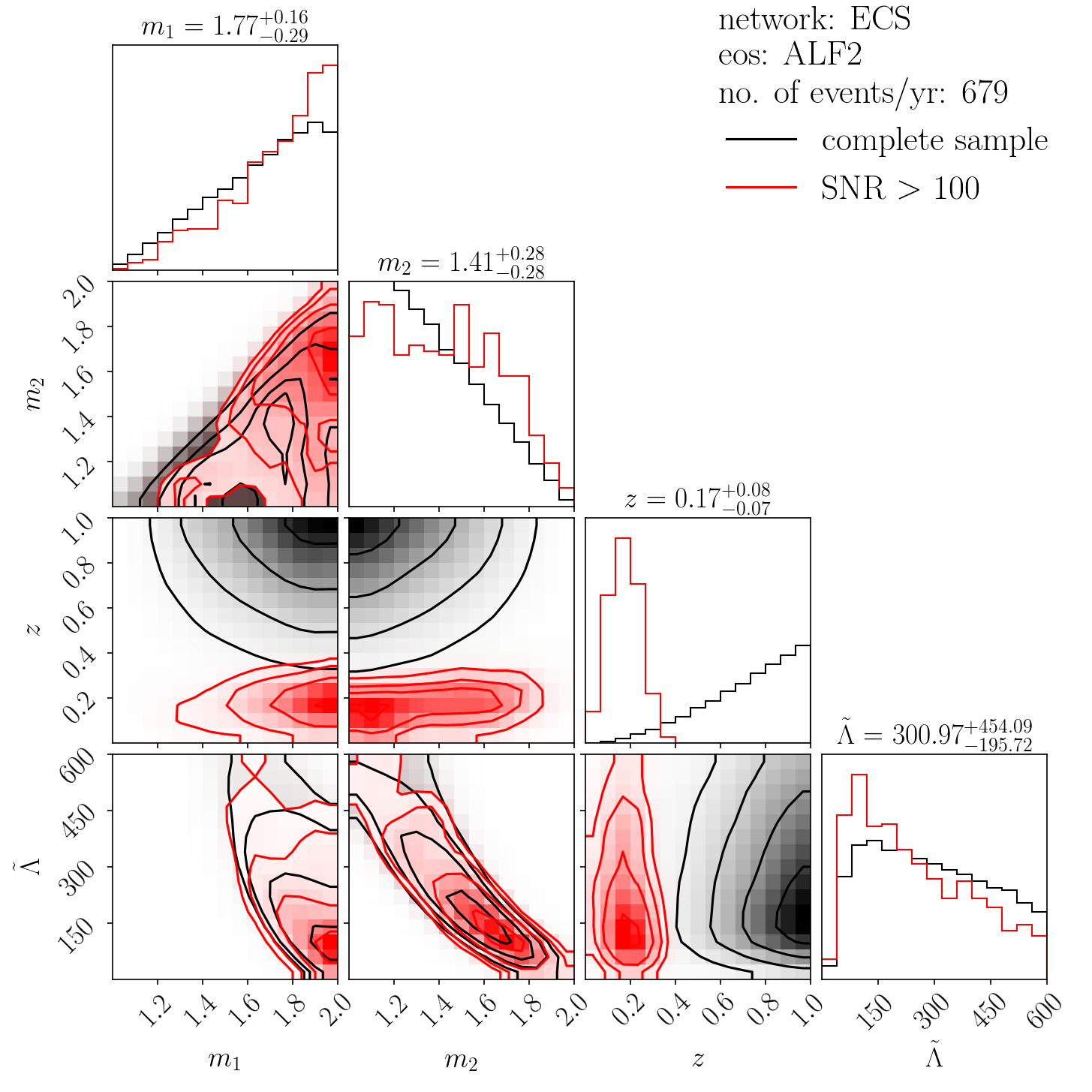}
    \end{center}
    \caption{Distribution of selected parameters in our population for different SNR thresholds in the two detector networks considered. These are the sub-populations that are used in the bias correction analysis. The HLVKI+ network is not expected to observe a significant number of events above an SNR of 30. The median values for the observed population are also quoted on top of each column}.
    \label{fig:parameter_distribution}
\end{figure*}

In Table~\ref{tab:det_num}, we show the number of BNS mergers per year above a certain signal-to-noise ratio (SNR) for our simulated populations in the two detector networks and the three EOSs considered in this study \cite{Borhanian:2022czq}. The variation of the detection rate across different EOS is directly related to the maximum mass allowed by the corresponding EOS. Note that for an SNR threshold of 10, which is a typical detection criterion for an event, the next-generation network detects almost all BNS mergers within a redshift of $z=1$ \cite{Borhanian:2022czq}. In Fig.~\ref{fig:parameter_distribution}, we also depict the distributions of various parameters of the binary given a threshold SNR. This shows the parameter space probed by the observed population compared to the astrophysical distribution of sources.

\begin{table}
    \begin{tabular}{c c|c|c}
        \toprule
        \multicolumn{2}{c|}{\diagbox{EOS}{Network}} & HLVKI+ & ECS \\
        \multicolumn{2}{c|}{} & events per year & events per year \\
        \midrule
        \multirow{3}{*}{\STAB{\rotatebox[origin=c]{90}{SNR = 10}}} & ALF2 & 260 & 74304\\ [7pt]
        & APR3 & 327 & 75868\\ [7pt]
        & APR4 & 299 & 75191\\ %[7pt]
        \midrule
        \multirow{3}{*}{\STAB{\rotatebox[origin=c]{90}{SNR = 30}}} & ALF2 & 10 & 22565\\ [7pt]
        & APR3 & 19 & 27620 \\ [7pt]
        & APR4 & 9 & 25252\\ %[7pt]
        \midrule
        \multirow{3}{*}{\STAB{\rotatebox[origin=c]{90}{SNR = 100}}} & ALF2 & - & 638\\ [7pt]
        & APR3 & - & 846 \\ [7pt]
        & APR4 & - & 767 \\ %[7pt]
        \bottomrule
    \end{tabular}
    \caption{The number of BNS mergers with an SNR greater than a threshold of 10, 30, and 100 out of a simulated population of 80,000 events expected to be detected annually by next generation GW detectors. We show the number of events in the two detector networks and the three EOS considered.}
    \label{tab:det_num}
\end{table}

\subsection{Parameter estimation}
\label{subsec:parameter_estimation}
\begin{figure*}[ht]
    \centering
        \includegraphics[width=0.45\textwidth]{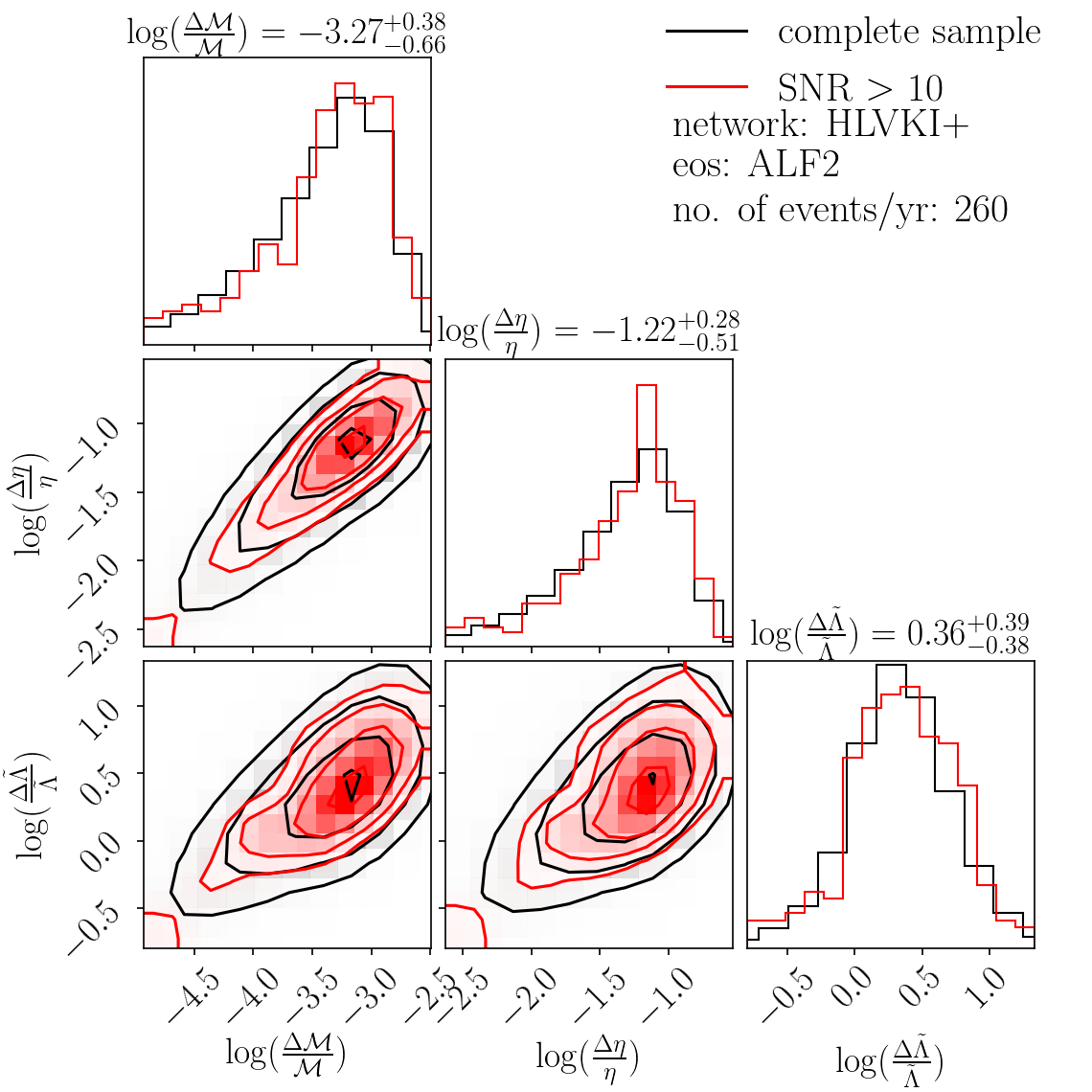}
        \includegraphics[width=0.45\textwidth]{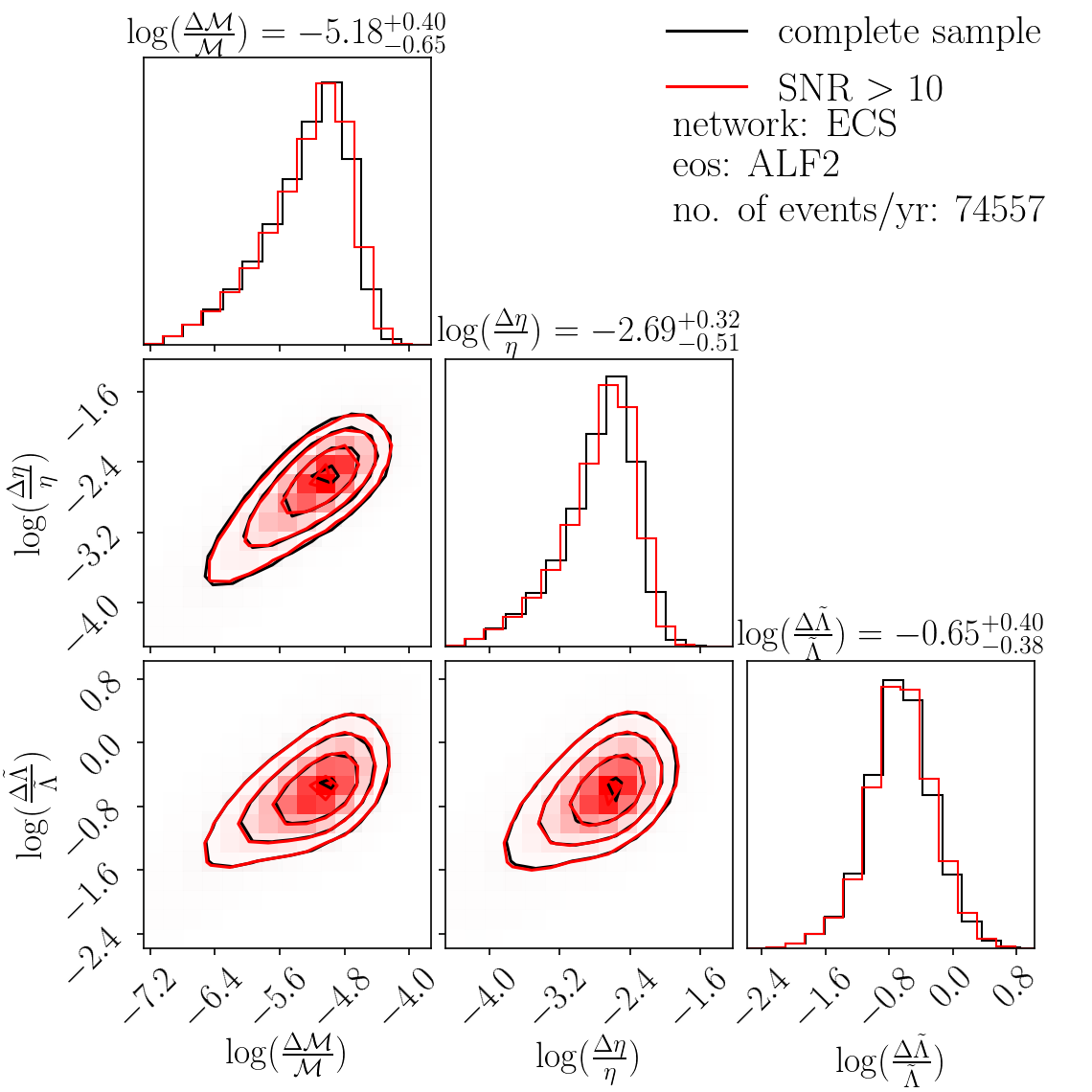}
    \caption{Distribution of relative errors on log scale for the key quantities for our population of BNS systems with ALF2 EOS. The full population is depicted in black while the sample that with SNR greater than 10 is shown in red. The right panel contains the results for ECS while the left panel is for the HLVKI+ network. The median values of error distributions in each of the quantities are mentioned on top of the each column.}
    \label{fig:errordist_snr10}
\end{figure*}

We simulate gravitational waves from the population of binaries distributed as detailed in the present section using the TaylorF2-tidal waveform model \cite{Wade:2014vqa}. The signal spans a frequency range of $[f_{\rm min}, f_{\rm max}$] where $f_{\rm min}$ is 5 Hz for ECS while 10 Hz for A+ detectors and  $f_{\rm max} = \rm min(f_{\rm ISCO}, 1024 \,\rm Hz)$. $f_{\rm ISCO}=1/(6^{3/2}\pi M)$ is the orbital frequency of the inner-most stable circular orbit for a point particle in an effective black hole spacetime, where $M$ is the total mass of the binary \cite{Chandrasekhar:1985kt}. The signal is truncated at 1024 Hz because signals at higher frequencies do not contribute to the SNR~\cite{Borhanian:2022czq}. 

The errors on the parameters of the gravitational-wave signal are calculated using the Fisher approximation for the likelihood of a signal using the publicly available code \texttt{gwbench}~\cite{Borhanian:2020ypi}. The Fisher matrix is defined using frequency domain inspiral gravitational waveform, $h(f)$ as
\begin{equation}
    \mathcal{F}_{ij} = \left<\frac{\partial h(f)}{\partial \theta^i}, \frac{\partial h(f)}{\partial \theta^j} \right>
\end{equation}
with a vector in this space given by $\vec{\theta}=(\mathcal{M}, \eta, \tilde{\Lambda}, D_L, \iota, \alpha, \delta, \psi, \phi_0, t_c)$ where $\mathcal{M}$, $\eta$, $\tilde{\Lambda}$, $D_L$ and $t_c$ are the chirp mass, symmetric mass ratio, combined tidal deformability, luminosity distance and the time of coalescence, respectively, and the other parameters are defined in the previous section. The covariance matrix for the errors on these parameters is then $\mathcal{C}_{ij}=\mathcal{F}_{ij}^{-1}$. The inner product in the above expression is given by
\begin{equation}
    \left<a(f),b(f)\right> = 2\int_{f_{\rm min}}^{f_{\rm max}} \frac{\tilde{a}(f)^*\tilde{b}(f) + \tilde{a}(f)\tilde{b}(f)^*}{S_n(f)} df
\end{equation}
where $S_n(f)$ is the noise spectral density of the detectors. 

In Fig.~\ref{fig:errordist_snr10}, we show the normalised distribution of $1\sigma$ statistical errors for our population in the parameters $\mathcal{M}$, $\eta$, and $\tilde{\Lambda}$. The error distributions are shown for only these parameters because we will use them to calculate the individual tidal deformabilities and radii of the NSs. Furthermore, only the population having ALF2 as its EOS is shown and they are similar for the other EOS. The results for the full population are shown in black while the sub-population with a minimum SNR of 10 is shown in red. The right panel shows the capabilities of the next-generation network of ECS whereas the left panel depicts the HLVKI+ network. The 1D histograms show that the error distributions for the detected population and the full population are similar in the parameters shown with the median errors for the detected population tabulated above the histograms corresponding to the relevant parameters. %(\arnab{Rahul, confirm that the numbers are quoted for the detected population and not the full.} \rk{confirmed that numbers on title of both Fig \ref{fig:parameter_distribution} and \ref{fig:errordist_snr10}; I can extract the numbers for the whole population if we need.})
This implies that the errors on these parameters are weak functions of the SNR of the signal, which is used to demarcate a detection from a non-detection. The ellipses show how the errors on the different parameters are correlated. As expected, we see a positive correlation between the error distributions of the intrinsic parameters signifying that if one of the parameters is well-measured then so are the others. Nevertheless, it can also be seen that the correlation of $\mathcal{M}$ and $\eta$ with $\tilde{\Lambda}$ is less than that between them. This is because the $\tilde{\Lambda}$ errors are determined by the high-frequency sensitivity of the network while the $\mathcal{M}$ and $\eta$ errors are primarily controlled by the low-frequency sensitivity \cite{Harry:2018hke, Dietrich:2020eud}. Note that the ellipses are not the correlations among the different parameters for a given event but rather the distribution of the errors for a population of events.

\begin{figure*}[t!]
    {\centering
    \includegraphics[width=1.0\linewidth]{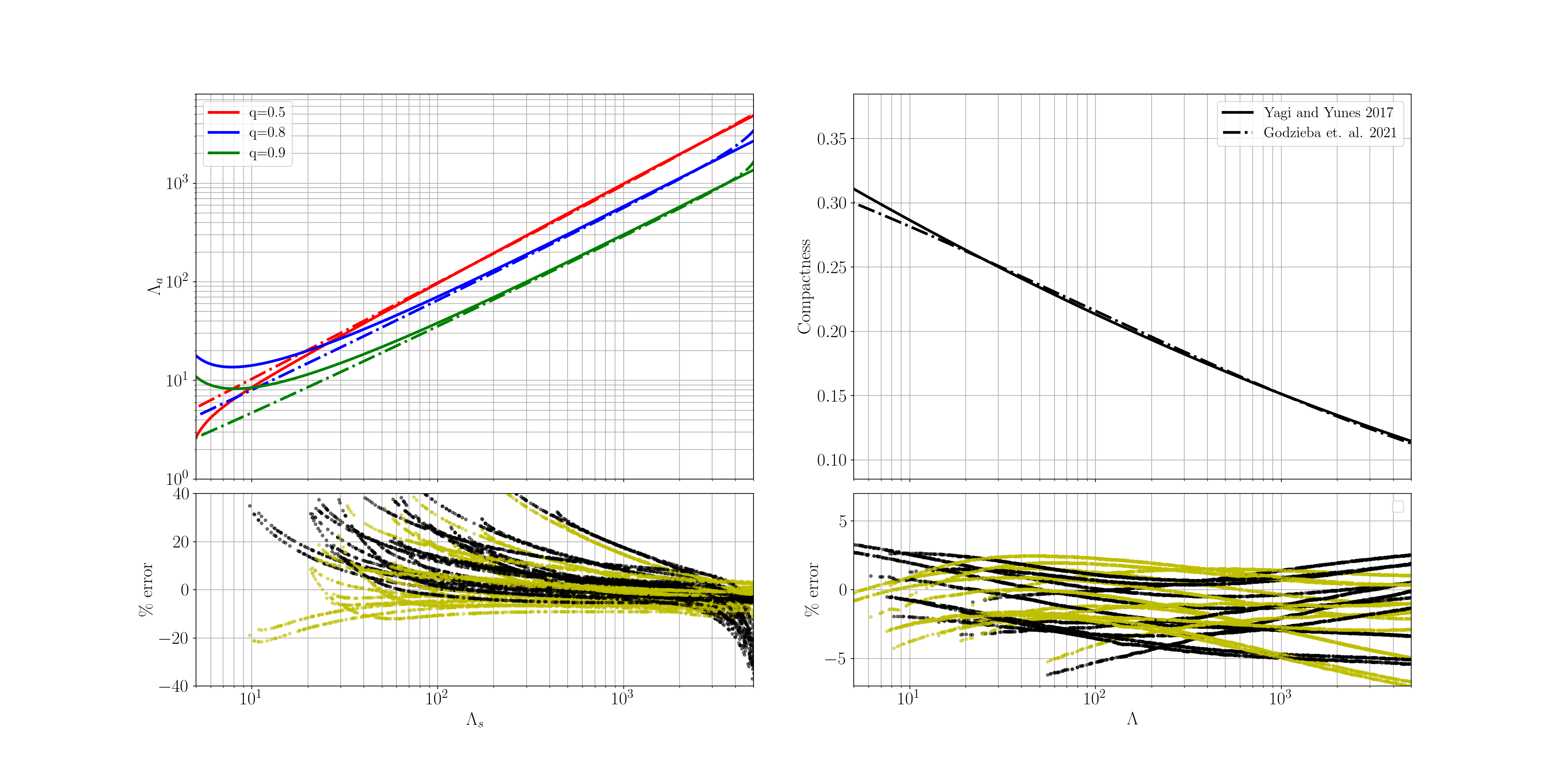} }
    %trim={<left> <lower> <right> <upper>}
    \caption{The top left panel plots the universal relation between the asymmetric and symmetric combinations, $\Lambda_a$ and $\Lambda_s,$ respectively, of the individual tidal deformabilities $\Lambda_1$ and $\Lambda_2$ for several EOS as constructed in \textcite{Yagi:2016bkt} and \textcite{Godzieba:2020bbz}.  The (small) difference in the two versions of the universal relations is due to the error in the fitting function and the EOS data. In the lower-left panel, we plot the deviation of the universal relations from the TOV sequences corresponding to the EOS models used in this work provided by \textcite{Godzieba:2020bbz} (in black) and \textcite{Yagi:2016bkt} (in yellow), respectively.  These residuals can be as large as 25\% for the largest and smallest tidal deformabilities ($\Lambda_s$) corresponding to smallest and largest neutron star masses, respectively.  The top right panel plots the second universal relation between a neutron star's compactness ${\cal C}$ and its tidal deformability $\Lambda$ as constructed by the same two authors \cite{Yagi:2016bkt, Godzieba:2020bbz}. Please note that the left panel shows universal relations for a pair of NSs while the universal relation shown in the right panel is for single NS. We use these exactly known deviations to correct the radius and tidal deformability posteriors at the time of model selection.} 
    \label{fig:unirel}
\end{figure*}

\section{\label{sec:method}Use of quasi-universal relations in model selection}

This section begins by discussing the two universal relations that relate the symmetric and asymmetric combinations of the tidal deformabilities on the one hand and the compactness and tidal deformability on the other. This is followed by a brief outline of the current method for inferring the tidal deformabilities and radii of individual neutron stars from gravitational-wave observations \citep{LIGOScientific:2018cki}. We point out how this approach can lead to biased estimation of the tidal deformability parameters and the radii due to systematic errors in the universal relations and hence lead to erroneous EOS model selection. For the ECS network of next-generation observatories, the systematics can dominate over statistical uncertainties even for individual events. For the HLVKI+ network, the systematics for individual events are smaller than statistical uncertainties; however, model selection with a population of 30 events or more can be biased even for this network. We describe how to rectify systematic uncertainties at the time of model selection and introduce several statistical measures to show that the bias-corrected estimates of the tidal deformability and radius converge to the correct EOS. Marginalizing over the errors due to universal relations can account for the biases but that comes at the expense of increased errors in the inferred quantities. 

\subsection{Universal Relations and Residuals}
\subsubsection{Universal relation between tidal deformabilities of a pair of neutron stars}
%\pagebreak
The structure of neutron stars is determined by their nuclear EOS via the TOV equations \cite{Oppenheimer:1939ne, Tolman:1939jz}. The EOS of neutron stars is currently unknown and there are numerous models describing the pressure-density (equivalently, mass-radius) curves of neutron stars (for a review see \textcite{Lattimer:2000nx} and references therein). Although X-ray and gravitational-wave observations severely constrain the family of viable EOS models, many of them are still consistent with data. 

\paragraph{First universal relation} In spite of the huge variation in the relationship between the masses and radii of neutron stars amongst different EOS models, \textcite{Yagi:2016bkt} found the remarkable result that the asymmetric combination $\Lambda_a$ of the tidal deformabilities of two neutron stars, defined by $2\Lambda_a \equiv (\Lambda_2 - \Lambda_1),$ is uniquely related to the symmetric combination, $\Lambda_s$, defined by $2\Lambda_s \equiv (\Lambda_2+\Lambda_1),$ depending only on the ratio of their masses $q \equiv m_2/m_1 \le 1.$ This universal relation is given by \cite{Yagi:2013sva,Chatziioannou:2018vzf}:
\begin{subequations} 
    \label{eq:unirel}
    \begin{equation}
        {\Lambda}_a =  F_{{n}}(q){\Lambda_s}\frac{1+\sum_{i, j}^{3,2}b_{ij}q^j/{{\Lambda}_s}^{i/5}} {1+\sum_{i,j}^{3,2}c_{ij}q^j/{\Lambda_s}^{i/5}},
    \label{eq:unirel1}
    \end{equation}
    \begin{equation}
        F_n(q) = \frac{1-q^{10/(3-n)}}{1+q^{10/(3-n)}},
    \label{eq:unirel2}
    \end{equation}
\end{subequations}
where $\Lambda_1$ and $\Lambda_2$ are the individual tidal deformabilities of the companion stars, $b_{ij},$ $c_{ij}$ and $n$ are the fitting parameters given in Table 3 of \textcite{Yagi:2013sva} (see also \textcite{Chatziioannou:2018vzf} and \textcite{Godzieba:2020bbz}). 

\begin{figure*}[ht!]
    {\centering
    \includegraphics[width=1.0\linewidth, trim={0cm 0cm 0cm 0cm},clip]{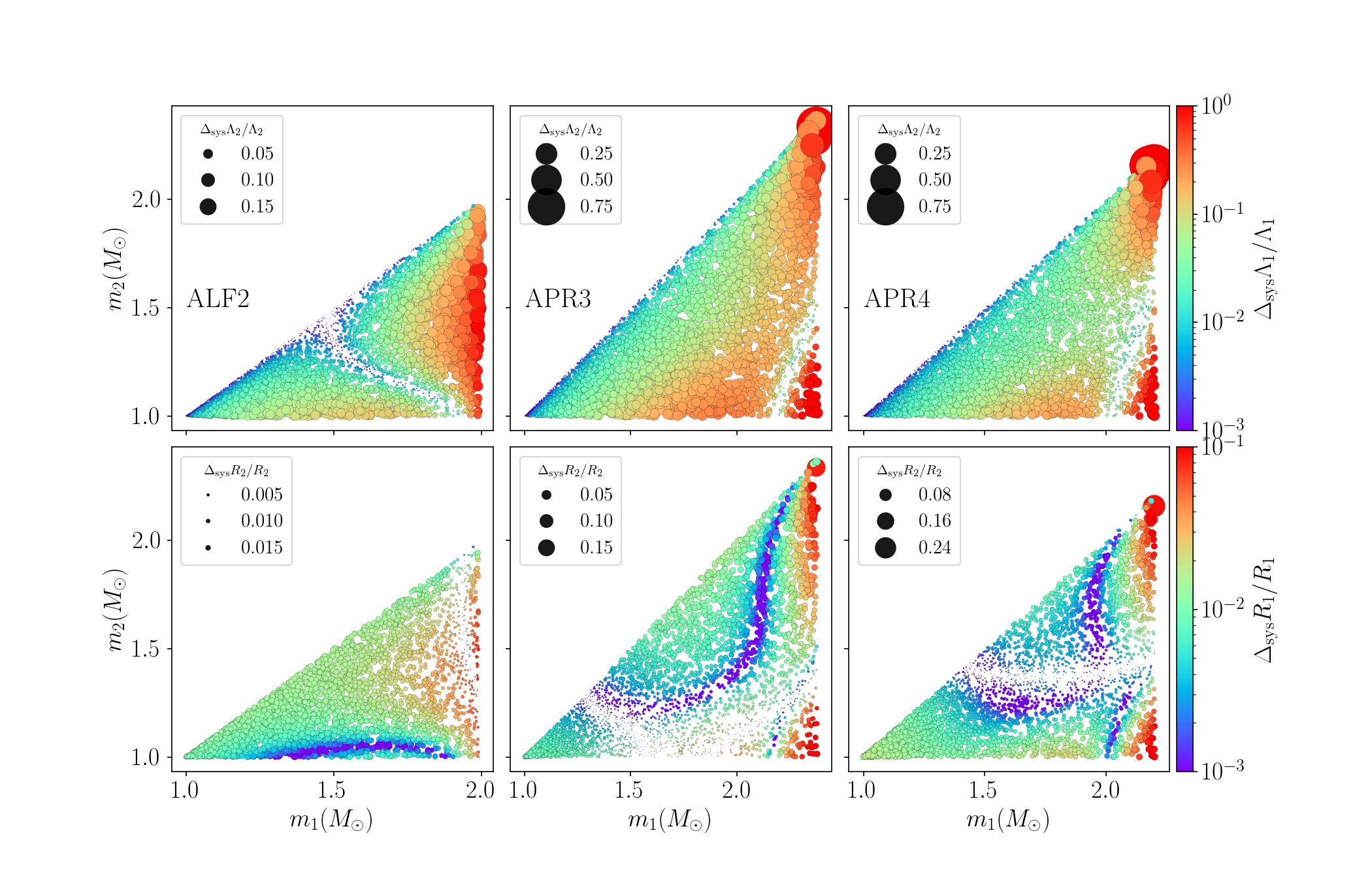} }
    \caption{This figure shows the relative error in the reconstruction of individual tidal deformabilities of NSs in the space of component masses of BNS systems for three example EOS. The component masses are sampled uniformly over the range allowed by the respective EOS without imposing any restriction on the mass ratio. Colors represent the systematic error in the tidal deformability of the primary companion (i.e., the heavier NS) while the size of the circle represents the systematic error in the tidal deformability of the lighter companion. The errors are greater for more massive NSs and softer EOS.}
    \label{fig:syst_bias}
\end{figure*}

\paragraph{Residuals in the first universal relation} The universal relation between $\Lambda_a$ and $\Lambda_s$ is plotted in the top-left panel of Fig.~\ref{fig:unirel} for three different values of the mass ratio $q=0.5,$ 0.8 and 0.9. Solid and dot-dashed curves use the fitting formulas in \textcite{Yagi:2013sva} and \textcite{Godzieba:2020bbz}, respectively. The two versions of the universal relations agree with each other pretty well over a wide range of $\Lambda_s$ for $q=0.5$ but less so for larger masses (smaller $\Lambda_s$) and larger mass ratios.

The bottom left panel plots the residual of the fits with respect to the exact TOV sequences for the EOS models used in this work. 
The residuals remain below $\sim 10\%$ for many EOS models and over a wide range of $\Lambda_s;$ the fits of \textcite{Godzieba:2020bbz} have smaller residuals than those of \textcite{Yagi:2013sva} but can still be as large as 25\% when the tidal deformabilities are small, corresponding to heavier neutron stars and/or softer EOSs. 

\paragraph{Individual tidal deformabilities} In spite of the residuals, Eq.\,(\ref{eq:unirel}) is still very useful in inferring the individual tidal deformabilities and hence assist in the process of EOS model selection. The PN expansion of the gravitational-wave phase contains the individual tidal deformabilities as linear combinations in the PN coefficients at the fifth and sixth PN orders [i.e., corrections in the phase evolution at orders ${\cal O}(v/c)^{10}$ and ${\cal O}(v/c)^{12}$ beyond the leading order quadrupole term], respectively. Tidal effects are encoded in parameters $\tilde\Lambda$ and $\delta\tilde\Lambda$ defined by: 
\begin{subequations}
\label{eq:def_lams}
    \begin{equation}
    %\tilde{\Lambda} = \frac{1}{26}\left[ (11m_2+M)\frac{\Lambda_1}{m_1}+(11m_1+M)\frac{\Lambda_2}{m_2}  \right]
    %\tilde{\Lambda} = \frac{1}{26} \left[ (1+12q)\Lambda_1 + (1+12/q) \Lambda_2 \right]
    \tilde{\Lambda} = \frac{1}{26} \left[ (1+12q)(\Lambda_{s}-\Lambda_a) + \left (1+\frac{12}{q} \right) (\Lambda_{s}+\Lambda_a) \right]
        \label{eq:lambdatilde}
    \end{equation}
    \begin{equation}
    \begin{aligned}
    \delta \tilde{\Lambda} &= \sqrt{1-4  \eta}\left( 1-\frac{13272}{1319} \eta+\frac{8944}{1319} \eta^{2} \right) \Lambda_{s} \\ 
     &+\left( 1-\frac{15910}{1319} \eta+\frac{32850}{1319} \eta^{2}+\frac{3380}{1319} \eta^{3} \right)  \Lambda_{a}.
    \end{aligned}
    \end{equation}
\end{subequations}
The high PN orders at which they appear imply that the effect of $\tilde\Lambda$ is only important for frequencies $f_{\rm GW}\gsim 100\, \rm Hz$ and that of $\delta\tilde\Lambda$ at even larger frequencies \cite{Harry:2021hls} (see also \textcite{Dietrich:2020eud}). Thus, only the final few cycles of the waveform before the merger contain significant tidal effects and only $\tilde\Lambda$ is measurable to a good accuracy even when the SNRs are $\sim 100$ \cite{Smith:2021bqc}. Thus, there is no hope of inferring the individual tidal deformabilities from gravitational-wave observations alone. But universal relation Eq.\,(\ref{eq:unirel}) can be of help albeit inferred $\Lambda$ values will be biased.

\paragraph{Systematic errors in $\Lambda$} Suppose $(q, \tilde\Lambda)$ are known exactly. We can numerically solve the pair of Eqs.\,(\ref{eq:unirel})  and (\ref{eq:lambdatilde}) for $\Lambda_s$ and $\Lambda_a$ and hence infer the individual tidal deformabilities $\Lambda_k,$ $k=1,2.$ The numerical inversion algorithm is found to be accurate to a relative error of $\sim 10^{-8}$ while the interpolation of TOV tables are accurate to better than $\sim 10^{-4}$. Since the universal relations are not exact, however, the inferred values of $\Lambda_k$ will not be the same as the ones that went into computing $\tilde \Lambda$. We assess the systematics incurred using a population of BNS systems uniformly sampled in the $m_1$--$m_2$ plane. 

For each pair of masses, and for a given EOS, we find $\Lambda_1$ and $\Lambda_2$ by solving the TOV equations using the TOV solver developed in \citet{Damour:2009vw,Bernuzzi:2008fu} \footnote{\href{https://bitbucket.org/bernuzzi/tov/src/master/}{https://bitbucket.org/bernuzzi/tov/src/master/}}, which are then used to obtain $\tilde\Lambda.$ For a given EOS, we solve TOV equations for a discreet set of masses to obtain the tidal deformabilities and radii and interpolate the solutions to obtain these parameters for arbitrary NS masses. The interpolated values of tidal deformabilities and radii agree with the exact numerical solution of TOV equations to within a fractional error of $10^{-4},$ which, as we shall see below, is far smaller than the systematic errors due to universal relations.

For each binary pair in the population, the bottom panels of Fig.\,\ref{fig:syst_bias} show the fractional difference in the true and reconstructed tidal deformabilities $|\Delta_{\rm sys}\Lambda_{1,2}|/\Lambda_{1,2}$, the color-bar representing the systematic error in $\Lambda_1$ (i.e., the tidal deformability of the heavier companion) and the size of the circles representing the systematic error in $\Lambda_2$ (i.e., the tidal deformability of the lighter companion). The biases are shown for three example EOSs: ALF2, APR3, and APR4, corresponding to increasingly softer EOS from left to right. The biases are particularly large $(\sim 75\%)$ for softer EOS and heavier NSs whose $\Lambda$ values are around few tens to hundreds.  For intermediate masses and stiffer EOS, the errors are lower but could still be few to ten percent. This is consistent with the fact that the fit residuals of the universal relation (\ref{eq:unirel}) are smaller for stiffer EOSs \cite{Yagi:2016bkt}.  

Thus, while the tidal deformabilities obtained using universal relations are EOS-agnostic the inferred values are biased. Consequently, a model selection algorithm that compares the mass-tidal deformability curve obtained from a collection of BNS events with the ones obtained from solving TOV equations could lead to a greater evidence for an incorrect EOS model. 

\subsubsection{Universal relation between tidal deformability and compactness of a single neutron star}
The first universal relation helps infer the individual tidal deformabilities of neutron stars from gravitational-wave observations but this does not help in inferring their radii. This is because the tidal deformability of a neutron star is related to its radius via:
\begin{equation}
    \Lambda \equiv \frac{2}{3}k_2 \left ( \frac{R}{m} \right )^5,
\end{equation}
where $k_2,$ $R$ and $m$ are the star's tidal Love number, radius and mass \cite{Flanagan:2007ix}. A knowledge of the Love number $k_2,$ also determined by the EOS, is necessary to deduce the radius of a neutron star from its mass and tidal deformability.  Although $k_2$ varies quite a bit from one EOS to another for a given neutron-star mass, there seems to be a second universal relation \cite{Yagi:2016bkt, Damour:2009vw} that connects $\Lambda$ to the star's compactness defined by ${\cal C} \equiv GM/(Rc^2$): 
\begin{equation} \label{eq:cvslambda}
    {\cal C} (\Lambda)= \sum_{k=0}^{2} a_k (\ln{\Lambda} )^k,
\end{equation}
where $a_k$s are the fitting coefficients in \textcite{Godzieba:2020bbz}.  The right panel of Fig.\,\ref{fig:unirel} plots this second universal relation constructed in \textcite{Godzieba:2020bbz} and \textcite{Yagi:2016bkt}. Both versions agree with each other pretty well over a wide range of tidal deformability. The residual between the universal relation and the true dependence of the compactness on the tidal deformability for the collection of EOS considered in this work (bottom right panel) is at most 5\% over the entire range of $\Lambda.$ This error is tolerable given the large measurement uncertainties of reduced tidal deformability expected in the near future (i.e., for the HLVKI+ network). However, an error of 5\% is too large for the ECS netowrk which has the potential to measure the radii and deformabilities at the sub-percent level.

This second universal relation, together with the posterior probabilities of the component masses and tidal deformabilities, allows the construction of the posterior probabilities of the radii. While the $\Lambda$ posteriors are only affected by the systematics in the first universal relation, the radii posteriors are affected by the systematics of both universal relations. 

In the next section, we summarize the analysis pipeline used to infer the mass-tidal deformability and mass-radius curves using the two universal relations. A similar pipeline was applied to GW170817 to obtain EOS-independent posterior distribution of masses and radii \citep{LIGOScientific:2018cki}. 

\begin{figure}[t!]
    {\centering \includegraphics[width=1.0\linewidth, trim={8cm 6cm 2cm 2cm},clip]{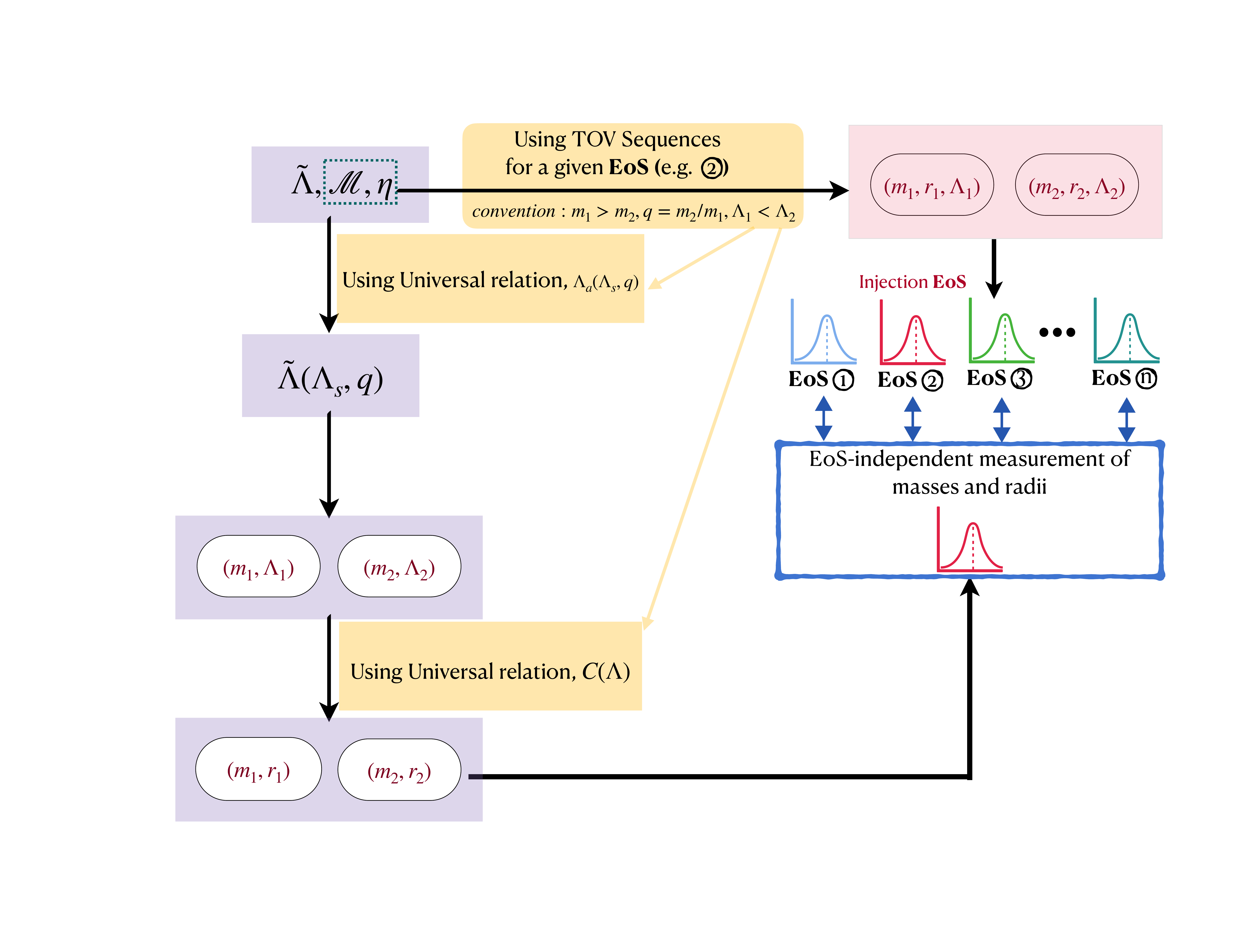}}
    \caption{This is a sketch of the data analysis pipeline. The posterior samples for $M_c, \eta, \tilde{\Lambda}$ are created using the Fisher information matrix which is in the left-uppermost box. Yellow boxes represent information coming out from the construction of TOV sequences using multiple choice of EOS. We use quasi-universal relations, \ref{eq:unirel1} and \ref{eq:cvslambda} to obtain the component mass-radii posterior probability distribution, which is then compared to a set of EOS models as represented by a vertical double-arrow.} 
    \label{fig:flowchart}
\end{figure}

\subsection{Inference of tidal deformability parameters and neutron star radii without the residuals}
\label{subsec:curr_proc}
The current approach for the calculation of EOS-agnostic radii of neutron stars from gravitational-wave observations of BNS mergers, as described in  \textcite{LIGOScientific:2018cki}, involves the use of the two universal relations given in Eqs.\,(\ref{eq:unirel}) and (\ref{eq:cvslambda}). We developed an alternative, but equivalent, approach, the flowchart for which is shown in Fig.\,\ref{fig:flowchart}. 
\begin{enumerate}
    \item In the first step, posterior distributions are obtained for the chirp mass, symmetric mass ratio and $\tilde{\Lambda}$  for a catalog of expected events as shown in the box at the top left of the flowchart.  To this end, we use the \texttt{GWBENCH} implementation of Fisher information matrix \cite{Borhanian:2020ypi} for fast computation of the posteriors but in a real data analysis problem posteriors would be obtained using a Bayesian inference algorithm. 

    \item From the posterior distribution of the chirp mass and the symmetric mass ratio we can derive the posterior distribution of the companion masses $m_1$ and $m_2$ and the mass ratio $q.$ 

    \item In the next step (first downward arrow from top left), the first universal relation Eq.\,(\ref{eq:unirel1}) is used to eliminate $\Lambda_a$ from Eq.\,(\ref{eq:lambdatilde}) to arrive at an expression for $\tilde\Lambda$ that depends only on the symmetric combination of the tidal parameters and the mass ratio $q,$ namely $\tilde\Lambda = \tilde\Lambda(\Lambda_s, q).$

    \item The transcendental equation $\tilde\Lambda = \tilde\Lambda(\Lambda_s, q),$ together with the posterior distributions of $q$ and $\tilde\Lambda,$ is then solved to obtain the posterior of $\Lambda_s$ and the first universal relation is deployed once again to derive the posterior of $\Lambda_a.$ From the $\Lambda_s$ and $\Lambda_a$ posteriors, it is straightforward to deduce the posteriors of the individual tidal deformabilities (second downward arrow from top left). 

    \item In the next step, the second universal relation Eq.\,(\ref{eq:cvslambda}), together with the $m_1$ and $m_2$ posteriors, is used to infer the posteriors of the neutron star radii (third downward arrow from top left). 

    \item The above steps essentially give us posteriors in the mass-tidal deformability ($m$--$\Lambda$) or mass-radius ($m$-$R$) plane (bottom, horizontal, upward turning arrow).

    \item Given an EOS, the mass-radius and mass-tidal deformability curves can also be obtained by solving the TOV equations (top horizontal arrow) and can be compared with the posterior distributions.
\end{enumerate}

The above pipeline essentially compares the measured parameters (e.g., the mass-radius curve) with the predictions of a set of models to choose the best EOS consistent with the data. However, the systematic biases in the tidal parameters (arising from the approximate nature of the first universal relation) and radii (arising from both the first and second universal relations) could favor the wrong model. We next discuss a strategy to remedy the biases by incorporating the residuals as part of the universal relation. We differ from Ref.~\cite{LIGOScientific:2019eut} in that we do not sample the tidal deformability of the companions but instead only the reduced tidal deformability. Additionally, instead of marginalizing over the residuals while inferring radii and individual tidal deformabilities we correct for them at the time of model selection, which differs in one key aspect from Ref.~\cite{LIGOScientific:2018cki}. Oue method is equivalent to calculating $\Lambda_s-\Lambda_a$ relation for each EOS.

\subsection{Mitigating systematic errors in the tidal deformability and radius}
\label{subsec:bias_corr}
In this section, we elucidate our method to correct for the systematic biases incurred in the estimation of the individual tidal deformabilities and radii due to the use of \emph{quasi}-universal relations. The biases, being systematic and not statistical, do not asymptote to zero with increasing SNR but become the dominant source of error as the statistical errors decrease. 

\paragraph{Using residuals to correct systematic biases}
Our proposal is to correct for systematic errors at the time of model selection. To do so, we begin with a specific EOS model and compare its predictions of the mass-$\Lambda$ or mass-radius curve with the posterior distributions of the same but obtained from gravitational-wave observations, typically computing the $\chi^2$ or Bayesian evidence for the model given the data. At this point, we know not only the \emph{prior probability} for the model but also the \emph{residuals of the model} with respect to the universal relations. We can, therefore, subtract the known residuals for the model from the measured tidal deformabilities and radii before calculating the $\chi^2$ or evidence for the model. We emphasise that the residuals \emph{cannot} be used to obtain EOS-agnostic tidal deformabilities or radii as the corrections are specific to the EOS model chosen. 

In more detail, we start with an EOS-agnostic estimate as calculated using the procedure outlined in the previous section.
%We find that this introduces an error in the determination of the correct EOS model. We suggest a procedure that can be used to correct these biases given an EOS model (as is the case during model selection). Such corrections also improve the variance of a distribution as described in detail in~\citep{Feigelson:2012xg}. 
Given an EOS model and the measured masses from gravitational wave observations, one can infer the corresponding distributions of $\Lambda$s and radii using the $m$--$\Lambda$ or the $M$--$R$ curves for the EOS model. In the presence of systematic biases these curves would shift from those inferred using universal relations, which also broaden due to the statistical uncertainties in their measurements. Depending on the SNR of an event, statistical uncertainties can be larger than systematic biases. In that case, correcting for the systematics would not significantly improve the evidence for the correct model. But as we will see, the ECS network will observe events where such a correction would be critical. Moreover, as mentioned before, biases accumulate as the evidences from multiple events are combined and, therefore, the correction would be important even for the less sensitive HLVKI+ network.

%\arnab{We can add the part about point estimates and their variance but it needs to be explained in a bit more detail. Currently it feels out of place with the rest of the description.}
% biases (defined as $B[\hat{\theta}] = E[\hat{\theta}] - \theta $ where $\hat{\theta}$ is point estimate of the parameter in question and $\theta$ is the true value.)
% \begin{equation}
%     E[(\hat{\theta}-\theta)^2] = Var(\hat{\theta}) + (E[\hat{\theta}]-\theta)^2 
% \end{equation}

\begin{figure}[h]
    {\centering
    \includegraphics[width=\columnwidth]{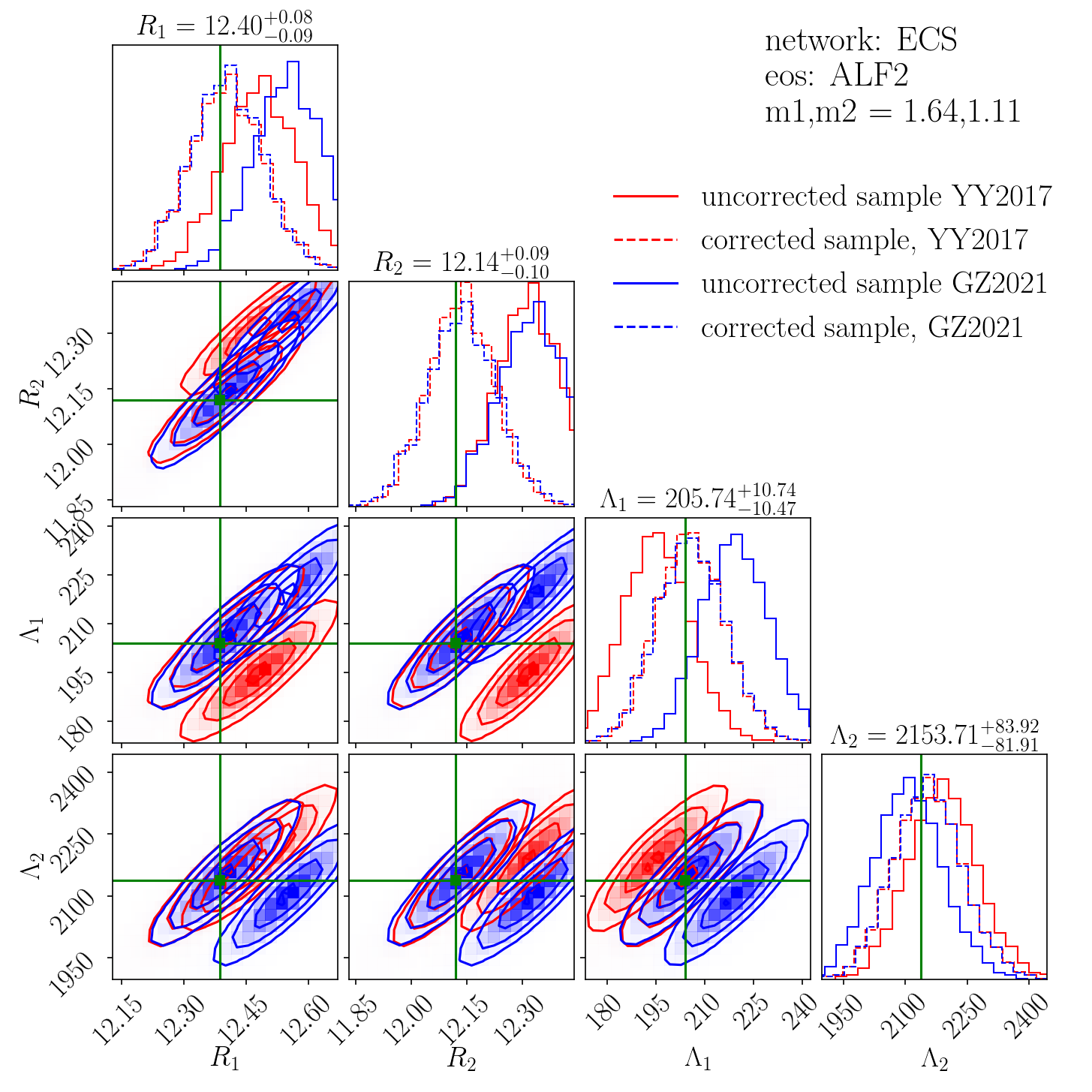}}
    \caption{The comparison between the uncorrected and corrected distributions of radii as well as the tidal deformabilities of the component neutron stars are shown for a fiducial event in our population. The uncorrected and respective corrected distributions are shown with respect to two universal relations shown in the figure -- YY2017 \citep{Yagi:2016bkt} and GZ2021 \citep{Godzieba:2020bbz}. The event was simulated in the ECS network with the ALF2 EOS. The true values corresponding to the injected masses of $m_1=1.64 \rm M_{\odot}$ and $m_2=1.11 \rm M_{\odot}$ are shown with green horizontal and vertical lines and values on top of each subplot is the median values of the corresponding distributions. Note that both components need not be massive for the bias to be significant.} 
    \label{fig:pipeline}
\end{figure}

\paragraph{Bias correction for a specific event}
In Fig.~\ref{fig:pipeline} we show the impact of bias correction for a fiducial event in our population. Here, an event with the ALF2 EOS is chosen from the ECS network. The component masses of the binary system are $m_1=1.63 \rm M_{\odot}$ and $m_2=1.11 \rm M_{\odot}$ and the event had an SNR of 103. 

The green horizontal and vertical lines show the true tidal deformabilities and radii corresponding to these masses. The EOS agnostic measurements for the event are depicted by solid lines in red and blue for the universal relations proposed in \textcite{Yagi:2016bkt,Chatziioannou:2018vzf} and \textcite{Godzieba:2020bbz}, respectively. The corrected distributions are shown with the same colors but in dashed lines. Evidently, the shift in the estimation of the radii with respect to their true values vanishes upon correction.

\begin{figure*}[ht]
    {\centering
    \includegraphics[width=0.9\textwidth]{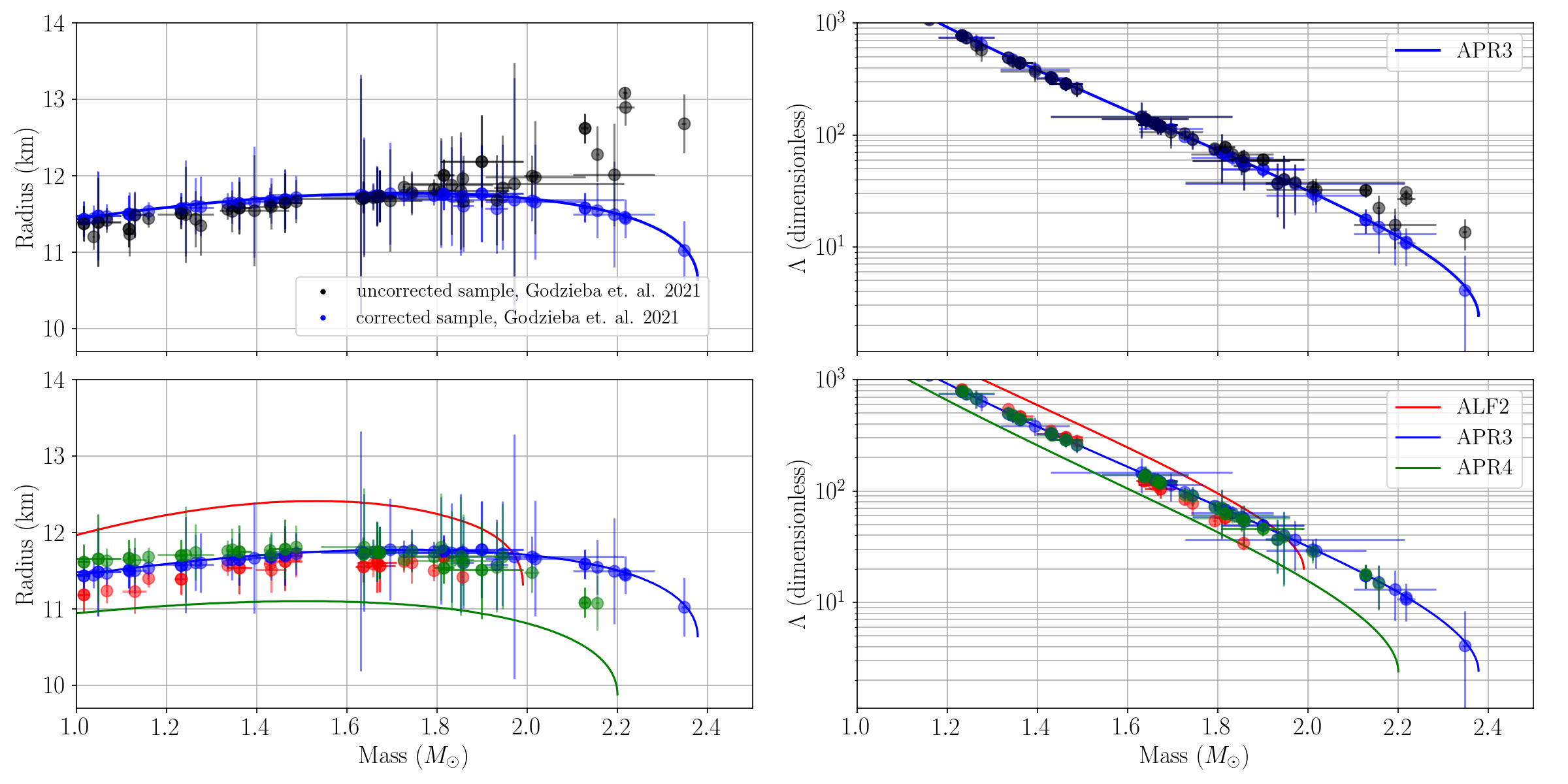}}
\caption{Inferred radii and $\Lambda$ before and after bias correction for 30 random events with SNR values larger than 100 in the ECS network, where injection EOS is APR3. Such corrections can also be included separately for two universal relations for tidal deformability and compactness correction. However we find the second universal relation to be quite accurate. Hence, we report the joint correction due to the relevance of radii as key parameter constraining EOS. We find the recovery of $\tilde{\Lambda}$ be quite accurate, however individual tidal deformabilities show larger variance. We report the impact of correction on individual $\Lambda$.}\label{fig:radiuserror}
\end{figure*}

\paragraph{Bias correction for a population of events}
In the top row of Fig.~\ref{fig:radiuserror}, we show in gray filled circles the median values of the radii (left panel) and tidal deformabilities (right panel) as a function of the component mass, for a randomly selected population of 30 events with SNRs greater than 100 in the ECS network. Tidal deformabilities and radii for this set of events were generated for the APR3 EOS model shown as blue solid line. Note that the median values of $\Lambda$ and radii differ significantly from the corresponding EOS curves for the largest and the smallest masses as expected from the systematic bias plot in Fig.~\ref{fig:syst_bias}. Note that the smallest masses are part of a small mass ratio system.

Correcting the radii and tidal deformabilities using residuals appropriate for APR3 yield filled blue circles. Note that the corrected values are now much closer to the true EOS curves. Since the SNRs are pretty large, statistical uncertainties for these events are far smaller than the systematic errors, demonstrating the extent of the biases and effectiveness of the corrections. Note also that systematic biases in the case mass-radius curve, affected by two universal relations, are far greater than the mass-$\Lambda$ curve, affected by only the first universal relation.

\subsection{Model Selection Criteria}
\label{subsec:model_sel}
We now describe our model selection method for a population of events satisfying a given SNR threshold. 
% First, for a given population of events we construct bias-corrected distributions of the tidal deformabilities and radii for each model in a set of EOS models for which evidence is required. 
As before, the population contains 50 events each having an SNR of at least 100 and the events were generated assuming APR3 to be the true EOS. To select a model among a set of EOSs, say, ALF2, APR3, and APR4, we correct for biases in  EOS-agnostic distributions using residuals appropriate to each EOS in the set. The bottom panels of Fig.~\ref{fig:radiuserror} show how uncorrected median values (gray circles in the top panels) shift to new positions when bias corrections appropriate for the ALF2 (red circles), APR3 (blue circles) and APR4 (green circles) models were applied. Also shown are the mass-$\Lambda$ (right panel) and mass-radius (left panel) curves corresponding to the three models. Evidently, the population matches better with an EOS model when the bias correction corresponds to the true EOS. 

This procedure can be repeated for radii and tidal deformabilities sampled from the posterior distributions of tidal deformabilities and radii, which can then be used in computing the $\chi^2$ relative to each model (see below).  We emphasize, however, that in this process we did not consider the errors in the distributions of masses as they are negligibly small but it is straightforward to account for statistical uncertainties in masses. 

\paragraph{Model selection with chi-square} Next, to quantify how well the inferred mass-radius curve matches with an EOS model we calculate the $\chi^2$ between the two using 
\begin{equation}
\label{eq:chisq}
    \chi^2_{k,M} = \frac{1}{N} \sum_{n=1}^{N}  \frac{(x_{n}^{k}-x_{n}^{M})^2}{\sigma_{n}^{2}}
\end{equation}
where $x_n^M$ is the tidal deformability (or the radius) of the EOS model $M$ corresponding to the $n$th event and $x_n^k$ is a sample drawn from the bias-corrected posterior distribution (for model $M$) of the $n$th event whose standard deviation is $\sigma_n$. The set of values $\{m_n, x_n^k\},$ $n=1,\ldots,N,$ form the $k$th realisation of the mass-$\Lambda$ (or the mass-radius) curve, giving the chi-square value $\chi^2_{k,M}$ for model $M.$ We compute $\chi^2$ for 1000 realizations of the mass-$\Lambda$ (or the mass-radius) curves to obtain the $\chi^2$ distribution for a given model.

% \begin{figure*}[ht]
%     \centering
%     \includegraphics[width=2\columnwidth]{figs/eventfigs_correction/mrlam_sampling.png}
%     \caption{Sampling mass-radius and mass-$\Lambda$ curves from observed events and their uncertainty. Each of the curves are then used to calculate $\chi^2$.}
%     \label{fig:radlamsampling}
% \end{figure*}

% In Fig.~\ref{fig:radlamsampling}, we show 1000 realizations of the inferred M-$\Lambda$ or M-R curve using the above population which are then used to construct the $\chi^2$ distribution. 
% \arnab{I don't know if this figure is really needed. I feel it is not adding much.}

\begin{figure*}[ht]
    {\centering
    \includegraphics[width=2\columnwidth]{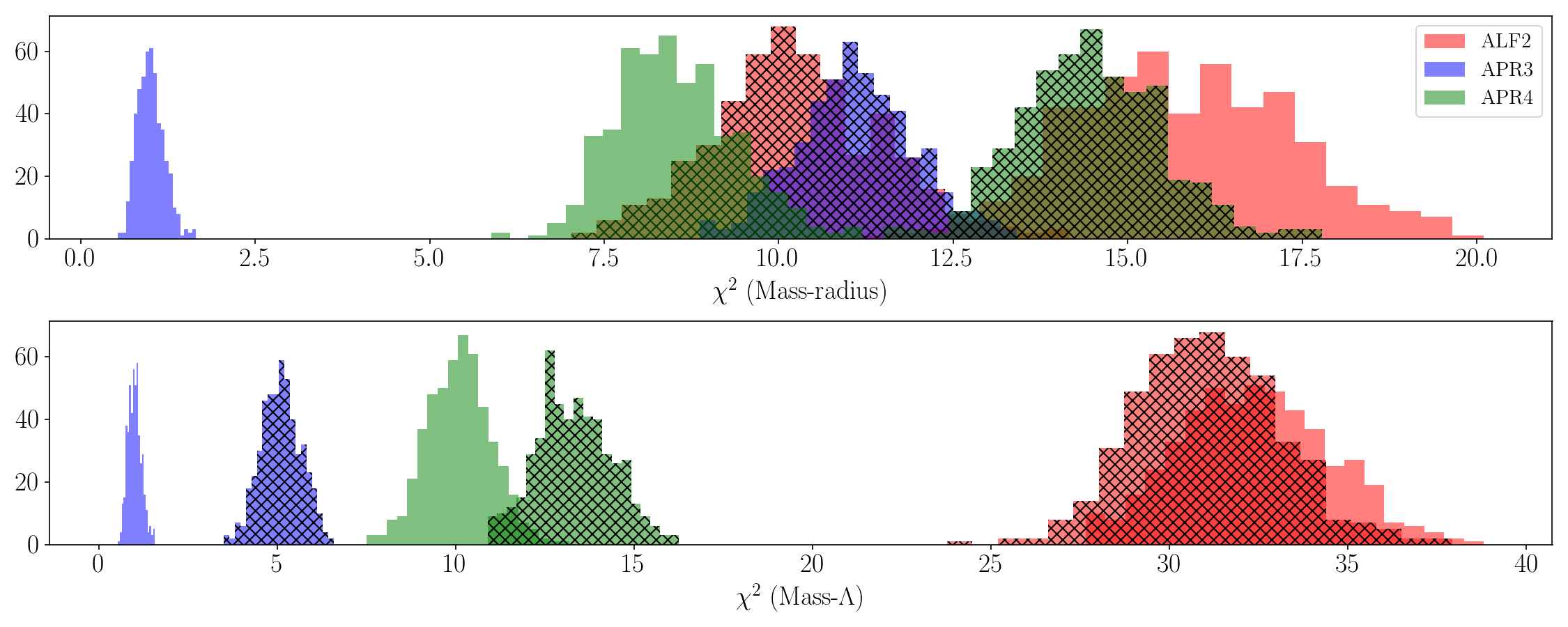}}
    \caption{Distribution of $\chi^2$ for a sample of 30 random events with SNR larger than 100 in the ECS network where the true EOS model is APR3. Bias-corrected $\chi^2$ distributions are shown as unhatched histograms while uncorrected ones are shown as hatched histograms, for radii (top panel) and $\Lambda$ (bottom panel). Low values of $\chi^2$ at the left end of the panels for bias-corrected $\chi^2$ distributions imply that the reference EOS model is distinguishable from the rest. Similar values of $\chi^2$ for the reference and alternative EOS models for bias-uncorrected $\chi^2$ distributions in the middle of the panels implies that it is hard to distinguish the reference EOS model from an alternative. Note that the distributions are not normalized which accounts for why some look thinner than others.}
    \label{fig:chisq_lamrad}
\end{figure*}

We construct such $\chi^2$ distributions for each model in the set of all EOSs. For an unbiased model selection, the $\chi^2$ distribution constructed from a population of bias-corrected posteriors should have the smallest $\chi^2$ value for the true EOS. In Fig.~\ref{fig:chisq_lamrad}, we plot $\chi^2$ histograms before (hatched) and after (unhatched) bias correction.  APR3 was chosen as the true EOS model for the simulated population in the ECS network and 30 events with SNR larger than 100 were chosen at random from the full population for model selection. It is evident from the plots that bias correction vastly improves model selection, giving lower $\chi^2$ values for the correct EOS model for both the mass-radius (top panel) and mass-$\Lambda$ comparisons. Furthermore, for this sample population of events model selection without bias correction will either be inconclusive or lead to the selection of an incorrect model as the true model, once again highlighting the efficacy of bias correction. 

\paragraph{KS test for model selection} We use the directional Kolmogorov-Smirnov test statistic~\cite{Feigelson:2012xg} to distinguish two (near-by) $\chi^2$ distributions. Let $\{\chi^2_{k,T}\}$ and $\{\chi^2_{k,A}\}$ be the $\chi^2$ distributions for two EOS models $T$ and $A$, respectively, and let $\Phi_{T}$ and $\Phi_{A}$ be the corresponding cumulative distribution functions (CDFs). We have chosen $T$ to be the true model and $A$ to be an alternative. Our null ($\mathcal{H}_0$) and alternative ($\mathcal{H}_1$) hypotheses are defined as: 
\begin{itemize}
    \item $\mathcal{H}_0$: $\Phi_T(\chi^2) \leq \Phi_A(\chi^2)$ for all $\chi^2$, and
    \item $\mathcal{H}_1$: $\Phi_T(\chi^2) > \Phi_A(\chi^2)$ for at least one $\chi^2$.
\end{itemize}
This implies that for an unbiased model selection, where the $\chi^2$ values for the true EOS are expected to be smaller and mostly non-overlapping with the $\chi^2$ values for the false EOS, the KS statistic will be close to 1. On the contrary, if two $\chi^2$ distributions are indistinguishable or the model selection is biased meaning that the $\chi^2$ distribution for the incorrect EOS has lower values, the KS statistic will approach 0. In order to not be misled by the fluctuations of a specific realization of a population of events, we bootstrap over 500 distinct realizations of the population. We verified that this is large enough to describe the variation in different realizations of the population. 

\subsection{$L_2$ distance between EOS models}
\begin{figure*}%[H]
    {\centering
    \includegraphics[width=0.9\textwidth]{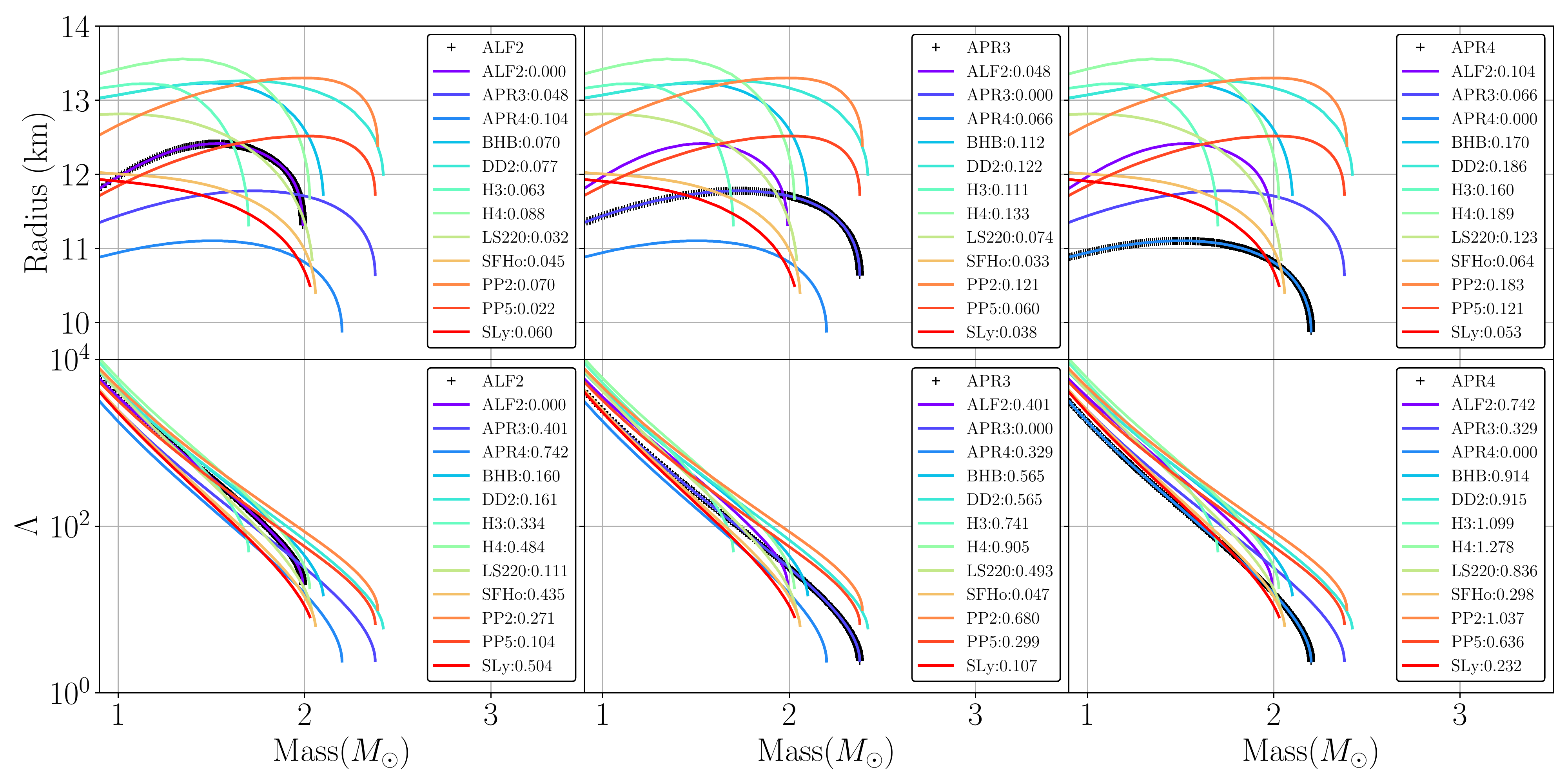}}
    \caption{In this figure we plot mass-radius (top panels) and mass-$\Lambda$ (bottom panels) for a selection of 12 different EOS models and their $L_2$ distance from a reference EOS model thick black line, ALF2 (left panels), APR3 (middle panels) and APR4 (right panels). The range of integration in the $L_2$ norm is taken from 1 $M_\odot$ up to the smaller of the maximum mass of the two EOSs.} 
    \label{fig:L2eos}
\end{figure*}

We exemplify the limitations of any model selection method in the following. First, the mass-radius curves of EOSs are not unique for all masses of NSs, but only a subset of them (see Fig.~\ref{fig:L2eos}). Many of the mass-radius or mass-tidal deformability EOS pairs intersect each other making those pairs identical at and around the point of intersections (for example ALF2-SLy around $1.4 M_\odot$). Some pairs of EOSs are similar, or even identical, over a large range of masses, which means they are distinguishable only for events in the non-overlapping region (for example, BHB and DD2 overlap for masses $< 1.5 M_\odot$ in mass-radius plane). Hence, identification of the correct model also depends on the component masses of NSs in the catalog of BNS events, and only events at which a pair of EOSs do not intersect can distinguish them. Luckily, there are many EOS pairs that do not intersect and differ over the entire allowed range of masses. It is easier to discriminate between such pairs.

The $L_2$-distance between the mass-radius or mass-$\Lambda$ curves can be used as a measure of the distance between a pair of EOS models. We propose that two models are more easily distinguishable greater is the $L_2$-distance between the corresponding curves defined as: 
\begin{equation}
    L_{2,R}(A,B) \equiv N_R \int_{m_l}^{m_u} [R_A(m)-R_B(m)]^2 dm 
\end{equation}
\begin{equation}
    L_{2,\Lambda}(A,B) \equiv N_\Lambda \int_{m_l}^{m_u} [\Lambda_A(m)-\Lambda_B(m)]^2 dm 
\end{equation}
where  $N_R$ and $N_\Lambda$ are normalization constants to render the distances dimensionless chosen to be:
\begin{eqnarray}
    N_R \equiv \left [ \int_{m_l}^{m_u} R_A^2 dm  \int_{m_l}^{m_u} R_B^2 dm \right]^{-1/2}, \cr
    N_\Lambda \equiv \left [ \int_{m_l}^{m_u} \Lambda_A^2 dm \int_{m_l}^{m_u} \Lambda_B^2 dm \right ]^{-1/2}, \nonumber
\end{eqnarray}
$R_A(m)$ and $\Lambda_A(m)$ ($R_B(m)$ and $\Lambda_B(m)$) are the mass-radius and mass-$\Lambda$ curves corresponding to model $A$ ($B$), $m_l$ is the smallest NS mass in the observed population and $m_u$ is the smaller of the maximum mass allowed by models $A$ and $B.$  Fig. \ref{fig:L2eos} plots  mass-radius (top panels) and mass-$\Lambda$ (bottom panels) curves for 12 different EOS models and their  $L_2$ distance from  ALF2 (left panels), APR3 (middle panels) and APR4 (right panels). 

It is evident from the figure that mass-$\Lambda$ and mass-radius curves are complementary in measuring the distance between different EOS models. For example, DD2 and PP2 are most distant from APR3 in the mass-radius plane (top-middle panel) while H4 happens to be most distant in the mass-$\Lambda$ plane (bottom-middle panel). Likewise, APR4 is most distant from ALF2 in both the mass-radius (top-left) and mass-$\Lambda$ (bottom-left) planes.  This feature highlights the importance of model selection in both, mass-radius and mass-tidal deformability plane. On the contrary, the model closest to the reference EOS happens to be the same no matter the parameter space. For example, SFHo has the smallest distance from APR3 in both mass-radius and mass-$\Lambda$ planes. 

\section{Model selection with A+ and XG networks}
\label{sec:results}
In this section, we will determine the ability of the ECS network to distinguish between different EOS models using the measures introduced in the previous section. We will also consider the HLVKI+ network as a fiducial. Results for mass-$\Lambda$ curves are presented here; conclusions drawn from mass-radius curves are similar.

\begin{figure*}[ht]
    {\centering \includegraphics[width=2\columnwidth]{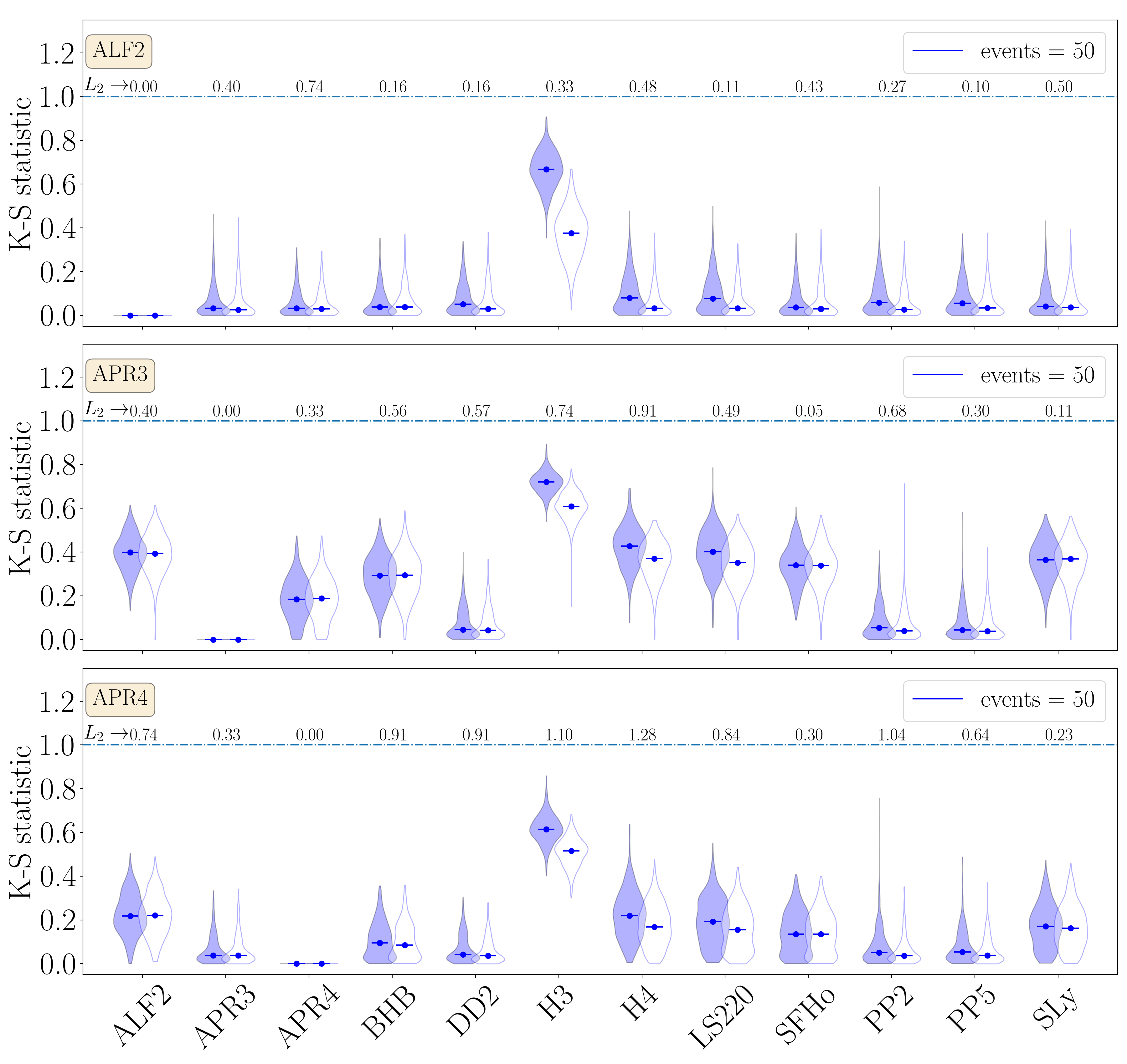}}
    \caption{The plot shows the distribution of the directional KS statistics between the reference model [ALF2 (top), APR3 (middle) and APR4 (bottom)] and an alternative listed along the horizontal axis for mass-$\Lambda$ curves. The statistic is computed for 50 simulated events with $\mbox{SNR}>10$ in the HLVKI+ network. Also listed along the top of each panel are the $L_2$ distances between the reference model and the alternatives. Blue horizon lines in violin plots are the median values of the KS statistic. Filled (empty) violins are the KS statistic for bias-corrected (uncorrected) posteriors of the tidal deformability. The median posterior is close to zero when the alternative model is the same as the reference model. However, the medians are small for many alternative models. Thus, the HLVKI+ network will not have the sensitivity to determine the right EOS.}
    \label{fig:ks-dist_mlam_HLVKI+_snr10_ALF2}
\end{figure*}

\begin{figure*}[ht]
    {\centering \includegraphics[width=2\columnwidth]{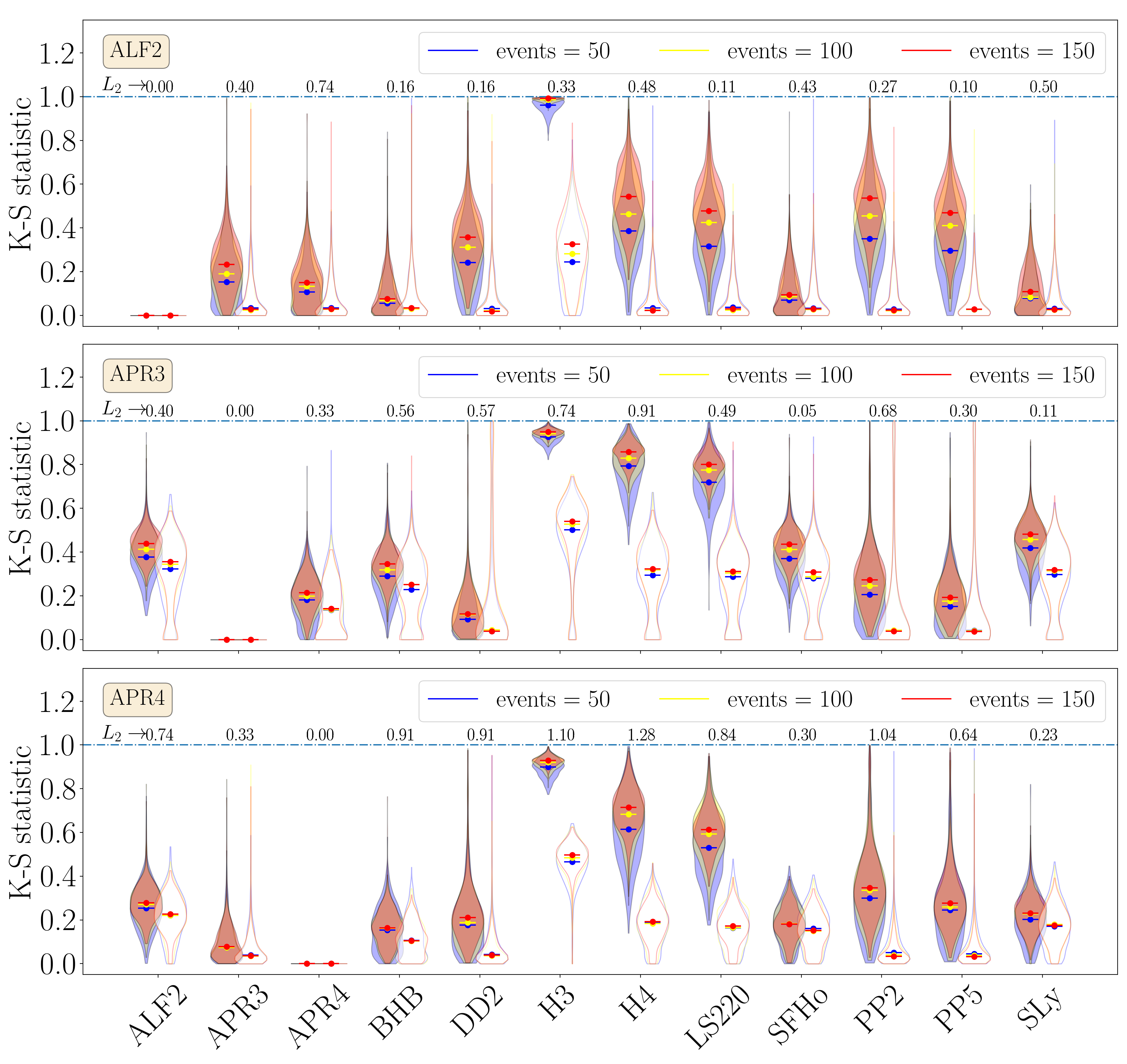}}
    \caption{Same as Fig.~\ref{fig:ks-dist_mlam_HLVKI+_snr10_ALF2} but for the ECS network with a random sample of 50 (blue violins), 100 (yellow violins) and 150 (red violins) events with SNR$>$10. It is evident that bias correction increases the KS statistic for incorrect models leaving the statistic unchanged for the reference EOS. As expected, the statistic has greater discriminatory power between different models for larger number of events.}
    \label{fig:ks-dist_mlam_ESa4cCa4c_snr10_ALF2}
\end{figure*}

\begin{figure*}[ht]
    {\centering \includegraphics[width=2\columnwidth]{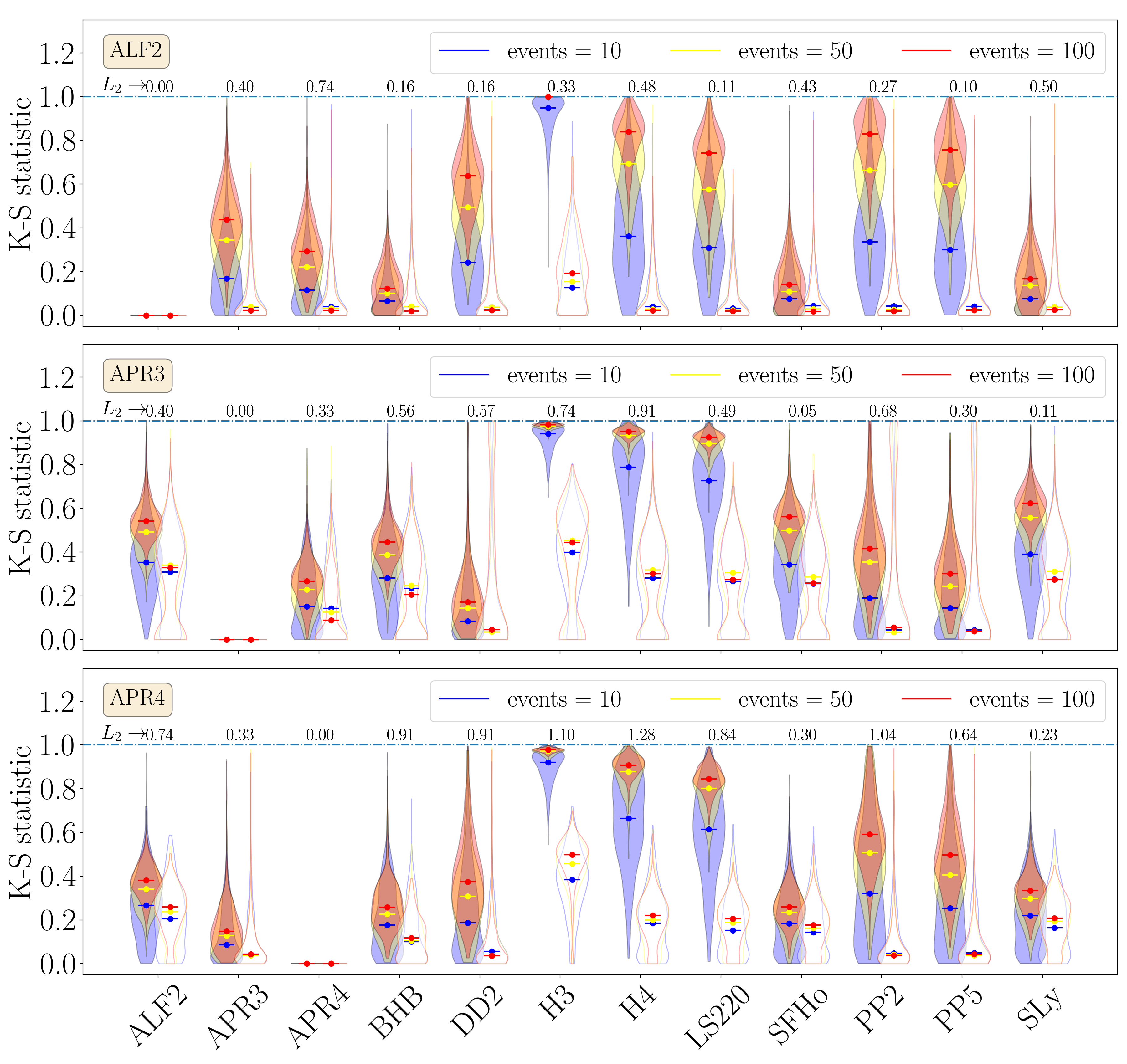}}
    \caption{Same as for Fig.~\ref{fig:ks-dist_mlam_ESa4cCa4c_snr10_ALF2} but for events with $\mbox{SNR}>$30. The reference model is now distinguishable from many more models than when the SNR threshold was 10. A catalog of 150 events with SNR$>30$ will be able to identify the right EOS model with good confidence.}
    \label{fig:ks-dist_mlam_ESa4cCa4c_snr30_ALF2}
\end{figure*}

\begin{figure*}[ht]
    {\centering \includegraphics[width=2\columnwidth]{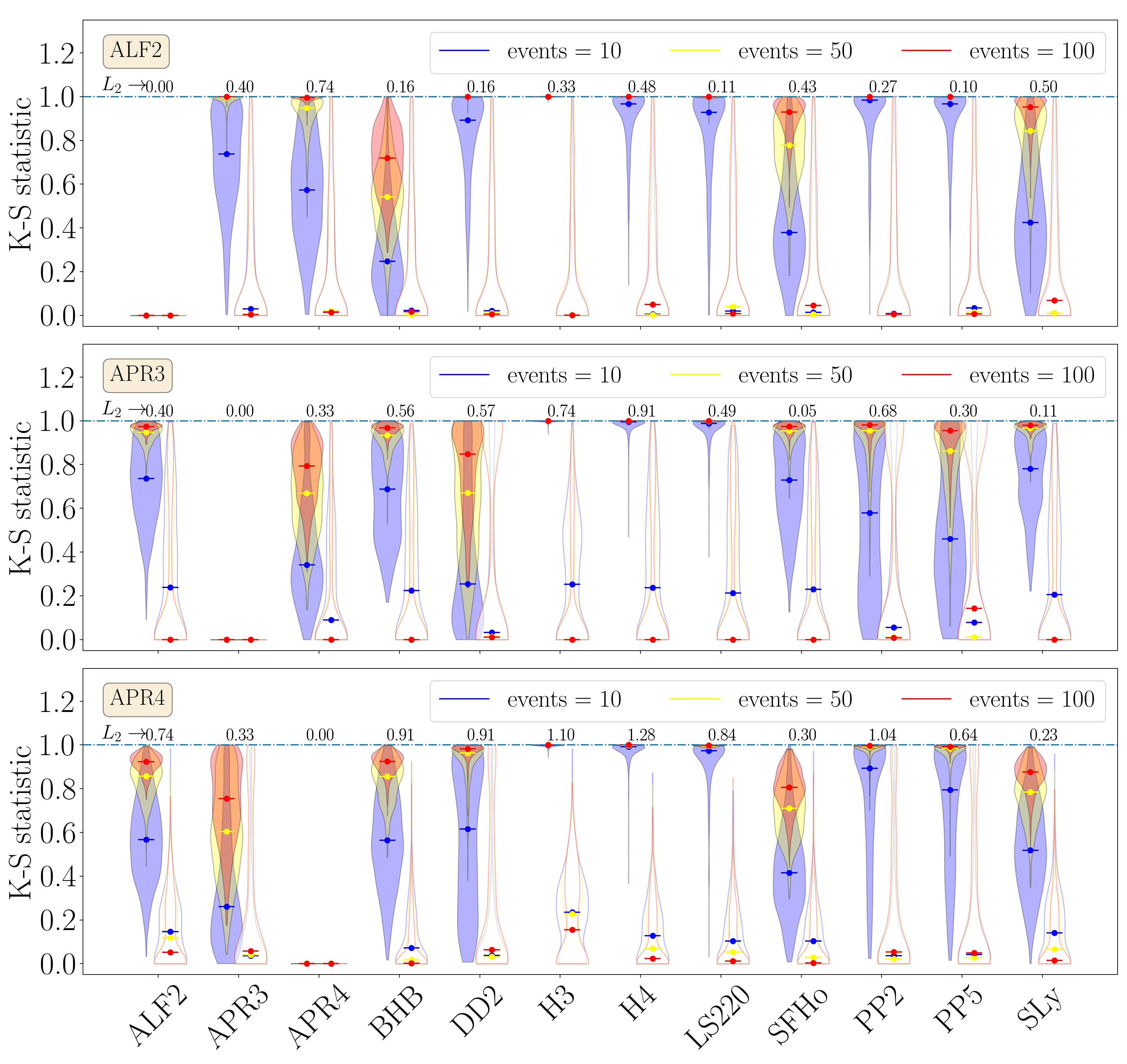}}
    \caption{Same as for Fig.~\ref{fig:ks-dist_mlam_ESa4cCa4c_snr10_ALF2} but for events with $\mbox{SNR}>100.$ Now, just 10 events suffice to select the correct model.} 
    \label{fig:ks-dist_mlam_ESa4cCa4c_snr100_ALF2}
\end{figure*}

As discussed in the previous section, the preferred model is the one for which the $\chi^2$ between the model and the inferred realization of an EOS curve is the smallest. 
Following the procedure described in the last section (i.e., construction of multiple realizations of the mass-$\Lambda$ curve for a given population by sampling from the posterior distribution of $\Lambda$ and bootstrapping over several populations of events) gives a distribution of the KS statistics, which is plotted in Figs.~\ref{fig:ks-dist_mlam_HLVKI+_snr10_ALF2}, \ref{fig:ks-dist_mlam_ESa4cCa4c_snr10_ALF2}, \ref{fig:ks-dist_mlam_ESa4cCa4c_snr30_ALF2}, and \ref{fig:ks-dist_mlam_ESa4cCa4c_snr100_ALF2}. In order to test the robustness of our method ALF2 (top panels), APR3 (middle panels), and APR4 (bottom panels) were in turn considered to be the true EOS model and model selection was performed over a set of 12 EOSs depicted in Fig.~\ref{fig:L2eos}. 

\paragraph{HLVKI+ Network}
Fig.~\ref{fig:ks-dist_mlam_HLVKI+_snr10_ALF2}, plots the distribution of the KS statistic for different realizations of a set of 50 events with SNR$>$10, observed in the HLVKI+ network over a one year observing period. The distribution of the KS statistic between the reference model (ALF2 top, APR3 middle and APR4 bottom) and an alternative (shown along the $x$ axis) is shown for both bias-corrected (filled violins) and uncorrected (empty violins) posteriors. Blue markers correspond to the KS statistic for the bias-corrected median realization. Also listed at the top of each panel are the $L_2$ distance between the reference EOS model and an alternative shown along the $x$ axis.

We see that in general bias correction increases the KS statistic for the incorrect models. 
This implies that the population realizations that were otherwise indistinguishable or giving rise to an incorrect model selection have better distinguishability. Regardless of the true EOS, H3 is more readily distinguishable from the reference models. Of further note is that APR3 is the easiest to distinguish among the three reference EOS. But otherwise it will be difficult to converge on the true EOS using the HLVKI+ network.

It may be tempting to conclude that the reference model does not have long tails towards KS statistic $>1$ but this is not correct since in a real experiment we would not know the true EOS model and hence likely to obtain violins with low medians and long posteriors for many models in the HLVKI+ detector network.

\paragraph{ECS Network}
Next, we turn to the next-generation detector network of ECS. Instead of using the full set of the detectable population of events, we choose a sub-population of events to distinguish between different EOS models. In Fig.~\ref{fig:ks-dist_mlam_ESa4cCa4c_snr10_ALF2}, we randomly choose 50, 100, and 150 events from the full population of events but with network $\mbox{SNR} > 10$. We again find that H3 is more readily distinguishable from the reference models. Moreover, distinguishability increases, as expected, with the number of events considered for model selection. We confirm once again that bias correction leads to better distinguishability and correct model selection. This is easily seen for the EOS models H3, H4, and LS220. When the reference EOS is ALF2, correcting the bias in the tidal deformability makes the median realization of a population of 150 events distinguishable. For an injected APR3 or APR4 EOS, it makes almost every realization of the population distinguishable. We note that the events in this population are expected to be predominantly low SNR.

In Figs.~\ref{fig:ks-dist_mlam_ESa4cCa4c_snr30_ALF2} and~\ref{fig:ks-dist_mlam_ESa4cCa4c_snr100_ALF2} we take the minimum SNR to be 30 and 100, respectively, and also consider three sub-populations of 10, 50, and 100 events. We find that higher SNR events increase the distinguishability across the set of EOS models considered. We report that the median realization of a population of 100 events having a minimum SNR of 100, as is expected from a year of observation, can distinguish the correct EOS model among all the models considered in this work. We further note a divergence in the KS statistic where the statistic for the uncorrected tidal deformability estimates decreases for the SNR threshold of 100 compared to an SNR threshold of 30 for the same number of events. In contrast, the statistic increases for bias-corrected distributions. We conclude that 
the ECS network, with its ability to frequently detect high SNR events, will have the ability to precisely determine the EOS of dense matter.

\section{Conclusions and Discussions}
\label{sec:conclusion_discussion}
The determination of the individual tidal deformabilities and radii of a BNS system using gravitational-wave observations suffers a systematic bias due to the use of universal relations. Traditionally, this bias is accounted for by marginalizing over the residuals in the universal relations and, in the process, trading in the systematic errors for statistical ones. This procedure, nevertheless, results in a biased estimate of the relevant quantities and, therefore, to a biased model selection. In this paper, we propose a different method for an unbiased model selection that does not involve residual marginalization and, hence, is, in principle, superior to the existing procedure.

Model selection is performed by calculating the $\chi^2$, or alternatively, the evidence, for a model given an observation. Since a specific model is assumed in the computation of the evidence, one has knowledge of the residual associated with the model and can rectify the measured quantities by their value at that point. We note that this method does not produce model-agnostic distributions of masses and tidal deformabilities since the corrections are for a specific model. 

Given an observed astrophysical population satisfying a minimum SNR condition, we create an EOS curve in the mass-$\Lambda$ and mass-radius planes by drawing a representative point from the posterior distributions. We repeat this process multiple times to get a band of values in the respective planes. This bootstrapping procedure helps to not bias the inferences to a specific realization of an EOS curve. Next, we consider a set of models to be hierarchically ranked based on their $\chi^2$ values. Here, we calculate the $\chi^2$ for every model in the set, first with the EOS-agnostic band and then with a band obtained after correcting for systematic errors specific to chosen models. This results in two sets of $\chi^2$ distributions which we call the \textit{uncorrected-}$\chi^2$ and \textit{corrected-}$\chi^2$ distribution sets. 

For an unbiased inference, the model corresponding to the distribution with smaller $\chi^2$ values is the preferred model. Unfortunately, this is not always the case since the bias-\textit{uncorrected-}$\chi^2$ distributions for the wrong model could have lower $\chi^2$ values. Luckily, the bias-\emph{corrected-}$\chi^2$ distributions are always the lowest for the true model. However, if two EOS curves are similar (i.e., small $\rm L_2$ distance) then their $\chi^2$ distributions overlap and the models might be indistinguishable. To quantify the distinguishability of two $\chi^2$ distributions and unbiased model selection, we use the directional KS statistic. A statistic for a pair of distributions close to 1 indicates an unbiased model selection and complete distinguishability of the pair. On the other hand, if the statistic is close to zero, the pair of distributions either overlap or model selection is biased\footnote{In this study, we restrict the calculation of the KS statistic to the case where one of the $\chi^2$ distributions is that of the reference model and so the KS statistic indicates how likely the reference model is to be chosen as the preferred one.}. When the statistic is close to zero it is not possible to discriminate overlapping distributions from biased ones. Specifically, one can have non-overlapping $\chi^2$ distributions with the incorrect model having smaller values, in which case the statistic will necessarily be zero. 

To ensure that our inferences are not biased by the specific realization of our population, we bootstrap over the observed population. We discussed the violin plots in the previous section, which show the distribution of the KS statistic between a reference model and an alternative model. We observe that the injected model is more likely to be recovered for bias-\textit{corrected-}$\chi^2$ distributions and for a greater number of observations. The two main advantages of our method over the residual marginalization method are as follows. First, we do not need to sample over the parameters that model the residuals and hence are computationally favored. Second, the statistical errors are mostly unaffected in our method and, therefore, our model selection has greater sensitivity. Though not explored in this work, it would be interesting to do a direct comparison of the model selection prowess of our method versus the method of residual marginalization. More specifically, this would reveal the effect of the latter method on the statistical errors and its subsequent effect on the unbiased distinguishability of nearby EOS curves. We leave this to future work.

\acknowledgements
We thank Philippe Landry for the internal review and his several suggestions to improve the manuscript. We thank Nathan Johnson-McDaniel, David Radice and Sukanta Bose for their comments and critical questions. We thank Monica Bapna and P. Ajith for useful discussions during a previous work showing the derivation of a more general methodology of model selection in an unpublished work. We thank Daniel Godzieba for sharing two EOS from his piecewise polytropic collection. AD and BSS were supported in part by NSF grant numbers PHY-2012083, PHY-2207638 and AST-2006384. This material is based upon work supported by NSF's LIGO Laboratory which is a major facility fully funded by the National Science Foundation.  

\bibliographystyle{apsrev4-1}
\bibliography{references}

\begin{thebibliography}{66}%
\makeatletter
\providecommand \@ifxundefined [1]{%
 \@ifx{#1\undefined}
}%
\providecommand \@ifnum [1]{%
 \ifnum #1\expandafter \@firstoftwo
 \else \expandafter \@secondoftwo
 \fi
}%
\providecommand \@ifx [1]{%
 \ifx #1\expandafter \@firstoftwo
 \else \expandafter \@secondoftwo
 \fi
}%
\providecommand \natexlab [1]{#1}%
\providecommand \enquote  [1]{``#1''}%
\providecommand \bibnamefont  [1]{#1}%
\providecommand \bibfnamefont [1]{#1}%
\providecommand \citenamefont [1]{#1}%
\providecommand \href@noop [0]{\@secondoftwo}%
\providecommand \href [0]{\begingroup \@sanitize@url \@href}%
\providecommand \@href[1]{\@@startlink{#1}\@@href}%
\providecommand \@@href[1]{\endgroup#1\@@endlink}%
\providecommand \@sanitize@url [0]{\catcode `\\12\catcode `\$12\catcode
  `\&12\catcode `\#12\catcode `\^12\catcode `\_12\catcode `\%12\relax}%
\providecommand \@@startlink[1]{}%
\providecommand \@@endlink[0]{}%
\providecommand \url  [0]{\begingroup\@sanitize@url \@url }%
\providecommand \@url [1]{\endgroup\@href {#1}{\urlprefix }}%
\providecommand \urlprefix  [0]{URL }%
\providecommand \Eprint [0]{\href }%
\providecommand \doibase [0]{http://dx.doi.org/}%
\providecommand \selectlanguage [0]{\@gobble}%
\providecommand \bibinfo  [0]{\@secondoftwo}%
\providecommand \bibfield  [0]{\@secondoftwo}%
\providecommand \translation [1]{[#1]}%
\providecommand \BibitemOpen [0]{}%
\providecommand \bibitemStop [0]{}%
\providecommand \bibitemNoStop [0]{.\EOS\space}%
\providecommand \EOS [0]{\spacefactor3000\relax}%
\providecommand \BibitemShut  [1]{\csname bibitem#1\endcsname}%
\let\auto@bib@innerbib\@empty
%</preamble>
\bibitem [{\citenamefont {Abbott}\ \emph {et~al.}(2017)\citenamefont {Abbott}
  \emph {et~al.}}]{LIGOScientific:2017vwq}%
  \BibitemOpen
  \bibfield  {author} {\bibinfo {author} {\bibfnamefont {B.~P.}\ \bibnamefont
  {Abbott}} \emph {et~al.} (\bibinfo {collaboration} {LIGO Scientific,
  Virgo}),\ }\href {\doibase 10.1103/PhysRevLett.119.161101} {\bibfield
  {journal} {\bibinfo  {journal} {Phys. Rev. Lett.}\ }\textbf {\bibinfo
  {volume} {119}},\ \bibinfo {pages} {161101} (\bibinfo {year} {2017})},\
  \Eprint {http://arxiv.org/abs/1710.05832} {arXiv:1710.05832 [gr-qc]}
  \BibitemShut {NoStop}%
\bibitem [{\citenamefont {Abbott}\ \emph
  {et~al.}(2019{\natexlab{a}})\citenamefont {Abbott} \emph
  {et~al.}}]{LIGOScientific:2018hze}%
  \BibitemOpen
  \bibfield  {author} {\bibinfo {author} {\bibfnamefont {B.~P.}\ \bibnamefont
  {Abbott}} \emph {et~al.} (\bibinfo {collaboration} {LIGO Scientific,
  Virgo}),\ }\href {\doibase 10.1103/PhysRevX.9.011001} {\bibfield  {journal}
  {\bibinfo  {journal} {Phys. Rev. X}\ }\textbf {\bibinfo {volume} {9}},\
  \bibinfo {pages} {011001} (\bibinfo {year} {2019}{\natexlab{a}})},\ \Eprint
  {http://arxiv.org/abs/1805.11579} {arXiv:1805.11579 [gr-qc]} \BibitemShut
  {NoStop}%
\bibitem [{\citenamefont {Abbott}\ \emph
  {et~al.}(2019{\natexlab{b}})\citenamefont {Abbott} \emph
  {et~al.}}]{LIGOScientific:2018mvr}%
  \BibitemOpen
  \bibfield  {author} {\bibinfo {author} {\bibfnamefont {B.~P.}\ \bibnamefont
  {Abbott}} \emph {et~al.} (\bibinfo {collaboration} {LIGO Scientific,
  Virgo}),\ }\href {\doibase 10.1103/PhysRevX.9.031040} {\bibfield  {journal}
  {\bibinfo  {journal} {Phys. Rev. X}\ }\textbf {\bibinfo {volume} {9}},\
  \bibinfo {pages} {031040} (\bibinfo {year} {2019}{\natexlab{b}})},\ \Eprint
  {http://arxiv.org/abs/1811.12907} {arXiv:1811.12907 [astro-ph.HE]}
  \BibitemShut {NoStop}%
\bibitem [{\citenamefont {Abbott}\ \emph
  {et~al.}(2021{\natexlab{a}})\citenamefont {Abbott} \emph
  {et~al.}}]{LIGOScientific:2020ibl}%
  \BibitemOpen
  \bibfield  {author} {\bibinfo {author} {\bibfnamefont {R.}~\bibnamefont
  {Abbott}} \emph {et~al.} (\bibinfo {collaboration} {LIGO Scientific,
  Virgo}),\ }\href {\doibase 10.1103/PhysRevX.11.021053} {\bibfield  {journal}
  {\bibinfo  {journal} {Phys. Rev. X}\ }\textbf {\bibinfo {volume} {11}},\
  \bibinfo {pages} {021053} (\bibinfo {year} {2021}{\natexlab{a}})},\ \Eprint
  {http://arxiv.org/abs/2010.14527} {arXiv:2010.14527 [gr-qc]} \BibitemShut
  {NoStop}%
\bibitem [{\citenamefont {Abbott}\ \emph
  {et~al.}(2021{\natexlab{b}})\citenamefont {Abbott} \emph
  {et~al.}}]{LIGOScientific:2021djp}%
  \BibitemOpen
  \bibfield  {author} {\bibinfo {author} {\bibfnamefont {R.}~\bibnamefont
  {Abbott}} \emph {et~al.} (\bibinfo {collaboration} {LIGO Scientific, VIRGO,
  KAGRA}),\ }\href@noop {} {\  (\bibinfo {year} {2021}{\natexlab{b}})},\
  \Eprint {http://arxiv.org/abs/2111.03606} {arXiv:2111.03606 [gr-qc]}
  \BibitemShut {NoStop}%
\bibitem [{\citenamefont {Aasi}\ \emph
  {et~al.}(2015{\natexlab{a}})\citenamefont {Aasi} \emph
  {et~al.}}]{TheLIGOScientific:2014jea}%
  \BibitemOpen
  \bibfield  {author} {\bibinfo {author} {\bibfnamefont {J.}~\bibnamefont
  {Aasi}} \emph {et~al.} (\bibinfo {collaboration} {LIGO Scientific}),\ }\href
  {\doibase 10.1088/0264-9381/32/7/074001} {\bibfield  {journal} {\bibinfo
  {journal} {Class. Quant. Grav.}\ }\textbf {\bibinfo {volume} {32}},\ \bibinfo
  {pages} {074001} (\bibinfo {year} {2015}{\natexlab{a}})},\ \Eprint
  {http://arxiv.org/abs/1411.4547} {arXiv:1411.4547 [gr-qc]} \BibitemShut
  {NoStop}%
\bibitem [{\citenamefont {Acernese}\ \emph
  {et~al.}(2015{\natexlab{a}})\citenamefont {Acernese} \emph
  {et~al.}}]{TheVirgo:2014hva}%
  \BibitemOpen
  \bibfield  {author} {\bibinfo {author} {\bibfnamefont {F.}~\bibnamefont
  {Acernese}} \emph {et~al.} (\bibinfo {collaboration} {VIRGO}),\ }\href
  {\doibase 10.1088/0264-9381/32/2/024001} {\bibfield  {journal} {\bibinfo
  {journal} {Class. Quant. Grav.}\ }\textbf {\bibinfo {volume} {32}},\ \bibinfo
  {pages} {024001} (\bibinfo {year} {2015}{\natexlab{a}})},\ \Eprint
  {http://arxiv.org/abs/1408.3978} {arXiv:1408.3978 [gr-qc]} \BibitemShut
  {NoStop}%
\bibitem [{\citenamefont {Abbott}\ \emph
  {et~al.}(2021{\natexlab{c}})\citenamefont {Abbott} \emph
  {et~al.}}]{LIGOScientific:2021psn}%
  \BibitemOpen
  \bibfield  {author} {\bibinfo {author} {\bibfnamefont {R.}~\bibnamefont
  {Abbott}} \emph {et~al.} (\bibinfo {collaboration} {LIGO Scientific, VIRGO,
  KAGRA}),\ }\href@noop {} {\  (\bibinfo {year} {2021}{\natexlab{c}})},\
  \Eprint {http://arxiv.org/abs/2111.03634} {arXiv:2111.03634 [astro-ph.HE]}
  \BibitemShut {NoStop}%
\bibitem [{\citenamefont {Flanagan}\ and\ \citenamefont
  {Hinderer}(2008)}]{Flanagan:2007ix}%
  \BibitemOpen
  \bibfield  {author} {\bibinfo {author} {\bibfnamefont {E.~E.}\ \bibnamefont
  {Flanagan}}\ and\ \bibinfo {author} {\bibfnamefont {T.}~\bibnamefont
  {Hinderer}},\ }\href {\doibase 10.1103/PhysRevD.77.021502} {\bibfield
  {journal} {\bibinfo  {journal} {Phys. Rev. D}\ }\textbf {\bibinfo {volume}
  {77}},\ \bibinfo {pages} {021502} (\bibinfo {year} {2008})},\ \Eprint
  {http://arxiv.org/abs/0709.1915} {arXiv:0709.1915 [astro-ph]} \BibitemShut
  {NoStop}%
\bibitem [{\citenamefont {Hinderer}(2008)}]{Hinderer:2007mb}%
  \BibitemOpen
  \bibfield  {author} {\bibinfo {author} {\bibfnamefont {T.}~\bibnamefont
  {Hinderer}},\ }\href {\doibase 10.1086/533487} {\bibfield  {journal}
  {\bibinfo  {journal} {Astrophys. J.}\ }\textbf {\bibinfo {volume} {677}},\
  \bibinfo {pages} {1216} (\bibinfo {year} {2008})},\ \Eprint
  {http://arxiv.org/abs/0711.2420} {arXiv:0711.2420 [astro-ph]} \BibitemShut
  {NoStop}%
\bibitem [{\citenamefont {Vines}\ \emph {et~al.}(2011)\citenamefont {Vines},
  \citenamefont {Flanagan},\ and\ \citenamefont {Hinderer}}]{Vines:2011ud}%
  \BibitemOpen
  \bibfield  {author} {\bibinfo {author} {\bibfnamefont {J.}~\bibnamefont
  {Vines}}, \bibinfo {author} {\bibfnamefont {E.~E.}\ \bibnamefont {Flanagan}},
  \ and\ \bibinfo {author} {\bibfnamefont {T.}~\bibnamefont {Hinderer}},\
  }\href {\doibase 10.1103/PhysRevD.83.084051} {\bibfield  {journal} {\bibinfo
  {journal} {Phys. Rev. D}\ }\textbf {\bibinfo {volume} {83}},\ \bibinfo
  {pages} {084051} (\bibinfo {year} {2011})},\ \Eprint
  {http://arxiv.org/abs/1101.1673} {arXiv:1101.1673 [gr-qc]} \BibitemShut
  {NoStop}%
\bibitem [{\citenamefont {Baym}\ \emph {et~al.}(1971)\citenamefont {Baym},
  \citenamefont {Pethick},\ and\ \citenamefont {Sutherland}}]{Baym:1971pw}%
  \BibitemOpen
  \bibfield  {author} {\bibinfo {author} {\bibfnamefont {G.}~\bibnamefont
  {Baym}}, \bibinfo {author} {\bibfnamefont {C.}~\bibnamefont {Pethick}}, \
  and\ \bibinfo {author} {\bibfnamefont {P.}~\bibnamefont {Sutherland}},\
  }\href {\doibase 10.1086/151216} {\bibfield  {journal} {\bibinfo  {journal}
  {Astrophys. J.}\ }\textbf {\bibinfo {volume} {170}},\ \bibinfo {pages} {299}
  (\bibinfo {year} {1971})}\BibitemShut {NoStop}%
\bibitem [{\citenamefont {Read}\ \emph {et~al.}(2009)\citenamefont {Read},
  \citenamefont {Lackey}, \citenamefont {Owen},\ and\ \citenamefont
  {Friedman}}]{Read:2008iy}%
  \BibitemOpen
  \bibfield  {author} {\bibinfo {author} {\bibfnamefont {J.~S.}\ \bibnamefont
  {Read}}, \bibinfo {author} {\bibfnamefont {B.~D.}\ \bibnamefont {Lackey}},
  \bibinfo {author} {\bibfnamefont {B.~J.}\ \bibnamefont {Owen}}, \ and\
  \bibinfo {author} {\bibfnamefont {J.~L.}\ \bibnamefont {Friedman}},\ }\href
  {\doibase 10.1103/PhysRevD.79.124032} {\bibfield  {journal} {\bibinfo
  {journal} {Phys. Rev. D}\ }\textbf {\bibinfo {volume} {79}},\ \bibinfo
  {pages} {124032} (\bibinfo {year} {2009})},\ \Eprint
  {http://arxiv.org/abs/0812.2163} {arXiv:0812.2163 [astro-ph]} \BibitemShut
  {NoStop}%
\bibitem [{\citenamefont {Lattimer}\ and\ \citenamefont
  {Prakash}(2016)}]{Lattimer:2015nhk}%
  \BibitemOpen
  \bibfield  {author} {\bibinfo {author} {\bibfnamefont {J.~M.}\ \bibnamefont
  {Lattimer}}\ and\ \bibinfo {author} {\bibfnamefont {M.}~\bibnamefont
  {Prakash}},\ }\href {\doibase 10.1016/j.physrep.2015.12.005} {\bibfield
  {journal} {\bibinfo  {journal} {Phys. Rept.}\ }\textbf {\bibinfo {volume}
  {621}},\ \bibinfo {pages} {127} (\bibinfo {year} {2016})},\ \Eprint
  {http://arxiv.org/abs/1512.07820} {arXiv:1512.07820 [astro-ph.SR]}
  \BibitemShut {NoStop}%
\bibitem [{\citenamefont {Lindblom}(2010)}]{Lindblom:2010bb}%
  \BibitemOpen
  \bibfield  {author} {\bibinfo {author} {\bibfnamefont {L.}~\bibnamefont
  {Lindblom}},\ }\href {\doibase 10.1103/PhysRevD.82.103011} {\bibfield
  {journal} {\bibinfo  {journal} {Phys. Rev. D}\ }\textbf {\bibinfo {volume}
  {82}},\ \bibinfo {pages} {103011} (\bibinfo {year} {2010})},\ \Eprint
  {http://arxiv.org/abs/1009.0738} {arXiv:1009.0738 [astro-ph.HE]} \BibitemShut
  {NoStop}%
\bibitem [{\citenamefont {Lindblom}(2022)}]{Lindblom:2022mkr}%
  \BibitemOpen
  \bibfield  {author} {\bibinfo {author} {\bibfnamefont {L.}~\bibnamefont
  {Lindblom}},\ }\href {\doibase 10.1103/PhysRevD.105.063031} {\bibfield
  {journal} {\bibinfo  {journal} {Phys. Rev. D}\ }\textbf {\bibinfo {volume}
  {105}},\ \bibinfo {pages} {063031} (\bibinfo {year} {2022})},\ \Eprint
  {http://arxiv.org/abs/2202.12285} {arXiv:2202.12285 [astro-ph.HE]}
  \BibitemShut {NoStop}%
\bibitem [{\citenamefont {O'Boyle}\ \emph {et~al.}(2020)\citenamefont
  {O'Boyle}, \citenamefont {Markakis}, \citenamefont {Stergioulas},\ and\
  \citenamefont {Read}}]{OBoyle:2020qvf}%
  \BibitemOpen
  \bibfield  {author} {\bibinfo {author} {\bibfnamefont {M.~F.}\ \bibnamefont
  {O'Boyle}}, \bibinfo {author} {\bibfnamefont {C.}~\bibnamefont {Markakis}},
  \bibinfo {author} {\bibfnamefont {N.}~\bibnamefont {Stergioulas}}, \ and\
  \bibinfo {author} {\bibfnamefont {J.~S.}\ \bibnamefont {Read}},\ }\href
  {\doibase 10.1103/PhysRevD.102.083027} {\bibfield  {journal} {\bibinfo
  {journal} {Phys. Rev. D}\ }\textbf {\bibinfo {volume} {102}},\ \bibinfo
  {pages} {083027} (\bibinfo {year} {2020})},\ \Eprint
  {http://arxiv.org/abs/2008.03342} {arXiv:2008.03342 [astro-ph.HE]}
  \BibitemShut {NoStop}%
\bibitem [{\citenamefont {Oppenheimer}\ and\ \citenamefont
  {Volkoff}(1939)}]{Oppenheimer:1939ne}%
  \BibitemOpen
  \bibfield  {author} {\bibinfo {author} {\bibfnamefont {J.~R.}\ \bibnamefont
  {Oppenheimer}}\ and\ \bibinfo {author} {\bibfnamefont {G.~M.}\ \bibnamefont
  {Volkoff}},\ }\href {\doibase 10.1103/PhysRev.55.374} {\bibfield  {journal}
  {\bibinfo  {journal} {Phys. Rev.}\ }\textbf {\bibinfo {volume} {55}},\
  \bibinfo {pages} {374} (\bibinfo {year} {1939})}\BibitemShut {NoStop}%
\bibitem [{\citenamefont {Tolman}(1939)}]{Tolman:1939jz}%
  \BibitemOpen
  \bibfield  {author} {\bibinfo {author} {\bibfnamefont {R.~C.}\ \bibnamefont
  {Tolman}},\ }\href {\doibase 10.1103/PhysRev.55.364} {\bibfield  {journal}
  {\bibinfo  {journal} {Phys. Rev.}\ }\textbf {\bibinfo {volume} {55}},\
  \bibinfo {pages} {364} (\bibinfo {year} {1939})}\BibitemShut {NoStop}%
\bibitem [{\citenamefont {Favata}(2014)}]{Favata:2013rwa}%
  \BibitemOpen
  \bibfield  {author} {\bibinfo {author} {\bibfnamefont {M.}~\bibnamefont
  {Favata}},\ }\href {\doibase 10.1103/PhysRevLett.112.101101} {\bibfield
  {journal} {\bibinfo  {journal} {Phys. Rev. Lett.}\ }\textbf {\bibinfo
  {volume} {112}},\ \bibinfo {pages} {101101} (\bibinfo {year} {2014})},\
  \Eprint {http://arxiv.org/abs/1310.8288} {arXiv:1310.8288 [gr-qc]}
  \BibitemShut {NoStop}%
\bibitem [{\citenamefont {Wade}\ \emph {et~al.}(2014)\citenamefont {Wade},
  \citenamefont {Creighton}, \citenamefont {Ochsner}, \citenamefont {Lackey},
  \citenamefont {Farr}, \citenamefont {Littenberg},\ and\ \citenamefont
  {Raymond}}]{Wade:2014vqa}%
  \BibitemOpen
  \bibfield  {author} {\bibinfo {author} {\bibfnamefont {L.}~\bibnamefont
  {Wade}}, \bibinfo {author} {\bibfnamefont {J.~D.~E.}\ \bibnamefont
  {Creighton}}, \bibinfo {author} {\bibfnamefont {E.}~\bibnamefont {Ochsner}},
  \bibinfo {author} {\bibfnamefont {B.~D.}\ \bibnamefont {Lackey}}, \bibinfo
  {author} {\bibfnamefont {B.~F.}\ \bibnamefont {Farr}}, \bibinfo {author}
  {\bibfnamefont {T.~B.}\ \bibnamefont {Littenberg}}, \ and\ \bibinfo {author}
  {\bibfnamefont {V.}~\bibnamefont {Raymond}},\ }\href {\doibase
  10.1103/PhysRevD.89.103012} {\bibfield  {journal} {\bibinfo  {journal} {Phys.
  Rev. D}\ }\textbf {\bibinfo {volume} {89}},\ \bibinfo {pages} {103012}
  (\bibinfo {year} {2014})},\ \Eprint {http://arxiv.org/abs/1402.5156}
  {arXiv:1402.5156 [gr-qc]} \BibitemShut {NoStop}%
\bibitem [{\citenamefont {Harry}\ and\ \citenamefont
  {Lundgren}(2021)}]{Harry:2021hls}%
  \BibitemOpen
  \bibfield  {author} {\bibinfo {author} {\bibfnamefont {I.}~\bibnamefont
  {Harry}}\ and\ \bibinfo {author} {\bibfnamefont {A.}~\bibnamefont
  {Lundgren}},\ }\href {\doibase 10.1103/PhysRevD.104.043008} {\bibfield
  {journal} {\bibinfo  {journal} {Phys. Rev. D}\ }\textbf {\bibinfo {volume}
  {104}},\ \bibinfo {pages} {043008} (\bibinfo {year} {2021})},\ \Eprint
  {http://arxiv.org/abs/2101.01091} {arXiv:2101.01091 [gr-qc]} \BibitemShut
  {NoStop}%
\bibitem [{\citenamefont {Dietrich}\ \emph {et~al.}(2021)\citenamefont
  {Dietrich}, \citenamefont {Hinderer},\ and\ \citenamefont
  {Samajdar}}]{Dietrich:2020eud}%
  \BibitemOpen
  \bibfield  {author} {\bibinfo {author} {\bibfnamefont {T.}~\bibnamefont
  {Dietrich}}, \bibinfo {author} {\bibfnamefont {T.}~\bibnamefont {Hinderer}},
  \ and\ \bibinfo {author} {\bibfnamefont {A.}~\bibnamefont {Samajdar}},\
  }\href {\doibase 10.1007/s10714-020-02751-6} {\bibfield  {journal} {\bibinfo
  {journal} {Gen. Rel. Grav.}\ }\textbf {\bibinfo {volume} {53}},\ \bibinfo
  {pages} {27} (\bibinfo {year} {2021})},\ \Eprint
  {http://arxiv.org/abs/2004.02527} {arXiv:2004.02527 [gr-qc]} \BibitemShut
  {NoStop}%
\bibitem [{\citenamefont {Buikema}\ \emph {et~al.}(2020)\citenamefont {Buikema}
  \emph {et~al.}}]{aLIGO:2020wna}%
  \BibitemOpen
  \bibfield  {author} {\bibinfo {author} {\bibfnamefont {A.}~\bibnamefont
  {Buikema}} \emph {et~al.} (\bibinfo {collaboration} {aLIGO}),\ }\href
  {\doibase 10.1103/PhysRevD.102.062003} {\bibfield  {journal} {\bibinfo
  {journal} {Phys. Rev. D}\ }\textbf {\bibinfo {volume} {102}},\ \bibinfo
  {pages} {062003} (\bibinfo {year} {2020})},\ \Eprint
  {http://arxiv.org/abs/2008.01301} {arXiv:2008.01301 [astro-ph.IM]}
  \BibitemShut {NoStop}%
\bibitem [{\citenamefont {Smith}\ \emph {et~al.}(2021)\citenamefont {Smith}
  \emph {et~al.}}]{Smith:2021bqc}%
  \BibitemOpen
  \bibfield  {author} {\bibinfo {author} {\bibfnamefont {R.}~\bibnamefont
  {Smith}} \emph {et~al.},\ }\href {\doibase 10.1103/PhysRevLett.127.081102}
  {\bibfield  {journal} {\bibinfo  {journal} {Phys. Rev. Lett.}\ }\textbf
  {\bibinfo {volume} {127}},\ \bibinfo {pages} {081102} (\bibinfo {year}
  {2021})},\ \Eprint {http://arxiv.org/abs/2103.12274} {arXiv:2103.12274
  [gr-qc]} \BibitemShut {NoStop}%
\bibitem [{\citenamefont {Yagi}\ and\ \citenamefont
  {Yunes}(2016)}]{Yagi:2015pkc}%
  \BibitemOpen
  \bibfield  {author} {\bibinfo {author} {\bibfnamefont {K.}~\bibnamefont
  {Yagi}}\ and\ \bibinfo {author} {\bibfnamefont {N.}~\bibnamefont {Yunes}},\
  }\href {\doibase 10.1088/0264-9381/33/13/13LT01} {\bibfield  {journal}
  {\bibinfo  {journal} {Class. Quant. Grav.}\ }\textbf {\bibinfo {volume}
  {33}},\ \bibinfo {pages} {13LT01} (\bibinfo {year} {2016})},\ \Eprint
  {http://arxiv.org/abs/1512.02639} {arXiv:1512.02639 [gr-qc]} \BibitemShut
  {NoStop}%
\bibitem [{\citenamefont {Damour}\ and\ \citenamefont
  {Nagar}(2009)}]{Damour:2009vw}%
  \BibitemOpen
  \bibfield  {author} {\bibinfo {author} {\bibfnamefont {T.}~\bibnamefont
  {Damour}}\ and\ \bibinfo {author} {\bibfnamefont {A.}~\bibnamefont {Nagar}},\
  }\href {\doibase 10.1103/PhysRevD.80.084035} {\bibfield  {journal} {\bibinfo
  {journal} {Phys. Rev.}\ }\textbf {\bibinfo {volume} {D80}},\ \bibinfo {pages}
  {084035} (\bibinfo {year} {2009})},\ \Eprint {http://arxiv.org/abs/0906.0096}
  {arXiv:0906.0096 [gr-qc]} \BibitemShut {NoStop}%
%%CITATION = ARXIV:0906.0096;%%
\bibitem [{\citenamefont {Saes}\ and\ \citenamefont
  {Mendes}(2021)}]{Saes:2021fzr}%
  \BibitemOpen
  \bibfield  {author} {\bibinfo {author} {\bibfnamefont {J.}~\bibnamefont
  {Saes}}\ and\ \bibinfo {author} {\bibfnamefont {R.~F.~P.}\ \bibnamefont
  {Mendes}},\ }\href@noop {} {\  (\bibinfo {year} {2021})},\ \Eprint
  {http://arxiv.org/abs/2109.11571} {arXiv:2109.11571 [gr-qc]} \BibitemShut
  {NoStop}%
\bibitem [{\citenamefont {Abbott}\ \emph {et~al.}(2018)\citenamefont {Abbott}
  \emph {et~al.}}]{LIGOScientific:2018cki}%
  \BibitemOpen
  \bibfield  {author} {\bibinfo {author} {\bibfnamefont {B.~P.}\ \bibnamefont
  {Abbott}} \emph {et~al.} (\bibinfo {collaboration} {LIGO Scientific,
  Virgo}),\ }\href {\doibase 10.1103/PhysRevLett.121.161101} {\bibfield
  {journal} {\bibinfo  {journal} {Phys. Rev. Lett.}\ }\textbf {\bibinfo
  {volume} {121}},\ \bibinfo {pages} {161101} (\bibinfo {year} {2018})},\
  \Eprint {http://arxiv.org/abs/1805.11581} {arXiv:1805.11581 [gr-qc]}
  \BibitemShut {NoStop}%
\bibitem [{\citenamefont {Chatziioannou}\ \emph {et~al.}(2018)\citenamefont
  {Chatziioannou}, \citenamefont {Haster},\ and\ \citenamefont
  {Zimmerman}}]{Chatziioannou:2018vzf}%
  \BibitemOpen
  \bibfield  {author} {\bibinfo {author} {\bibfnamefont {K.}~\bibnamefont
  {Chatziioannou}}, \bibinfo {author} {\bibfnamefont {C.-J.}\ \bibnamefont
  {Haster}}, \ and\ \bibinfo {author} {\bibfnamefont {A.}~\bibnamefont
  {Zimmerman}},\ }\href {\doibase 10.1103/PhysRevD.97.104036} {\bibfield
  {journal} {\bibinfo  {journal} {Phys. Rev. D}\ }\textbf {\bibinfo {volume}
  {97}},\ \bibinfo {pages} {104036} (\bibinfo {year} {2018})},\ \Eprint
  {http://arxiv.org/abs/1804.03221} {arXiv:1804.03221 [gr-qc]} \BibitemShut
  {NoStop}%
\bibitem [{\citenamefont {Kumar}\ and\ \citenamefont
  {Landry}(2019)}]{Kumar:2019xgp}%
  \BibitemOpen
  \bibfield  {author} {\bibinfo {author} {\bibfnamefont {B.}~\bibnamefont
  {Kumar}}\ and\ \bibinfo {author} {\bibfnamefont {P.}~\bibnamefont {Landry}},\
  }\href {\doibase 10.1103/PhysRevD.99.123026} {\bibfield  {journal} {\bibinfo
  {journal} {Phys. Rev. D}\ }\textbf {\bibinfo {volume} {99}},\ \bibinfo
  {pages} {123026} (\bibinfo {year} {2019})},\ \Eprint
  {http://arxiv.org/abs/1902.04557} {arXiv:1902.04557 [gr-qc]} \BibitemShut
  {NoStop}%
\bibitem [{\citenamefont {Carson}\ \emph {et~al.}(2019)\citenamefont {Carson},
  \citenamefont {Chatziioannou}, \citenamefont {Haster}, \citenamefont {Yagi},\
  and\ \citenamefont {Yunes}}]{Carson:2019rjx}%
  \BibitemOpen
  \bibfield  {author} {\bibinfo {author} {\bibfnamefont {Z.}~\bibnamefont
  {Carson}}, \bibinfo {author} {\bibfnamefont {K.}~\bibnamefont
  {Chatziioannou}}, \bibinfo {author} {\bibfnamefont {C.-J.}\ \bibnamefont
  {Haster}}, \bibinfo {author} {\bibfnamefont {K.}~\bibnamefont {Yagi}}, \ and\
  \bibinfo {author} {\bibfnamefont {N.}~\bibnamefont {Yunes}},\ }\href
  {\doibase 10.1103/PhysRevD.99.083016} {\bibfield  {journal} {\bibinfo
  {journal} {Phys. Rev. D}\ }\textbf {\bibinfo {volume} {99}},\ \bibinfo
  {pages} {083016} (\bibinfo {year} {2019})},\ \Eprint
  {http://arxiv.org/abs/1903.03909} {arXiv:1903.03909 [gr-qc]} \BibitemShut
  {NoStop}%
\bibitem [{\citenamefont {Biswas}(2022)}]{Biswas:2021pvm}%
  \BibitemOpen
  \bibfield  {author} {\bibinfo {author} {\bibfnamefont {B.}~\bibnamefont
  {Biswas}},\ }\href {\doibase 10.3847/1538-4357/ac447b} {\bibfield  {journal}
  {\bibinfo  {journal} {Astrophys. J.}\ }\textbf {\bibinfo {volume} {926}},\
  \bibinfo {pages} {75} (\bibinfo {year} {2022})},\ \Eprint
  {http://arxiv.org/abs/2106.02644} {arXiv:2106.02644 [astro-ph.HE]}
  \BibitemShut {NoStop}%
\bibitem [{\citenamefont {Yagi}\ and\ \citenamefont
  {Yunes}(2017)}]{Yagi:2016bkt}%
  \BibitemOpen
  \bibfield  {author} {\bibinfo {author} {\bibfnamefont {K.}~\bibnamefont
  {Yagi}}\ and\ \bibinfo {author} {\bibfnamefont {N.}~\bibnamefont {Yunes}},\
  }\href {\doibase 10.1016/j.physrep.2017.03.002} {\bibfield  {journal}
  {\bibinfo  {journal} {Phys. Rept.}\ }\textbf {\bibinfo {volume} {681}},\
  \bibinfo {pages} {1} (\bibinfo {year} {2017})},\ \Eprint
  {http://arxiv.org/abs/1608.02582} {arXiv:1608.02582 [gr-qc]} \BibitemShut
  {NoStop}%
\bibitem [{\citenamefont {Abbott}\ \emph {et~al.}(2020)\citenamefont {Abbott}
  \emph {et~al.}}]{LIGOScientific:2019eut}%
  \BibitemOpen
  \bibfield  {author} {\bibinfo {author} {\bibfnamefont {B.~P.}\ \bibnamefont
  {Abbott}} \emph {et~al.} (\bibinfo {collaboration} {LIGO Scientific,
  Virgo}),\ }\href {\doibase 10.1088/1361-6382/ab5f7c} {\bibfield  {journal}
  {\bibinfo  {journal} {Class. Quant. Grav.}\ }\textbf {\bibinfo {volume}
  {37}},\ \bibinfo {pages} {045006} (\bibinfo {year} {2020})},\ \Eprint
  {http://arxiv.org/abs/1908.01012} {arXiv:1908.01012 [gr-qc]} \BibitemShut
  {NoStop}%
\bibitem [{\citenamefont {Lackey}\ and\ \citenamefont
  {Wade}(2015)}]{Lackey:2014fwa}%
  \BibitemOpen
  \bibfield  {author} {\bibinfo {author} {\bibfnamefont {B.~D.}\ \bibnamefont
  {Lackey}}\ and\ \bibinfo {author} {\bibfnamefont {L.}~\bibnamefont {Wade}},\
  }\href {\doibase 10.1103/PhysRevD.91.043002} {\bibfield  {journal} {\bibinfo
  {journal} {Phys. Rev. D}\ }\textbf {\bibinfo {volume} {91}},\ \bibinfo
  {pages} {043002} (\bibinfo {year} {2015})},\ \Eprint
  {http://arxiv.org/abs/1410.8866} {arXiv:1410.8866 [gr-qc]} \BibitemShut
  {NoStop}%
\bibitem [{\citenamefont {Pacilio}\ \emph {et~al.}(2022)\citenamefont
  {Pacilio}, \citenamefont {Maselli}, \citenamefont {Fasano},\ and\
  \citenamefont {Pani}}]{Pacilio:2021jmq}%
  \BibitemOpen
  \bibfield  {author} {\bibinfo {author} {\bibfnamefont {C.}~\bibnamefont
  {Pacilio}}, \bibinfo {author} {\bibfnamefont {A.}~\bibnamefont {Maselli}},
  \bibinfo {author} {\bibfnamefont {M.}~\bibnamefont {Fasano}}, \ and\ \bibinfo
  {author} {\bibfnamefont {P.}~\bibnamefont {Pani}},\ }\href {\doibase
  10.1103/PhysRevLett.128.101101} {\bibfield  {journal} {\bibinfo  {journal}
  {Phys. Rev. Lett.}\ }\textbf {\bibinfo {volume} {128}},\ \bibinfo {pages}
  {101101} (\bibinfo {year} {2022})},\ \Eprint
  {http://arxiv.org/abs/2104.10035} {arXiv:2104.10035 [gr-qc]} \BibitemShut
  {NoStop}%
\bibitem [{\citenamefont {Ghosh}\ \emph {et~al.}(2021)\citenamefont {Ghosh},
  \citenamefont {Liu}, \citenamefont {Creighton}, \citenamefont {Kastaun},
  \citenamefont {Pratten},\ and\ \citenamefont {Hernandez}}]{Ghosh:2021eqv}%
  \BibitemOpen
  \bibfield  {author} {\bibinfo {author} {\bibfnamefont {S.}~\bibnamefont
  {Ghosh}}, \bibinfo {author} {\bibfnamefont {X.}~\bibnamefont {Liu}}, \bibinfo
  {author} {\bibfnamefont {J.}~\bibnamefont {Creighton}}, \bibinfo {author}
  {\bibfnamefont {W.}~\bibnamefont {Kastaun}}, \bibinfo {author} {\bibfnamefont
  {G.}~\bibnamefont {Pratten}}, \ and\ \bibinfo {author} {\bibfnamefont
  {I.~M.}\ \bibnamefont {Hernandez}},\ }\href {\doibase
  10.1103/PhysRevD.104.083003} {\bibfield  {journal} {\bibinfo  {journal}
  {Phys. Rev. D}\ }\textbf {\bibinfo {volume} {104}},\ \bibinfo {pages}
  {083003} (\bibinfo {year} {2021})}\BibitemShut {NoStop}%
\bibitem [{\citenamefont {Aasi}\ \emph
  {et~al.}(2015{\natexlab{b}})\citenamefont {Aasi} \emph
  {et~al.}}]{LIGOScientific:2014pky}%
  \BibitemOpen
  \bibfield  {author} {\bibinfo {author} {\bibfnamefont {J.}~\bibnamefont
  {Aasi}} \emph {et~al.} (\bibinfo {collaboration} {LIGO Scientific}),\ }\href
  {\doibase 10.1088/0264-9381/32/7/074001} {\bibfield  {journal} {\bibinfo
  {journal} {Class. Quant. Grav.}\ }\textbf {\bibinfo {volume} {32}},\ \bibinfo
  {pages} {074001} (\bibinfo {year} {2015}{\natexlab{b}})},\ \Eprint
  {http://arxiv.org/abs/1411.4547} {arXiv:1411.4547 [gr-qc]} \BibitemShut
  {NoStop}%
\bibitem [{\citenamefont {Acernese}\ \emph
  {et~al.}(2015{\natexlab{b}})\citenamefont {Acernese} \emph
  {et~al.}}]{VIRGO:2014yos}%
  \BibitemOpen
  \bibfield  {author} {\bibinfo {author} {\bibfnamefont {F.}~\bibnamefont
  {Acernese}} \emph {et~al.} (\bibinfo {collaboration} {VIRGO}),\ }\href
  {\doibase 10.1088/0264-9381/32/2/024001} {\bibfield  {journal} {\bibinfo
  {journal} {Class. Quant. Grav.}\ }\textbf {\bibinfo {volume} {32}},\ \bibinfo
  {pages} {024001} (\bibinfo {year} {2015}{\natexlab{b}})},\ \Eprint
  {http://arxiv.org/abs/1408.3978} {arXiv:1408.3978 [gr-qc]} \BibitemShut
  {NoStop}%
\bibitem [{\citenamefont {Akutsu}\ \emph {et~al.}(2019)\citenamefont {Akutsu}
  \emph {et~al.}}]{KAGRA:2018plz}%
  \BibitemOpen
  \bibfield  {author} {\bibinfo {author} {\bibfnamefont {T.}~\bibnamefont
  {Akutsu}} \emph {et~al.} (\bibinfo {collaboration} {KAGRA}),\ }\href
  {\doibase 10.1038/s41550-018-0658-y} {\bibfield  {journal} {\bibinfo
  {journal} {Nature Astron.}\ }\textbf {\bibinfo {volume} {3}},\ \bibinfo
  {pages} {35} (\bibinfo {year} {2019})},\ \Eprint
  {http://arxiv.org/abs/1811.08079} {arXiv:1811.08079 [gr-qc]} \BibitemShut
  {NoStop}%
\bibitem [{\citenamefont {Reitze}\ \emph {et~al.}(2019)\citenamefont {Reitze}
  \emph {et~al.}}]{Reitze:2019iox}%
  \BibitemOpen
  \bibfield  {author} {\bibinfo {author} {\bibfnamefont {D.}~\bibnamefont
  {Reitze}} \emph {et~al.},\ }\href@noop {} {\bibfield  {journal} {\bibinfo
  {journal} {Bull. Am. Astron. Soc.}\ }\textbf {\bibinfo {volume} {51}},\
  \bibinfo {pages} {035} (\bibinfo {year} {2019})},\ \Eprint
  {http://arxiv.org/abs/1907.04833} {arXiv:1907.04833 [astro-ph.IM]}
  \BibitemShut {NoStop}%
\bibitem [{\citenamefont {Punturo}\ \emph {et~al.}(2010)\citenamefont {Punturo}
  \emph {et~al.}}]{Punturo:2010zza}%
  \BibitemOpen
  \bibfield  {author} {\bibinfo {author} {\bibfnamefont {M.}~\bibnamefont
  {Punturo}} \emph {et~al.},\ }\href {\doibase 10.1088/0264-9381/27/8/084007}
  {\bibfield  {journal} {\bibinfo  {journal} {Class. Quant. Grav.}\ }\textbf
  {\bibinfo {volume} {27}},\ \bibinfo {pages} {084007} (\bibinfo {year}
  {2010})}\BibitemShut {NoStop}%
\bibitem [{\citenamefont {Borhanian}\ and\ \citenamefont
  {Sathyaprakash}(2022)}]{Borhanian:2022czq}%
  \BibitemOpen
  \bibfield  {author} {\bibinfo {author} {\bibfnamefont {S.}~\bibnamefont
  {Borhanian}}\ and\ \bibinfo {author} {\bibfnamefont {B.~S.}\ \bibnamefont
  {Sathyaprakash}},\ }\href@noop {} {\  (\bibinfo {year} {2022})},\ \Eprint
  {http://arxiv.org/abs/2202.11048} {arXiv:2202.11048 [gr-qc]} \BibitemShut
  {NoStop}%
\bibitem [{\citenamefont {de~Freitas~Pacheco}(1997)}]{deFreitasPacheco:1997fr}%
  \BibitemOpen
  \bibfield  {author} {\bibinfo {author} {\bibfnamefont {J.~A.}\ \bibnamefont
  {de~Freitas~Pacheco}},\ }\href {\doibase 10.1016/S0927-6505(97)00040-6}
  {\bibfield  {journal} {\bibinfo  {journal} {Astropart. Phys.}\ }\textbf
  {\bibinfo {volume} {8}},\ \bibinfo {pages} {21} (\bibinfo {year}
  {1997})}\BibitemShut {NoStop}%
\bibitem [{\citenamefont {Vangioni}\ \emph {et~al.}(2015)\citenamefont
  {Vangioni}, \citenamefont {Olive}, \citenamefont {Prestegard}, \citenamefont
  {Silk}, \citenamefont {Petitjean},\ and\ \citenamefont
  {Mandic}}]{Vangioni:2014axa}%
  \BibitemOpen
  \bibfield  {author} {\bibinfo {author} {\bibfnamefont {E.}~\bibnamefont
  {Vangioni}}, \bibinfo {author} {\bibfnamefont {K.~A.}\ \bibnamefont {Olive}},
  \bibinfo {author} {\bibfnamefont {T.}~\bibnamefont {Prestegard}}, \bibinfo
  {author} {\bibfnamefont {J.}~\bibnamefont {Silk}}, \bibinfo {author}
  {\bibfnamefont {P.}~\bibnamefont {Petitjean}}, \ and\ \bibinfo {author}
  {\bibfnamefont {V.}~\bibnamefont {Mandic}},\ }\href {\doibase
  10.1093/mnras/stu2600} {\bibfield  {journal} {\bibinfo  {journal} {Mon. Not.
  Roy. Astron. Soc.}\ }\textbf {\bibinfo {volume} {447}},\ \bibinfo {pages}
  {2575} (\bibinfo {year} {2015})},\ \Eprint {http://arxiv.org/abs/1409.2462}
  {arXiv:1409.2462 [astro-ph.GA]} \BibitemShut {NoStop}%
\bibitem [{\citenamefont {Beniamini}\ and\ \citenamefont
  {Piran}(2019)}]{Beniamini:2019iop}%
  \BibitemOpen
  \bibfield  {author} {\bibinfo {author} {\bibfnamefont {P.}~\bibnamefont
  {Beniamini}}\ and\ \bibinfo {author} {\bibfnamefont {T.}~\bibnamefont
  {Piran}},\ }\href {\doibase 10.1093/mnras/stz1589} {\bibfield  {journal}
  {\bibinfo  {journal} {Mon. Not. R. Astron. Soc.}\ }\textbf {\bibinfo {volume}
  {487}},\ \bibinfo {pages} {4847} (\bibinfo {year} {2019})}\BibitemShut
  {NoStop}%
\bibitem [{\citenamefont {Aghanim}\ \emph
  {et~al.}(2020{\natexlab{a}})\citenamefont {Aghanim}, \citenamefont {Akrami},
  \citenamefont {Ashdown}, \citenamefont {Aumont}, \citenamefont {Baccigalupi},
  \citenamefont {Ballardini}, \citenamefont {Banday}, \citenamefont {Barreiro},
  \citenamefont {Bartolo}, \citenamefont {Basak}, \citenamefont {Battye},
  \citenamefont {Benabed}, \citenamefont {Bernard}, \citenamefont {Bersanelli},
  \citenamefont {Bielewicz}, \citenamefont {Bock}, \citenamefont {Bond},
  \citenamefont {Borrill}, \citenamefont {Bouchet}, \citenamefont {Boulanger},
  \citenamefont {Bucher}, \citenamefont {Burigana}, \citenamefont {Butler},
  \citenamefont {Calabrese}, \citenamefont {Cardoso}, \citenamefont {Carron},
  \citenamefont {Challinor}, \citenamefont {Chiang}, \citenamefont {Chluba},
  \citenamefont {Colombo}, \citenamefont {Combet}, \citenamefont {Contreras},
  \citenamefont {Crill}, \citenamefont {Cuttaia}, \citenamefont {de~Bernardis},
  \citenamefont {de~Zotti}, \citenamefont {Delabrouille}, \citenamefont
  {Delouis}, \citenamefont {Di~Valentino}, \citenamefont {Diego}, \citenamefont
  {Dor{\'e}}, \citenamefont {Douspis}, \citenamefont {Ducout}, \citenamefont
  {Dupac}, \citenamefont {Dusini}, \citenamefont {Efstathiou}, \citenamefont
  {Elsner}, \citenamefont {En{\ss}lin}, \citenamefont {Eriksen}, \citenamefont
  {Fantaye}, \citenamefont {Farhang}, \citenamefont {Fergusson}, \citenamefont
  {Fernandez-Cobos}, \citenamefont {Finelli}, \citenamefont {Forastieri},
  \citenamefont {Frailis}, \citenamefont {Fraisse}, \citenamefont {Franceschi},
  \citenamefont {Frolov}, \citenamefont {Galeotta}, \citenamefont {Galli},
  \citenamefont {Ganga}, \citenamefont {G{\'e}nova-Santos}, \citenamefont
  {Gerbino}, \citenamefont {Ghosh}, \citenamefont {Gonz{\'a}lez-Nuevo},
  \citenamefont {G{\'o}rski}, \citenamefont {Gratton}, \citenamefont
  {Gruppuso}, \citenamefont {Gudmundsson}, \citenamefont {Hamann},
  \citenamefont {Handley}, \citenamefont {Hansen}, \citenamefont {Herranz},
  \citenamefont {Hildebrandt}, \citenamefont {Hivon}, \citenamefont {Huang},
  \citenamefont {Jaffe}, \citenamefont {Jones}, \citenamefont {Karakci},
  \citenamefont {Keih{\"a}nen}, \citenamefont {Keskitalo}, \citenamefont
  {Kiiveri}, \citenamefont {Kim}, \citenamefont {Kisner}, \citenamefont {Knox},
  \citenamefont {Krachmalnicoff}, \citenamefont {Kunz}, \citenamefont
  {Kurki-Suonio}, \citenamefont {Lagache}, \citenamefont {Lamarre},
  \citenamefont {Lasenby}, \citenamefont {Lattanzi}, \citenamefont {Lawrence},
  \citenamefont {Le~Jeune}, \citenamefont {Lemos}, \citenamefont {Lesgourgues},
  \citenamefont {Levrier}, \citenamefont {Lewis}, \citenamefont {Liguori},
  \citenamefont {Lilje}, \citenamefont {Lilley}, \citenamefont {Lindholm},
  \citenamefont {L{\'o}pez-Caniego}, \citenamefont {Lubin}, \citenamefont {Ma},
  \citenamefont {Mac{\'\i}as-P{\'e}rez}, \citenamefont {Maggio}, \citenamefont
  {Maino}, \citenamefont {Mandolesi}, \citenamefont {Mangilli}, \citenamefont
  {Marcos-Caballero}, \citenamefont {Maris}, \citenamefont {Martin},
  \citenamefont {Martinelli}, \citenamefont {Mart{\'\i}nez-Gonz{\'a}lez},
  \citenamefont {Matarrese}, \citenamefont {Mauri}, \citenamefont {McEwen},
  \citenamefont {Meinhold}, \citenamefont {Melchiorri}, \citenamefont
  {Mennella}, \citenamefont {Migliaccio}, \citenamefont {Millea}, \citenamefont
  {Mitra}, \citenamefont {Miville-Desch{\^e}nes}, \citenamefont {Molinari},
  \citenamefont {Montier}, \citenamefont {Morgante}, \citenamefont {Moss},
  \citenamefont {Natoli}, \citenamefont {N{\o}rgaard-Nielsen}, \citenamefont
  {Pagano}, \citenamefont {Paoletti}, \citenamefont {Partridge}, \citenamefont
  {Patanchon}, \citenamefont {Peiris}, \citenamefont {Perrotta}, \citenamefont
  {Pettorino}, \citenamefont {Piacentini}, \citenamefont {Polastri},
  \citenamefont {Polenta}, \citenamefont {Puget}, \citenamefont {Rachen},
  \citenamefont {Reinecke}, \citenamefont {Remazeilles}, \citenamefont {Renzi},
  \citenamefont {Rocha}, \citenamefont {Rosset}, \citenamefont {Roudier},
  \citenamefont {Rubi{\~n}o-Mart{\'\i}n}, \citenamefont {Ruiz-Granados},
  \citenamefont {Salvati}, \citenamefont {Sandri}, \citenamefont {Savelainen},
  \citenamefont {Scott}, \citenamefont {Shellard}, \citenamefont {Sirignano},
  \citenamefont {Sirri}, \citenamefont {Spencer}, \citenamefont {Sunyaev},
  \citenamefont {Suur-Uski}, \citenamefont {Tauber}, \citenamefont
  {Tavagnacco}, \citenamefont {Tenti}, \citenamefont {Toffolatti},
  \citenamefont {Tomasi}, \citenamefont {Trombetti}, \citenamefont
  {Valenziano}, \citenamefont {Valiviita}, \citenamefont {Van~Tent},
  \citenamefont {Vibert}, \citenamefont {Vielva}, \citenamefont {Villa},
  \citenamefont {Vittorio}, \citenamefont {Wandelt}, \citenamefont {Wehus},
  \citenamefont {White}, \citenamefont {White}, \citenamefont {Zacchei},\ and\
  \citenamefont {Zonca}}]{Aghanim2020-tl}%
  \BibitemOpen
  \bibfield  {author} {\bibinfo {author} {\bibfnamefont {N.}~\bibnamefont
  {Aghanim}}, \bibinfo {author} {\bibfnamefont {Y.}~\bibnamefont {Akrami}},
  \bibinfo {author} {\bibfnamefont {M.}~\bibnamefont {Ashdown}}, \bibinfo
  {author} {\bibfnamefont {J.}~\bibnamefont {Aumont}}, \bibinfo {author}
  {\bibfnamefont {C.}~\bibnamefont {Baccigalupi}}, \bibinfo {author}
  {\bibfnamefont {M.}~\bibnamefont {Ballardini}}, \bibinfo {author}
  {\bibfnamefont {A.~J.}\ \bibnamefont {Banday}}, \bibinfo {author}
  {\bibfnamefont {R.~B.}\ \bibnamefont {Barreiro}}, \bibinfo {author}
  {\bibfnamefont {N.}~\bibnamefont {Bartolo}}, \bibinfo {author} {\bibfnamefont
  {S.}~\bibnamefont {Basak}}, \bibinfo {author} {\bibfnamefont
  {R.}~\bibnamefont {Battye}}, \bibinfo {author} {\bibfnamefont
  {K.}~\bibnamefont {Benabed}}, \bibinfo {author} {\bibfnamefont {J.-P.}\
  \bibnamefont {Bernard}}, \bibinfo {author} {\bibfnamefont {M.}~\bibnamefont
  {Bersanelli}}, \bibinfo {author} {\bibfnamefont {P.}~\bibnamefont
  {Bielewicz}}, \bibinfo {author} {\bibfnamefont {J.~J.}\ \bibnamefont {Bock}},
  \bibinfo {author} {\bibfnamefont {J.~R.}\ \bibnamefont {Bond}}, \bibinfo
  {author} {\bibfnamefont {J.}~\bibnamefont {Borrill}}, \bibinfo {author}
  {\bibfnamefont {F.~R.}\ \bibnamefont {Bouchet}}, \bibinfo {author}
  {\bibfnamefont {F.}~\bibnamefont {Boulanger}}, \bibinfo {author}
  {\bibfnamefont {M.}~\bibnamefont {Bucher}}, \bibinfo {author} {\bibfnamefont
  {C.}~\bibnamefont {Burigana}}, \bibinfo {author} {\bibfnamefont {R.~C.}\
  \bibnamefont {Butler}}, \bibinfo {author} {\bibfnamefont {E.}~\bibnamefont
  {Calabrese}}, \bibinfo {author} {\bibfnamefont {J.-F.}\ \bibnamefont
  {Cardoso}}, \bibinfo {author} {\bibfnamefont {J.}~\bibnamefont {Carron}},
  \bibinfo {author} {\bibfnamefont {A.}~\bibnamefont {Challinor}}, \bibinfo
  {author} {\bibfnamefont {H.~C.}\ \bibnamefont {Chiang}}, \bibinfo {author}
  {\bibfnamefont {J.}~\bibnamefont {Chluba}}, \bibinfo {author} {\bibfnamefont
  {L.~P.~L.}\ \bibnamefont {Colombo}}, \bibinfo {author} {\bibfnamefont
  {C.}~\bibnamefont {Combet}}, \bibinfo {author} {\bibfnamefont
  {D.}~\bibnamefont {Contreras}}, \bibinfo {author} {\bibfnamefont {B.~P.}\
  \bibnamefont {Crill}}, \bibinfo {author} {\bibfnamefont {F.}~\bibnamefont
  {Cuttaia}}, \bibinfo {author} {\bibfnamefont {P.}~\bibnamefont
  {de~Bernardis}}, \bibinfo {author} {\bibfnamefont {G.}~\bibnamefont
  {de~Zotti}}, \bibinfo {author} {\bibfnamefont {J.}~\bibnamefont
  {Delabrouille}}, \bibinfo {author} {\bibfnamefont {J.-M.}\ \bibnamefont
  {Delouis}}, \bibinfo {author} {\bibfnamefont {E.}~\bibnamefont
  {Di~Valentino}}, \bibinfo {author} {\bibfnamefont {J.~M.}\ \bibnamefont
  {Diego}}, \bibinfo {author} {\bibfnamefont {O.}~\bibnamefont {Dor{\'e}}},
  \bibinfo {author} {\bibfnamefont {M.}~\bibnamefont {Douspis}}, \bibinfo
  {author} {\bibfnamefont {A.}~\bibnamefont {Ducout}}, \bibinfo {author}
  {\bibfnamefont {X.}~\bibnamefont {Dupac}}, \bibinfo {author} {\bibfnamefont
  {S.}~\bibnamefont {Dusini}}, \bibinfo {author} {\bibfnamefont
  {G.}~\bibnamefont {Efstathiou}}, \bibinfo {author} {\bibfnamefont
  {F.}~\bibnamefont {Elsner}}, \bibinfo {author} {\bibfnamefont {T.~A.}\
  \bibnamefont {En{\ss}lin}}, \bibinfo {author} {\bibfnamefont {H.~K.}\
  \bibnamefont {Eriksen}}, \bibinfo {author} {\bibfnamefont {Y.}~\bibnamefont
  {Fantaye}}, \bibinfo {author} {\bibfnamefont {M.}~\bibnamefont {Farhang}},
  \bibinfo {author} {\bibfnamefont {J.}~\bibnamefont {Fergusson}}, \bibinfo
  {author} {\bibfnamefont {R.}~\bibnamefont {Fernandez-Cobos}}, \bibinfo
  {author} {\bibfnamefont {F.}~\bibnamefont {Finelli}}, \bibinfo {author}
  {\bibfnamefont {F.}~\bibnamefont {Forastieri}}, \bibinfo {author}
  {\bibfnamefont {M.}~\bibnamefont {Frailis}}, \bibinfo {author} {\bibfnamefont
  {A.~A.}\ \bibnamefont {Fraisse}}, \bibinfo {author} {\bibfnamefont
  {E.}~\bibnamefont {Franceschi}}, \bibinfo {author} {\bibfnamefont
  {A.}~\bibnamefont {Frolov}}, \bibinfo {author} {\bibfnamefont
  {S.}~\bibnamefont {Galeotta}}, \bibinfo {author} {\bibfnamefont
  {S.}~\bibnamefont {Galli}}, \bibinfo {author} {\bibfnamefont
  {K.}~\bibnamefont {Ganga}}, \bibinfo {author} {\bibfnamefont {R.~T.}\
  \bibnamefont {G{\'e}nova-Santos}}, \bibinfo {author} {\bibfnamefont
  {M.}~\bibnamefont {Gerbino}}, \bibinfo {author} {\bibfnamefont
  {T.}~\bibnamefont {Ghosh}}, \bibinfo {author} {\bibfnamefont
  {J.}~\bibnamefont {Gonz{\'a}lez-Nuevo}}, \bibinfo {author} {\bibfnamefont
  {K.~M.}\ \bibnamefont {G{\'o}rski}}, \bibinfo {author} {\bibfnamefont
  {S.}~\bibnamefont {Gratton}}, \bibinfo {author} {\bibfnamefont
  {A.}~\bibnamefont {Gruppuso}}, \bibinfo {author} {\bibfnamefont {J.~E.}\
  \bibnamefont {Gudmundsson}}, \bibinfo {author} {\bibfnamefont
  {J.}~\bibnamefont {Hamann}}, \bibinfo {author} {\bibfnamefont
  {W.}~\bibnamefont {Handley}}, \bibinfo {author} {\bibfnamefont {F.~K.}\
  \bibnamefont {Hansen}}, \bibinfo {author} {\bibfnamefont {D.}~\bibnamefont
  {Herranz}}, \bibinfo {author} {\bibfnamefont {S.~R.}\ \bibnamefont
  {Hildebrandt}}, \bibinfo {author} {\bibfnamefont {E.}~\bibnamefont {Hivon}},
  \bibinfo {author} {\bibfnamefont {Z.}~\bibnamefont {Huang}}, \bibinfo
  {author} {\bibfnamefont {A.~H.}\ \bibnamefont {Jaffe}}, \bibinfo {author}
  {\bibfnamefont {W.~C.}\ \bibnamefont {Jones}}, \bibinfo {author}
  {\bibfnamefont {A.}~\bibnamefont {Karakci}}, \bibinfo {author} {\bibfnamefont
  {E.}~\bibnamefont {Keih{\"a}nen}}, \bibinfo {author} {\bibfnamefont
  {R.}~\bibnamefont {Keskitalo}}, \bibinfo {author} {\bibfnamefont
  {K.}~\bibnamefont {Kiiveri}}, \bibinfo {author} {\bibfnamefont
  {J.}~\bibnamefont {Kim}}, \bibinfo {author} {\bibfnamefont {T.~S.}\
  \bibnamefont {Kisner}}, \bibinfo {author} {\bibfnamefont {L.}~\bibnamefont
  {Knox}}, \bibinfo {author} {\bibfnamefont {N.}~\bibnamefont
  {Krachmalnicoff}}, \bibinfo {author} {\bibfnamefont {M.}~\bibnamefont
  {Kunz}}, \bibinfo {author} {\bibfnamefont {H.}~\bibnamefont {Kurki-Suonio}},
  \bibinfo {author} {\bibfnamefont {G.}~\bibnamefont {Lagache}}, \bibinfo
  {author} {\bibfnamefont {J.-M.}\ \bibnamefont {Lamarre}}, \bibinfo {author}
  {\bibfnamefont {A.}~\bibnamefont {Lasenby}}, \bibinfo {author} {\bibfnamefont
  {M.}~\bibnamefont {Lattanzi}}, \bibinfo {author} {\bibfnamefont {C.~R.}\
  \bibnamefont {Lawrence}}, \bibinfo {author} {\bibfnamefont {M.}~\bibnamefont
  {Le~Jeune}}, \bibinfo {author} {\bibfnamefont {P.}~\bibnamefont {Lemos}},
  \bibinfo {author} {\bibfnamefont {J.}~\bibnamefont {Lesgourgues}}, \bibinfo
  {author} {\bibfnamefont {F.}~\bibnamefont {Levrier}}, \bibinfo {author}
  {\bibfnamefont {A.}~\bibnamefont {Lewis}}, \bibinfo {author} {\bibfnamefont
  {M.}~\bibnamefont {Liguori}}, \bibinfo {author} {\bibfnamefont {P.~B.}\
  \bibnamefont {Lilje}}, \bibinfo {author} {\bibfnamefont {M.}~\bibnamefont
  {Lilley}}, \bibinfo {author} {\bibfnamefont {V.}~\bibnamefont {Lindholm}},
  \bibinfo {author} {\bibfnamefont {M.}~\bibnamefont {L{\'o}pez-Caniego}},
  \bibinfo {author} {\bibfnamefont {P.~M.}\ \bibnamefont {Lubin}}, \bibinfo
  {author} {\bibfnamefont {Y.-Z.}\ \bibnamefont {Ma}}, \bibinfo {author}
  {\bibfnamefont {J.~F.}\ \bibnamefont {Mac{\'\i}as-P{\'e}rez}}, \bibinfo
  {author} {\bibfnamefont {G.}~\bibnamefont {Maggio}}, \bibinfo {author}
  {\bibfnamefont {D.}~\bibnamefont {Maino}}, \bibinfo {author} {\bibfnamefont
  {N.}~\bibnamefont {Mandolesi}}, \bibinfo {author} {\bibfnamefont
  {A.}~\bibnamefont {Mangilli}}, \bibinfo {author} {\bibfnamefont
  {A.}~\bibnamefont {Marcos-Caballero}}, \bibinfo {author} {\bibfnamefont
  {M.}~\bibnamefont {Maris}}, \bibinfo {author} {\bibfnamefont {P.~G.}\
  \bibnamefont {Martin}}, \bibinfo {author} {\bibfnamefont {M.}~\bibnamefont
  {Martinelli}}, \bibinfo {author} {\bibfnamefont {E.}~\bibnamefont
  {Mart{\'\i}nez-Gonz{\'a}lez}}, \bibinfo {author} {\bibfnamefont
  {S.}~\bibnamefont {Matarrese}}, \bibinfo {author} {\bibfnamefont
  {N.}~\bibnamefont {Mauri}}, \bibinfo {author} {\bibfnamefont {J.~D.}\
  \bibnamefont {McEwen}}, \bibinfo {author} {\bibfnamefont {P.~R.}\
  \bibnamefont {Meinhold}}, \bibinfo {author} {\bibfnamefont {A.}~\bibnamefont
  {Melchiorri}}, \bibinfo {author} {\bibfnamefont {A.}~\bibnamefont
  {Mennella}}, \bibinfo {author} {\bibfnamefont {M.}~\bibnamefont
  {Migliaccio}}, \bibinfo {author} {\bibfnamefont {M.}~\bibnamefont {Millea}},
  \bibinfo {author} {\bibfnamefont {S.}~\bibnamefont {Mitra}}, \bibinfo
  {author} {\bibfnamefont {M.-A.}\ \bibnamefont {Miville-Desch{\^e}nes}},
  \bibinfo {author} {\bibfnamefont {D.}~\bibnamefont {Molinari}}, \bibinfo
  {author} {\bibfnamefont {L.}~\bibnamefont {Montier}}, \bibinfo {author}
  {\bibfnamefont {G.}~\bibnamefont {Morgante}}, \bibinfo {author}
  {\bibfnamefont {A.}~\bibnamefont {Moss}}, \bibinfo {author} {\bibfnamefont
  {P.}~\bibnamefont {Natoli}}, \bibinfo {author} {\bibfnamefont {H.~U.}\
  \bibnamefont {N{\o}rgaard-Nielsen}}, \bibinfo {author} {\bibfnamefont
  {L.}~\bibnamefont {Pagano}}, \bibinfo {author} {\bibfnamefont
  {D.}~\bibnamefont {Paoletti}}, \bibinfo {author} {\bibfnamefont
  {B.}~\bibnamefont {Partridge}}, \bibinfo {author} {\bibfnamefont
  {G.}~\bibnamefont {Patanchon}}, \bibinfo {author} {\bibfnamefont {H.~V.}\
  \bibnamefont {Peiris}}, \bibinfo {author} {\bibfnamefont {F.}~\bibnamefont
  {Perrotta}}, \bibinfo {author} {\bibfnamefont {V.}~\bibnamefont {Pettorino}},
  \bibinfo {author} {\bibfnamefont {F.}~\bibnamefont {Piacentini}}, \bibinfo
  {author} {\bibfnamefont {L.}~\bibnamefont {Polastri}}, \bibinfo {author}
  {\bibfnamefont {G.}~\bibnamefont {Polenta}}, \bibinfo {author} {\bibfnamefont
  {J.-L.}\ \bibnamefont {Puget}}, \bibinfo {author} {\bibfnamefont {J.~P.}\
  \bibnamefont {Rachen}}, \bibinfo {author} {\bibfnamefont {M.}~\bibnamefont
  {Reinecke}}, \bibinfo {author} {\bibfnamefont {M.}~\bibnamefont
  {Remazeilles}}, \bibinfo {author} {\bibfnamefont {A.}~\bibnamefont {Renzi}},
  \bibinfo {author} {\bibfnamefont {G.}~\bibnamefont {Rocha}}, \bibinfo
  {author} {\bibfnamefont {C.}~\bibnamefont {Rosset}}, \bibinfo {author}
  {\bibfnamefont {G.}~\bibnamefont {Roudier}}, \bibinfo {author} {\bibfnamefont
  {J.~A.}\ \bibnamefont {Rubi{\~n}o-Mart{\'\i}n}}, \bibinfo {author}
  {\bibfnamefont {B.}~\bibnamefont {Ruiz-Granados}}, \bibinfo {author}
  {\bibfnamefont {L.}~\bibnamefont {Salvati}}, \bibinfo {author} {\bibfnamefont
  {M.}~\bibnamefont {Sandri}}, \bibinfo {author} {\bibfnamefont
  {M.}~\bibnamefont {Savelainen}}, \bibinfo {author} {\bibfnamefont
  {D.}~\bibnamefont {Scott}}, \bibinfo {author} {\bibfnamefont {E.~P.~S.}\
  \bibnamefont {Shellard}}, \bibinfo {author} {\bibfnamefont {C.}~\bibnamefont
  {Sirignano}}, \bibinfo {author} {\bibfnamefont {G.}~\bibnamefont {Sirri}},
  \bibinfo {author} {\bibfnamefont {L.~D.}\ \bibnamefont {Spencer}}, \bibinfo
  {author} {\bibfnamefont {R.}~\bibnamefont {Sunyaev}}, \bibinfo {author}
  {\bibfnamefont {A.-S.}\ \bibnamefont {Suur-Uski}}, \bibinfo {author}
  {\bibfnamefont {J.~A.}\ \bibnamefont {Tauber}}, \bibinfo {author}
  {\bibfnamefont {D.}~\bibnamefont {Tavagnacco}}, \bibinfo {author}
  {\bibfnamefont {M.}~\bibnamefont {Tenti}}, \bibinfo {author} {\bibfnamefont
  {L.}~\bibnamefont {Toffolatti}}, \bibinfo {author} {\bibfnamefont
  {M.}~\bibnamefont {Tomasi}}, \bibinfo {author} {\bibfnamefont
  {T.}~\bibnamefont {Trombetti}}, \bibinfo {author} {\bibfnamefont
  {L.}~\bibnamefont {Valenziano}}, \bibinfo {author} {\bibfnamefont
  {J.}~\bibnamefont {Valiviita}}, \bibinfo {author} {\bibfnamefont
  {B.}~\bibnamefont {Van~Tent}}, \bibinfo {author} {\bibfnamefont
  {L.}~\bibnamefont {Vibert}}, \bibinfo {author} {\bibfnamefont
  {P.}~\bibnamefont {Vielva}}, \bibinfo {author} {\bibfnamefont
  {F.}~\bibnamefont {Villa}}, \bibinfo {author} {\bibfnamefont
  {N.}~\bibnamefont {Vittorio}}, \bibinfo {author} {\bibfnamefont {B.~D.}\
  \bibnamefont {Wandelt}}, \bibinfo {author} {\bibfnamefont {I.~K.}\
  \bibnamefont {Wehus}}, \bibinfo {author} {\bibfnamefont {M.}~\bibnamefont
  {White}}, \bibinfo {author} {\bibfnamefont {S.~D.~M.}\ \bibnamefont {White}},
  \bibinfo {author} {\bibfnamefont {A.}~\bibnamefont {Zacchei}}, \ and\
  \bibinfo {author} {\bibfnamefont {A.}~\bibnamefont {Zonca}},\ }\href
  {\doibase 10.1051/0004-6361/201833910} {\bibfield  {journal} {\bibinfo
  {journal} {Astron. Astrophys. Suppl. Ser.}\ }\textbf {\bibinfo {volume}
  {641}},\ \bibinfo {pages} {A6} (\bibinfo {year}
  {2020}{\natexlab{a}})}\BibitemShut {NoStop}%
\bibitem [{\citenamefont {Abbott}\ \emph
  {et~al.}(2021{\natexlab{d}})\citenamefont {Abbott} \emph
  {et~al.}}]{LIGOScientific:2020kqk}%
  \BibitemOpen
  \bibfield  {author} {\bibinfo {author} {\bibfnamefont {R.}~\bibnamefont
  {Abbott}} \emph {et~al.} (\bibinfo {collaboration} {LIGO Scientific,
  Virgo}),\ }\href {\doibase 10.3847/2041-8213/abe949} {\bibfield  {journal}
  {\bibinfo  {journal} {Astrophys. J. Lett.}\ }\textbf {\bibinfo {volume}
  {913}},\ \bibinfo {pages} {L7} (\bibinfo {year} {2021}{\natexlab{d}})},\
  \Eprint {http://arxiv.org/abs/2010.14533} {arXiv:2010.14533 [astro-ph.HE]}
  \BibitemShut {NoStop}%
\bibitem [{\citenamefont {Alford}\ \emph {et~al.}(2005)\citenamefont {Alford},
  \citenamefont {Braby}, \citenamefont {Paris},\ and\ \citenamefont
  {Reddy}}]{Alford:2004pf}%
  \BibitemOpen
  \bibfield  {author} {\bibinfo {author} {\bibfnamefont {M.}~\bibnamefont
  {Alford}}, \bibinfo {author} {\bibfnamefont {M.}~\bibnamefont {Braby}},
  \bibinfo {author} {\bibfnamefont {M.~W.}\ \bibnamefont {Paris}}, \ and\
  \bibinfo {author} {\bibfnamefont {S.}~\bibnamefont {Reddy}},\ }\href
  {\doibase 10.1086/430902} {\bibfield  {journal} {\bibinfo  {journal}
  {Astrophys. J.}\ }\textbf {\bibinfo {volume} {629}},\ \bibinfo {pages} {969}
  (\bibinfo {year} {2005})},\ \Eprint {http://arxiv.org/abs/nucl-th/0411016}
  {arXiv:nucl-th/0411016} \BibitemShut {NoStop}%
\bibitem [{\citenamefont {Akmal}\ \emph {et~al.}(1998)\citenamefont {Akmal},
  \citenamefont {Pandharipande},\ and\ \citenamefont
  {Ravenhall}}]{Akmal:1998cf}%
  \BibitemOpen
  \bibfield  {author} {\bibinfo {author} {\bibfnamefont {A.}~\bibnamefont
  {Akmal}}, \bibinfo {author} {\bibfnamefont {V.~R.}\ \bibnamefont
  {Pandharipande}}, \ and\ \bibinfo {author} {\bibfnamefont {D.~G.}\
  \bibnamefont {Ravenhall}},\ }\href {\doibase 10.1103/PhysRevC.58.1804}
  {\bibfield  {journal} {\bibinfo  {journal} {Phys. Rev. C}\ }\textbf {\bibinfo
  {volume} {58}},\ \bibinfo {pages} {1804} (\bibinfo {year} {1998})},\ \Eprint
  {http://arxiv.org/abs/nucl-th/9804027} {arXiv:nucl-th/9804027} \BibitemShut
  {NoStop}%
\bibitem [{\citenamefont {Banik}\ \emph {et~al.}(2014)\citenamefont {Banik},
  \citenamefont {Hempel},\ and\ \citenamefont {Bandyopadhyay}}]{Banik:2014qja}%
  \BibitemOpen
  \bibfield  {author} {\bibinfo {author} {\bibfnamefont {S.}~\bibnamefont
  {Banik}}, \bibinfo {author} {\bibfnamefont {M.}~\bibnamefont {Hempel}}, \
  and\ \bibinfo {author} {\bibfnamefont {D.}~\bibnamefont {Bandyopadhyay}},\
  }\href {\doibase 10.1088/0067-0049/214/2/22} {\bibfield  {journal} {\bibinfo
  {journal} {Astrophys. J. Suppl.}\ }\textbf {\bibinfo {volume} {214}},\
  \bibinfo {pages} {22} (\bibinfo {year} {2014})},\ \Eprint
  {http://arxiv.org/abs/1404.6173} {arXiv:1404.6173 [astro-ph.HE]} \BibitemShut
  {NoStop}%
\bibitem [{\citenamefont {Typel}\ \emph {et~al.}(2010)\citenamefont {Typel},
  \citenamefont {Ropke}, \citenamefont {Klahn}, \citenamefont {Blaschke},\ and\
  \citenamefont {Wolter}}]{Typel:2009sy}%
  \BibitemOpen
  \bibfield  {author} {\bibinfo {author} {\bibfnamefont {S.}~\bibnamefont
  {Typel}}, \bibinfo {author} {\bibfnamefont {G.}~\bibnamefont {Ropke}},
  \bibinfo {author} {\bibfnamefont {T.}~\bibnamefont {Klahn}}, \bibinfo
  {author} {\bibfnamefont {D.}~\bibnamefont {Blaschke}}, \ and\ \bibinfo
  {author} {\bibfnamefont {H.~H.}\ \bibnamefont {Wolter}},\ }\href {\doibase
  10.1103/PhysRevC.81.015803} {\bibfield  {journal} {\bibinfo  {journal} {Phys.
  Rev. C}\ }\textbf {\bibinfo {volume} {81}},\ \bibinfo {pages} {015803}
  (\bibinfo {year} {2010})},\ \Eprint {http://arxiv.org/abs/0908.2344}
  {arXiv:0908.2344 [nucl-th]} \BibitemShut {NoStop}%
\bibitem [{\citenamefont {Lackey}\ \emph {et~al.}(2006)\citenamefont {Lackey},
  \citenamefont {Nayyar},\ and\ \citenamefont {Owen}}]{Lackey:2005tk}%
  \BibitemOpen
  \bibfield  {author} {\bibinfo {author} {\bibfnamefont {B.~D.}\ \bibnamefont
  {Lackey}}, \bibinfo {author} {\bibfnamefont {M.}~\bibnamefont {Nayyar}}, \
  and\ \bibinfo {author} {\bibfnamefont {B.~J.}\ \bibnamefont {Owen}},\ }\href
  {\doibase 10.1103/PhysRevD.73.024021} {\bibfield  {journal} {\bibinfo
  {journal} {Phys. Rev. D}\ }\textbf {\bibinfo {volume} {73}},\ \bibinfo
  {pages} {024021} (\bibinfo {year} {2006})},\ \Eprint
  {http://arxiv.org/abs/astro-ph/0507312} {arXiv:astro-ph/0507312} \BibitemShut
  {NoStop}%
\bibitem [{\citenamefont {Lattimer}\ and\ \citenamefont
  {Swesty}(1991)}]{Lattimer:1991nc}%
  \BibitemOpen
  \bibfield  {author} {\bibinfo {author} {\bibfnamefont {J.~M.}\ \bibnamefont
  {Lattimer}}\ and\ \bibinfo {author} {\bibfnamefont {F.~D.}\ \bibnamefont
  {Swesty}},\ }\href {\doibase 10.1016/0375-9474(91)90452-C} {\bibfield
  {journal} {\bibinfo  {journal} {Nucl. Phys. A}\ }\textbf {\bibinfo {volume}
  {535}},\ \bibinfo {pages} {331} (\bibinfo {year} {1991})}\BibitemShut
  {NoStop}%
\bibitem [{\citenamefont {Steiner}\ \emph {et~al.}(2013)\citenamefont
  {Steiner}, \citenamefont {Hempel},\ and\ \citenamefont
  {Fischer}}]{Steiner:2012rk}%
  \BibitemOpen
  \bibfield  {author} {\bibinfo {author} {\bibfnamefont {A.~W.}\ \bibnamefont
  {Steiner}}, \bibinfo {author} {\bibfnamefont {M.}~\bibnamefont {Hempel}}, \
  and\ \bibinfo {author} {\bibfnamefont {T.}~\bibnamefont {Fischer}},\ }\href
  {\doibase 10.1088/0004-637X/774/1/17} {\bibfield  {journal} {\bibinfo
  {journal} {Astrophys. J.}\ }\textbf {\bibinfo {volume} {774}},\ \bibinfo
  {pages} {17} (\bibinfo {year} {2013})},\ \Eprint
  {http://arxiv.org/abs/1207.2184} {arXiv:1207.2184 [astro-ph.SR]} \BibitemShut
  {NoStop}%
\bibitem [{\citenamefont {Douchin}\ and\ \citenamefont
  {Haensel}(2001)}]{Douchin:2001sv}%
  \BibitemOpen
  \bibfield  {author} {\bibinfo {author} {\bibfnamefont {F.}~\bibnamefont
  {Douchin}}\ and\ \bibinfo {author} {\bibfnamefont {P.}~\bibnamefont
  {Haensel}},\ }\href {\doibase 10.1051/0004-6361:20011402} {\bibfield
  {journal} {\bibinfo  {journal} {Astron. Astrophys.}\ }\textbf {\bibinfo
  {volume} {380}},\ \bibinfo {pages} {151} (\bibinfo {year} {2001})},\ \Eprint
  {http://arxiv.org/abs/astro-ph/0111092} {arXiv:astro-ph/0111092} \BibitemShut
  {NoStop}%
\bibitem [{\citenamefont {Godzieba}\ \emph {et~al.}(2021)\citenamefont
  {Godzieba}, \citenamefont {Gamba}, \citenamefont {Radice},\ and\
  \citenamefont {Bernuzzi}}]{Godzieba:2020bbz}%
  \BibitemOpen
  \bibfield  {author} {\bibinfo {author} {\bibfnamefont {D.~A.}\ \bibnamefont
  {Godzieba}}, \bibinfo {author} {\bibfnamefont {R.}~\bibnamefont {Gamba}},
  \bibinfo {author} {\bibfnamefont {D.}~\bibnamefont {Radice}}, \ and\ \bibinfo
  {author} {\bibfnamefont {S.}~\bibnamefont {Bernuzzi}},\ }\href {\doibase
  10.1103/PhysRevD.103.063036} {\bibfield  {journal} {\bibinfo  {journal}
  {Phys. Rev. D}\ }\textbf {\bibinfo {volume} {103}},\ \bibinfo {pages}
  {063036} (\bibinfo {year} {2021})},\ \Eprint
  {http://arxiv.org/abs/2012.12151} {arXiv:2012.12151 [astro-ph.HE]}
  \BibitemShut {NoStop}%
\bibitem [{\citenamefont {Aghanim}\ \emph
  {et~al.}(2020{\natexlab{b}})\citenamefont {Aghanim} \emph
  {et~al.}}]{Planck:2018vyg}%
  \BibitemOpen
  \bibfield  {author} {\bibinfo {author} {\bibfnamefont {N.}~\bibnamefont
  {Aghanim}} \emph {et~al.} (\bibinfo {collaboration} {Planck}),\ }\href
  {\doibase 10.1051/0004-6361/201833910} {\bibfield  {journal} {\bibinfo
  {journal} {Astron. Astrophys.}\ }\textbf {\bibinfo {volume} {641}},\ \bibinfo
  {pages} {A6} (\bibinfo {year} {2020}{\natexlab{b}})},\ \bibinfo {note}
  {[Erratum: Astron.Astrophys. 652, C4 (2021)]},\ \Eprint
  {http://arxiv.org/abs/1807.06209} {arXiv:1807.06209 [astro-ph.CO]}
  \BibitemShut {NoStop}%
\bibitem [{\citenamefont {Chandrasekhar}(1985)}]{Chandrasekhar:1985kt}%
  \BibitemOpen
  \bibfield  {author} {\bibinfo {author} {\bibfnamefont {S.}~\bibnamefont
  {Chandrasekhar}},\ }\href@noop {} {\emph {\bibinfo {title} {{The mathematical
  theory of black holes}}}}\ (\bibinfo {year} {1985})\BibitemShut {NoStop}%
\bibitem [{\citenamefont {Borhanian}(2021)}]{Borhanian:2020ypi}%
  \BibitemOpen
  \bibfield  {author} {\bibinfo {author} {\bibfnamefont {S.}~\bibnamefont
  {Borhanian}},\ }\href {\doibase 10.1088/1361-6382/ac1618} {\bibfield
  {journal} {\bibinfo  {journal} {Class. Quant. Grav.}\ }\textbf {\bibinfo
  {volume} {38}},\ \bibinfo {pages} {175014} (\bibinfo {year} {2021})},\
  \Eprint {http://arxiv.org/abs/2010.15202} {arXiv:2010.15202 [gr-qc]}
  \BibitemShut {NoStop}%
\bibitem [{\citenamefont {Harry}\ and\ \citenamefont
  {Hinderer}(2018)}]{Harry:2018hke}%
  \BibitemOpen
  \bibfield  {author} {\bibinfo {author} {\bibfnamefont {I.}~\bibnamefont
  {Harry}}\ and\ \bibinfo {author} {\bibfnamefont {T.}~\bibnamefont
  {Hinderer}},\ }\href {\doibase 10.1088/1361-6382/aac7e3} {\bibfield
  {journal} {\bibinfo  {journal} {Class. Quant. Grav.}\ }\textbf {\bibinfo
  {volume} {35}},\ \bibinfo {pages} {145010} (\bibinfo {year} {2018})},\
  \Eprint {http://arxiv.org/abs/1801.09972} {arXiv:1801.09972 [gr-qc]}
  \BibitemShut {NoStop}%
\bibitem [{\citenamefont {Lattimer}\ and\ \citenamefont
  {Prakash}(2001)}]{Lattimer:2000nx}%
  \BibitemOpen
  \bibfield  {author} {\bibinfo {author} {\bibfnamefont {J.~M.}\ \bibnamefont
  {Lattimer}}\ and\ \bibinfo {author} {\bibfnamefont {M.}~\bibnamefont
  {Prakash}},\ }\href {\doibase 10.1086/319702} {\bibfield  {journal} {\bibinfo
   {journal} {Astrophys. J.}\ }\textbf {\bibinfo {volume} {550}},\ \bibinfo
  {pages} {426} (\bibinfo {year} {2001})},\ \Eprint
  {http://arxiv.org/abs/astro-ph/0002232} {arXiv:astro-ph/0002232} \BibitemShut
  {NoStop}%
\bibitem [{\citenamefont {Yagi}(2014)}]{Yagi:2013sva}%
  \BibitemOpen
  \bibfield  {author} {\bibinfo {author} {\bibfnamefont {K.}~\bibnamefont
  {Yagi}},\ }\href {\doibase 10.1103/PhysRevD.89.043011} {\bibfield  {journal}
  {\bibinfo  {journal} {Phys. Rev. D}\ }\textbf {\bibinfo {volume} {89}},\
  \bibinfo {pages} {043011} (\bibinfo {year} {2014})},\ \bibinfo {note}
  {[Erratum: Phys.Rev.D 96, 129904 (2017), Erratum: Phys.Rev.D 97, 129901
  (2018)]},\ \Eprint {http://arxiv.org/abs/1311.0872} {arXiv:1311.0872 [gr-qc]}
  \BibitemShut {NoStop}%
\bibitem [{\citenamefont {Bernuzzi}\ and\ \citenamefont
  {Nagar}(2008)}]{Bernuzzi:2008fu}%
  \BibitemOpen
  \bibfield  {author} {\bibinfo {author} {\bibfnamefont {S.}~\bibnamefont
  {Bernuzzi}}\ and\ \bibinfo {author} {\bibfnamefont {A.}~\bibnamefont
  {Nagar}},\ }\href {\doibase 10.1103/PhysRevD.78.024024} {\bibfield  {journal}
  {\bibinfo  {journal} {Phys. Rev.}\ }\textbf {\bibinfo {volume} {D78}},\
  \bibinfo {pages} {024024} (\bibinfo {year} {2008})},\ \Eprint
  {http://arxiv.org/abs/0803.3804} {arXiv:0803.3804 [gr-qc]} \BibitemShut
  {NoStop}%
%%CITATION = ARXIV:0803.3804;%%
\bibitem [{\citenamefont {Feigelson}\ and\ \citenamefont
  {Babu}(2012)}]{Feigelson:2012xg}%
  \BibitemOpen
  \bibfield  {author} {\bibinfo {author} {\bibfnamefont {E.~D.}\ \bibnamefont
  {Feigelson}}\ and\ \bibinfo {author} {\bibfnamefont {G.~J.}\ \bibnamefont
  {Babu}},\ }\href@noop {} {\  (\bibinfo {year} {2012})},\ \Eprint
  {http://arxiv.org/abs/1205.2064} {arXiv:1205.2064 [astro-ph.IM]} \BibitemShut
  {NoStop}%
\end{thebibliography}%
\end{document}